\documentclass[twoside,final,onecolumn,nolineno]{jfm}

\usepackage{xcolor}
\usepackage{graphicx}
\usepackage{booktabs}
\usepackage{tikz}
\usepackage[super]{nth}

\usepackage[lining,semibold]{libertine} 
\usepackage[T1]{fontenc} 
\usepackage[libertine,vvarbb]{newtxmath}
\useosf 

\DeclareRobustCommand\full  {\tikz[baseline=-0.6ex]\draw[thick] (0,0)--(0.5,0);}
\DeclareRobustCommand\dotted{\tikz[baseline=-0.6ex]\draw[thick,dotted] (0,0)--(0.54,0);}
\DeclareRobustCommand\dashed{\tikz[baseline=-0.6ex]\draw[thick,densely dashed] (0,0)--(0.54,0);}
\DeclareRobustCommand\dashdotted {\tikz[baseline=-0.6ex]\draw[thick, densely dash dot] (0,0)--(0.5,0);}

\usetikzlibrary{calc}
\usetikzlibrary{positioning, arrows.meta, fit}
\usetikzlibrary{mindmap}

\usepackage{natbib}
\usepackage{hyperref}
\hypersetup{
  colorlinks = true,
  urlcolor   = black,
  citecolor  = black,
  allcolors = green!50!black,
}

\usepackage{orcidlink}

\newcommand{\RomanNumeralCaps}[1]
\nolinenumbers
\DeclareMathOperator{\csch}{csch}

\usepackage{siunitx}
\renewcommand*{\vec}[1]{\boldsymbol{#1}}
\newcommand{\del}{{\nabla}}
\newcommand{\conj}[1]{{{#1}^*}}
\newcommand{\paramF  }{{\color{black}F}}

\newcommand{\richardson}{\mathrm{\color{black}Ri}}
\newcommand{\atwood}{\mathrm{\color{black}At}}
\newcommand{\reynolds}{\mathrm{\color{black}Re}}

\newcommand{\weber}{\mathrm{\color{black}We}}
\newcommand{\schmidt}{\mathrm{\color{black}Sc}}

\newcommand{\hstar}{{\color{black}h^*_\init}}

\newcommand{\detune}{\mathsf{\color{black}\Delta}}

\newcommand{\KE}{\mathsf{\color{black}E_k}}
\newcommand{\GPE}{\mathsf{\color{black}E_p}}

\newcommand{\BPE}{\mathsf{\color{black}E_b}}
\newcommand{\Kwave}{{\color{black}\widetilde{\mathrm{K}}}}
\newcommand{\Kturb}{{\color{black}\mathrm{K}'}}
\newcommand{\Shear}{{\color{black}\mathrm{S}}}
\newcommand{\ShearWave}{{\color{black}\widetilde{\mathrm{S}}}}
\newcommand{\ShearTurb}{{\color{black}\mathrm{S}'}}

\newcommand{\init}{{\color{black}\mathrm{init}}}
\newcommand{\final}{{\color{black}\mathrm{end}}}
\newcommand{\mix}{{\color{black}\mathrm{mix}}}
\newcommand{\mixed}{{\color{black}\mathrm{mixed}}}
\newcommand{\surf}{{\color{black}\mathrm{surf}}}

\newcommand{\bin}{{\color{black}\mathrm{bin}}}

\newcommand{\turb}{t}
\newcommand{\steady}{{\color{black}\mathrm{steady}}}

\newcommand{\cm}{\unit{\centi\meter}}
\newcommand{\kgm}{\unit{\kilogram\per\cubic\meter}}
\newcommand{\rad}{\unit{\radian\per\second}}

\newcommand{\tf}{t_\mathrm{f}}
\newcommand{\ts}{t_\mathrm{s}}
\newcommand{\ddfast}[1]{\frac{\partial^2 #1}{\partial\tf^2}}
\newcommand{\dfast}[1]{\frac{\partial #1 }{\partial\tf}}  
\newcommand{\dslow}[1]{\frac{\partial #1 }{\partial\ts}}  
\newcommand{\dmixed}[1]{\frac{\partial^2 #1}{\partial\tf\partial\ts}}

\newcommand{\Dslow}[1]{\frac{d #1 }{d \ts}}

\author{%
  Andrés Castillo-Castellanos\aff{1}
  \corresp{\email{andres.castillo\_castellanos@ens-paris-saclay.fr}},
  Benoît-Joseph Gréa\aff{2,3}
  \corresp{\email{benoit-joseph.grea@cea.fr}},
  Antoine Briard\aff{2}
  \and
  Louis Gostiaux\aff{4}
}

\affiliation{%
  \aff{1} Université Paris-Saclay, CNRS, ENS Paris-Saclay, Centre Borelli, LRC-MESO, F-91190, Gif-sur-Yvette, France
  \aff{2} CEA, DAM, DIF, 91297 Arpajon, France
  \aff{3} Université Paris-Saclay, CEA, Laboratoire Matière en Condition Extrême, 91680
  Bruyères-le-Châtel, France
  \aff{4} Univ Lyon, École Centrale de Lyon, INSA Lyon, Université Claude
  Bernard Lyon I, CNRS, Laboratoire de Mécanique des Fluides et d’Acoustique,
  UMR 5509, F-69134, Ecully, France
}
\usepackage{setspace}
\usepackage{longtable}
\title{Mixing induced by Faraday surface waves}

\begin{document}

\maketitle

\begin{abstract}
  We investigate how surface waves enhance mixing across the interface between
  two miscible fluids with a small density contrast. Imposing a vertical,
  time-periodic acceleration, we excite Faraday waves both experimentally and
  numerically. In systems with a shallow density gradient, these standing waves
  advect the interface and can trigger secondary instabilities. When driven
  beyond the linear regime, large Faraday crests collapse to form cavities,
  injecting bubbles and lighter fluid deep into the heavier layer. Together,
  these mechanisms gradually homogenize the upper layer, diminish the
  interfacial density jump, and drive the interface downward until it decouples
  from surface forcing. We report a non-monotonic mixing rate—first increasing
  as the interfacial energy barrier lowers, then decreasing as less energy is
  injected into the weakened surface—revealing a balance between barrier
  reduction and energy input. Based on these observations, we introduce a
  one-dimensional model incorporating a turbulent diffusivity coefficient that
  depends on depth and the internal Richardson number, which captures the
  qualitative evolution of the system.
\end{abstract}


\section{Introduction}
\label{section:introduction}

Mixing across fluid density interfaces plays a critical role in both natural
phenomena and industrial operations~\citep{Villermaux1999,Ibrahim_2005}. In
nature, breaking ocean surface waves generate turbulence, promoting mixing at
and below the surface~\citep{Thorpe2005, Mostert2022}. In industrial contexts,
sloshing during the storage and transport of liquefied natural gas (LNG) is
known to increase the evaporation rates by disrupting the stratification of
fluid layers~\citep{Behruzi2017}. This uncontrolled spontaneous mixing can lead
to a sudden rise in pressure~\citep{Scurlock2016}. While such mixing is
typically attributed to convective instabilities and double diffusive
convection, this study investigates an alternative mechanism where interfacial
waves, upon interacting with a free surface, break and promote turbulent mixing.

The interaction between a miscible and an immiscible fluid layer offers an
interesting setting, where the properties of the internal layer evolve due to
molecular diffusion and turbulent entrainment. \cite{Linden_1973} argued that
discrete turbulent motions may transfer their kinetic energy into potential
energy by deforming the interface and promoting mixing between the two layers.
For two-layer fluids, the interface acts as a barrier to turbulence,
transferring energy from the vertical to the horizontal scales (see, for
instance,~\cite{Jacobs2005, PoulainZarcos2022}). In this sense, the entrainment
process can be described using parameters local to the interface,
recontextualising previous experiments by~\cite{Turner_1965}.

Over time, the mixing layer deepens, driven by two competing mechanisms. First,
the density contrast across the interface decreases as the near-surface region
becomes increasingly homogeneous, reducing the energy required for irreversible
mixing, a process reminiscent of penetrative convection (see, for
instance,~\cite{Dorel2023}). Second, the energy injected into the interface
diminishes with distance from the free surface, analogous to the behaviour of
impinging jets, rising plumes, or oscillating grid turbulence in stratified
fluids~\citep{Rouse1955, Morton1956, Linden1975, Hopfinger1976, Cotel_1996,
Herault_2018, Vaux_2021}. We shall see that eventually, as the available energy
decreases faster than the energy needed for mixing, the interface is bound to
decouple from the surface dynamics and stall at a finite depth.

The thickening of the interface is predominantly caused by the transition to
turbulence, with molecular diffusion contributing to a lesser
extent~\citep{Zoueshtiagh2009, Benielli1998}. Modal competition may lead to
chaotic behaviour and generate mixing at small scales
\citep{Ciliberto1985,Briard2019,Briard2020}, while secondary instabilities often
trigger interfacial wave breaking, enhancing turbulent
mixing~\citep{Thorpe1968,Cavelier_2022}. We shall see that secondary
instabilities can break the interface's initial resistance to the surface
forcing, allowing the surface motion to more easily entrain the released fluid
upward, leading to short but intense mixing bursts.

\begin{figure}
\begin{center}
\begin{tikzpicture}
  \node[anchor=south] (clean) at (0,0.6) {
    \includegraphics[width=0.4\linewidth, trim=12 15 15 15, clip]{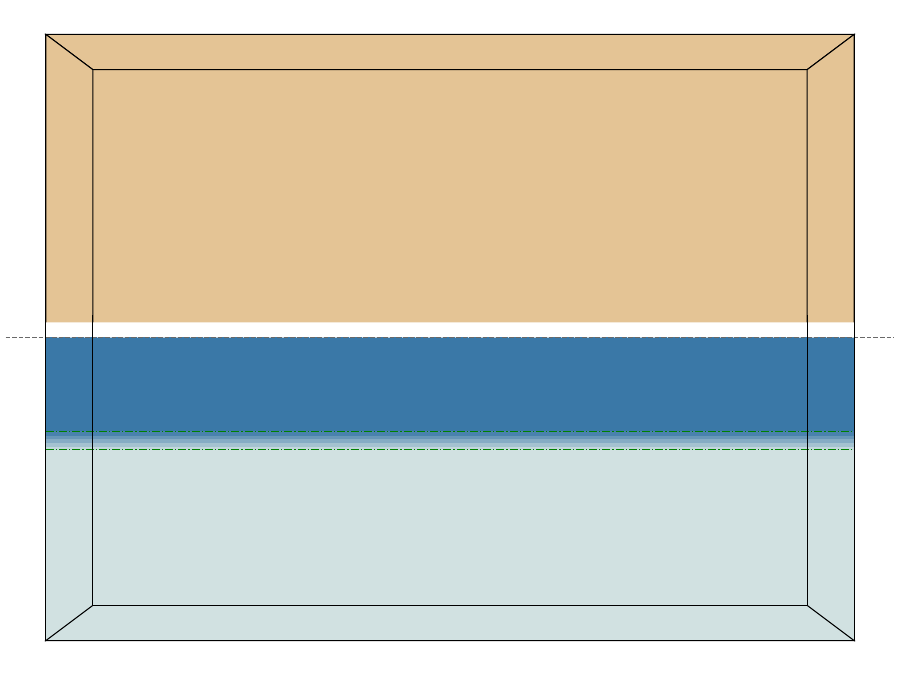}
  };
  \node[anchor=north] (scheme) at (0,0) {
    \includegraphics[width=0.4\linewidth, trim=15 15 15 15, clip]{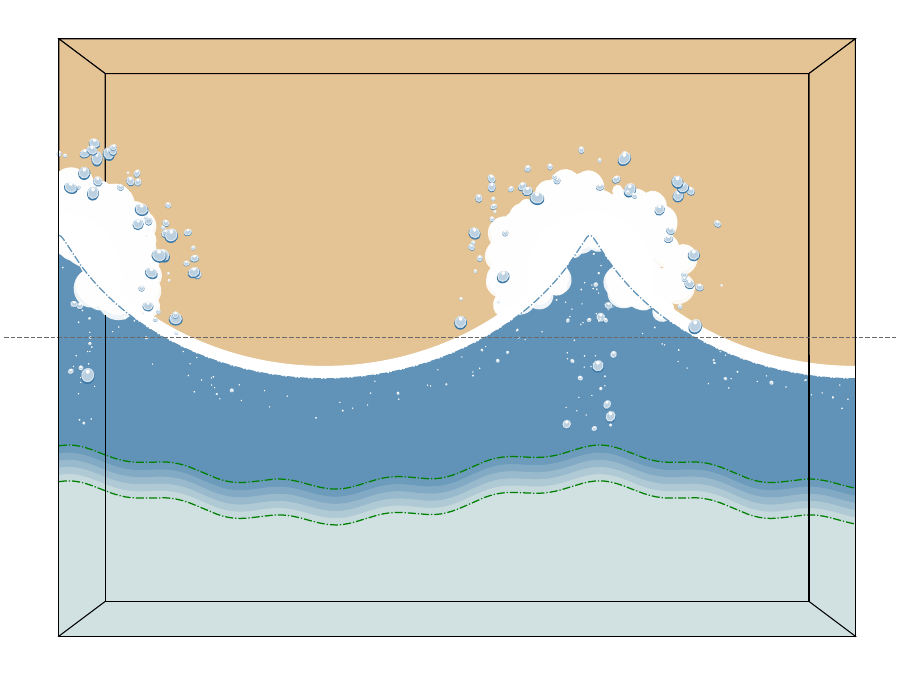}
  };
  \node[font=\scriptsize] at ($(scheme.center) + (0, 1.50)$) {air};
  \node[font=\scriptsize] at ($(scheme.center) + (0,-1.50)$) {salt-water};
  \node[font=\scriptsize] at ($(scheme.center) + (0,-1.05)$) {mixing layer};
  \node[font=\scriptsize, above] at ($(scheme.center) + (0,-0.65)$) {mixed layer};
  \node[font=\tiny] at ($(scheme.center) + (0,-0.65)$) {(mixture of fresh- and salt-water)};

  \node[font=\scriptsize, anchor=east] at ($(scheme.east) + (-0.35,-0.85)$) {$\zeta_\mix$};
  \node[font=\scriptsize, anchor=east] at ($(scheme.east) + (-0.35, 0.25)$) {$\zeta_\surf$};


  \node[font=\scriptsize] at ($(clean.center) + (0, 1.0)$) {air};
  \node[font=\scriptsize] at ($(clean.center) + (0,-0.2)$) {fresh-water};
  \node[font=\scriptsize] at ($(clean.center) + (0,-1.2)$) {salt-water};

  \node[font=\tiny, below=0.2em] at ($(clean.center) + (0,1.0)$) {$\rho_3, \mu_3$};
  \node[font=\tiny, below=0.2em] at ($(clean.center) + (0,-0.2)$) {$\rho_2, \mu_2$};
  \node[font=\tiny, below=0.2em] at ($(clean.center) + (0,-1.2)$) {$\rho_1, \mu_1$};

  \node[font=\scriptsize, anchor=east] at ($(clean.east) + (-0.8,-0.65+0.07)$) {$\blacktriangledown$};
  \node[font=\scriptsize, anchor=east] at ($(clean.east) + (-0.8, 0.00+0.07)$) {$\blacktriangledown$};
  \node[font=\scriptsize, anchor=east, above] at ($(clean.east) + (-0.8,-0.65+0.07)$) {$\zeta_\mix$};
  \node[font=\scriptsize, anchor=east, above] at ($(clean.east) + (-0.8, 0.00+0.07)$) {$\zeta_\surf$};

  \draw[-latex, thick] ($(clean.west) + (0, 0.0)$) -- ($(clean.east) + (0.1, 0.0)$) node[below] {$x$};
  \draw[-latex, thick] ($(clean.south west) + (0.25, 0.0)$) -- ($(clean.north west) + (0.25, 0.2)$) node[below right] {$z$};
  \draw[densely dashed, thick] ($(clean.west) + (0.25, -0.65)$) -- ($(clean.east) + (-0.2, -0.65)$) ;

  \draw[-latex, thick] ($(scheme.west) + (0, 0.0)$) -- ($(scheme.east) + (0.1, 0.0)$) node[below] {$x$};
  \draw[-latex, thick] ($(scheme.south west) + (0.28, 0.0)$) -- ($(scheme.north west) + (0.28, 0.2)$) node[below right] {$z$};

  \draw[-] ($(clean.south west) + ( 0.25,-0.05)$) -- ($(clean.south west) + ( 0.25, -0.25)$);
  \draw[-] ($(clean.south east) + (-0.20, 0.05)$) -- ($(clean.south east) + (-0.20, -0.25)$);
  \coordinate (left_dim) at ($(clean.south west) + (0.25, -0.20)$);
  \coordinate (right_dim) at ($(clean.south east) + (-0.20, -0.20)$);
  \coordinate (mid_dim) at ($(left_dim)!0.5!(right_dim)$);
  \draw[latex-] (left_dim) -- (mid_dim);
  \draw[latex-] (right_dim) -- (mid_dim);
  \node [anchor=center, fill=white, font=\scriptsize] (note5) at (mid_dim) {W};

  \draw[-] ($(clean.west) + (0.15,-0.65)$) -- ($(clean.west) + (-0.20,-0.65)$);
  \coordinate (top_H) at ($(clean.west) + (-0.15, 0.0)$);
  \coordinate (bot_H) at ($(clean.west) + (-0.15,-0.65)$);
  \coordinate (mid_H) at ($(top_H)!0.5!(bot_H)$);
  \draw[latex-] (top_H) -- ($(top_H) + (0,0.25)$);
  \draw[latex-] (bot_H) -- ($(bot_H) - (0,0.25)$);
  \node [anchor=center, fill=white, font=\scriptsize] at (mid_H) {$h_\init$};

  \draw[-] ($(scheme.west) + (0.18,-0.85)$) -- ($(scheme.west) + (-0.20,-0.85)$);
  \draw[-] ($(scheme.west) + (-0.05,-0.0)$) -- ($(scheme.west)  + (-0.20,-0.0)$);
  \coordinate (top_H) at ($(scheme.west) + (-0.16, 0.0)$);
  \coordinate (bot_H) at ($(scheme.west) + (-0.16,-0.85)$);
  \coordinate (mid_H) at ($(top_H)!0.5!(bot_H)$);
  \draw[latex-] (top_H) -- ($(top_H) + (0,0.25)$);
  \draw[latex-] (bot_H) -- ($(bot_H) - (0,0.25)$);
  \node [anchor=center, fill=white, font=\scriptsize] at (mid_H) {$h_\mixed$};

  \draw[-] ($(scheme.east) + (-0.10,-0.9)$) -- ($(scheme.east) + (0.15,-0.9)$);
  \draw[-] ($(scheme.east) + (-0.10,-1.2)$) -- ($(scheme.east) + (0.15,-1.2)$);
  \coordinate (top_H) at ($(scheme.east) + (0.10,-0.9)$);
  \coordinate (bot_H) at ($(scheme.east) + (0.10,-1.2)$);
  \coordinate (mid_H) at ($(top_H)!0.5!(bot_H)$);
  \draw[latex-] (top_H) -- ($(top_H) + (0,0.25)$);
  \draw[latex-] (bot_H) -- ($(bot_H) - (0,0.25)$);
  \node [anchor=center, font=\scriptsize] at (mid_H) {$L$};

  \draw[-] ($(clean.west) + (-0.05, 0.0)$) -- ($(clean.west) + (-0.55, 0.0)$);
  \draw[-] ($(clean.south west) + (0.2, 0.15)$) -- ($(clean.south west) + (-0.90, 0.15)$);
  \draw[-] ($(clean.north west) + (0.2,-0.15)$) -- ($(clean.north west) + (-0.90,-0.15)$);
  \coordinate (top_H) at ($(clean.west) + (-0.50,-0.0)$);
  \coordinate (bot_H) at ($(clean.south west) + (-0.50, 0.15)$);
  \coordinate (mid_H) at ($(top_H)!0.5!(bot_H)$);
  \draw[latex-] (top_H) -- (mid_H);
  \draw[latex-] (bot_H) -- (mid_H);
  \node [anchor=center, fill=white, font=\scriptsize] at (mid_H) {$H$};

  \coordinate (top_H) at ($(clean.north west) + (-0.85,-0.15)$);
  \coordinate (bot_H) at ($(clean.south west) + (-0.85, 0.15)$);
  \coordinate (mid_H) at ($(top_H)!0.5!(bot_H)$);
  \draw[latex-] (top_H) -- (mid_H);
  \draw[latex-] (bot_H) -- (mid_H);
  \node [anchor=center, fill=white, font=\scriptsize] at (mid_H) {$2H$};

  \draw[latex-latex, thick] ($(clean.south east) + (0.5,  0.0)$) -- ($(clean.south east) + (0.5, 2.0)$);
  \node [font=\tiny, align=center, fill=white] at ($(clean.south east) + (0.5, 1.0)$) {vertical external\\oscillations\\$a\cos(\omega t)$};

  \draw[latex-, thick] ($(clean.north east) + (0.5, -0.75)$) -- ($(clean.north east) + (0.5, -0.0)$);
  \node [anchor=west, font=\scriptsize] at ($(clean.north east) + (0.50, -0.35)$) {$g$};

\end{tikzpicture}\hspace{1em}
\begin{tikzpicture}[
    node distance=0.0cm and 0.0cm,
    block/.style={draw, fill=gray!10, thick, text width=0.3\linewidth, align=center, font=\fontsize{7pt}{8pt}\selectfont, inner sep=0.5em},
    container/.style={draw, thick, inner sep=0.25em},
    subcontainer/.style={draw, densely dotted, inner sep=0.5em},
    boxtitle/.style={font=\fontsize{8pt}{9pt}\selectfont\bfseries},
    subtitle/.style={font=\fontsize{7pt}{8pt}\selectfont\bfseries}
  ]
  \coordinate (base) at (0, 0);

  \node[boxtitle] (title_sd) at (base) {SURFACE DYNAMICS};
  \node[block, below= 0.0em of title_sd] (sd1) {Surface waves advect the internal stratification with $\omega/2$};
  \node[block, below= 0.5em of sd1] (sd2) {Breaking waves and splash-down inject turbulent energy into the system};
  \node[container, fit=(title_sd) (sd1) (sd2)] (sd_group) {};


  \node[boxtitle, below=1em of sd_group] (title_me) {STRATIFICATION DYNAMICS};

  \node[subtitle, below=0.5em of title_me] (title_st) {Over short timescales};
  \node[block, below=0.0em of title_st] (st1) {Secondary parametric instability due to local oscillating strain};
  \node[block, below=0.5em of st1] (st2) {For shallow depths, instability couples with surface dynamics (mixing bursts)};
  \node[subcontainer, fit=(title_st) (st1) (st2)] (st_group) {};

  \node[subtitle, below=0.75em of st_group] (title_lt) {Over longer timescales};
  \node[block, below=0.0em of title_lt] (lt1) {Gradual decoupling of the interface from the surface dynamics};
  \node[block, below=0.5em of lt1] (lt2) {Energy decays, entrainment barrier weakens, entrainment stalls};
  \node[subcontainer, fit=(title_lt) (lt1) (lt2)] (lt_group) {};

  \node[container, fit=(title_me) (st_group) (lt_group)] (me_group) {};


\end{tikzpicture}
\end{center}
\caption{
  Graphical abstract. 
  (Left panel) Experimental setup: three-layer stratified fluid subjected to
  vertical oscillations $a\cos(\omega t)$. Top image at rest; bottom image
  showing a breaking wave and the progression of the mixed layer. (Right
  panel) Summary of main findings: two mixing mechanisms (surface-injected
  turbulence and parametric instability) and short/long time effects on the 
  stratification. 
}
\label{Figure1:setup}
\end{figure}

In this study, we investigate the interaction between a free surface and a
miscible interface with a small density contrast through laboratory experiments
and direct numerical simulations (DNS) of a simplified configuration shown in
Figure \ref{Figure1:setup}. For this configuration, surface waves are generated
by applying a periodic vertical oscillation---of amplitude \(a\) and frequency
\(\omega\)---to the container as a result of the Faraday
instability~\citep{Faraday1831}. This configuration provides a large-amplitude
standing wave that can be tuned and characterised in detail, yet is persistent
enough to follow the slower evolution of the stratification. Although the
Faraday wave is not itself the focus of this work, understanding it is
essential to interpreting the mixing process. Below, we review the basic aspects
of the Faraday-wave literature necessary to state the problem and interpret our
results, while additional details are introduced alongside the corresponding
analyses (in Sections \ref{section:surface} to \ref{section:stratification:long}).

\begin{figure}
  \centering
  \begin{minipage}{0.63\linewidth}\color{white}
    \includegraphics[scale=0.28, trim=56.25 0 143.75 43.75, clip]{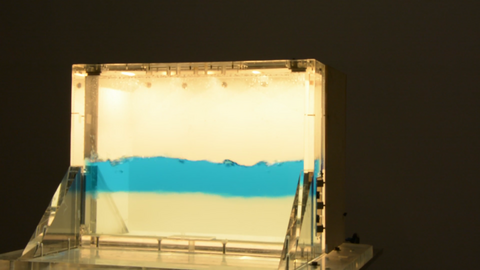}\hspace{-0.4em}
    \includegraphics[scale=0.28, trim=56.25 0 143.75 43.75, clip]{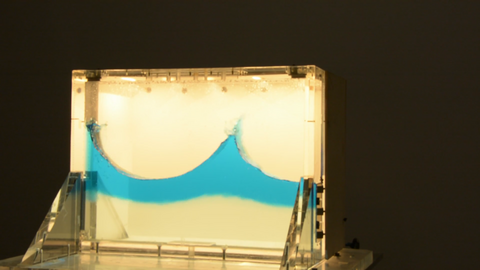}\hspace{-0.4em}
    \includegraphics[scale=0.28, trim=56.25 0 143.75 43.75, clip]{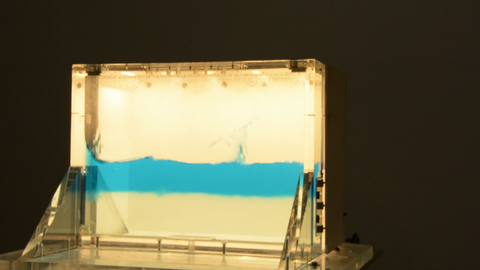}\hspace{-0.4em}

    \vspace{-2.3cm}
    ~(a) \hspace{7.5em} (b) \hspace{7.5em} (c) \par
    \vspace{1.75cm}

    \includegraphics[scale=0.28, trim=56.25 0 143.75 43.75, clip]{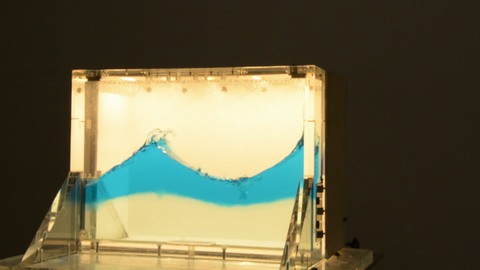}\hspace{-0.4em}
    \includegraphics[scale=0.28, trim=56.25 0 143.75 43.75, clip]{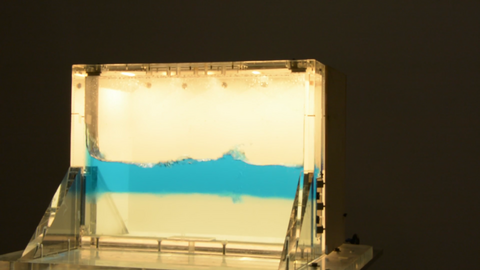}\hspace{-0.4em}
    \includegraphics[scale=0.28, trim=56.25 0 143.75 43.75, clip]{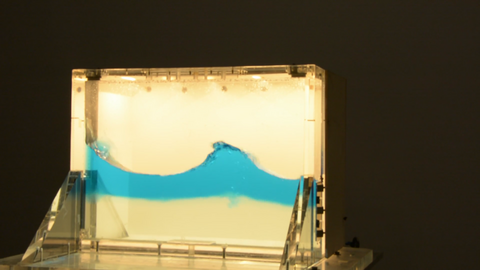}\hspace{-0.4em}

    \vspace{-2.3cm}
    ~(d) \hspace{7.5em} (e) \hspace{7.5em} (f) \par
    \vspace{1.8cm}
  \end{minipage}
  \begin{minipage}{0.25\linewidth}
    \includegraphics[width=\linewidth]{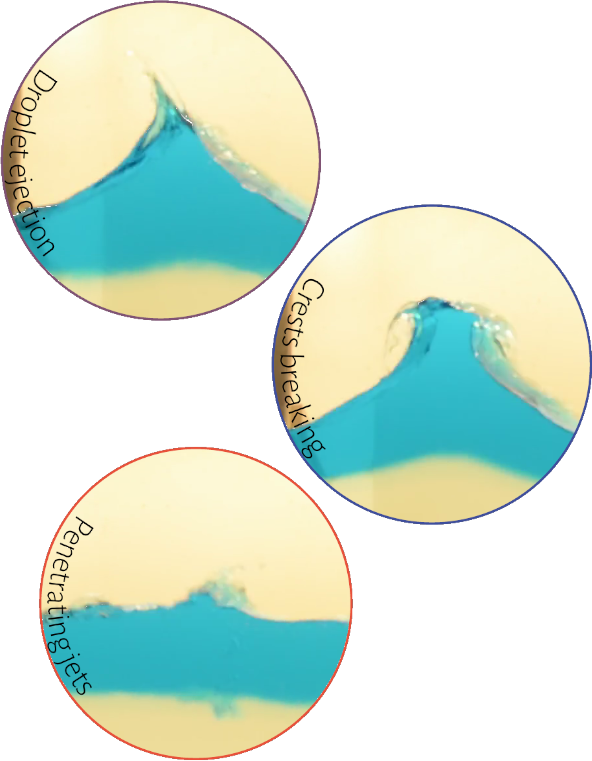}
  \end{minipage}
  \caption{
    A large-amplitude standing wave interacting with a stratified layer, with
    light fluid shown in blue. Subfigures (a) to (f) depict snapshots taken at
    regular intervals over three periods $T={2\pi}/{\omega}$. Insets in the
    right panel highlight characteristic features such as droplet ejection,
    breaking crests, and penetrating jets.
  }
  \label{Figure2:snapshot}
\end{figure}


In the classical Faraday experiment, free-surface motion is controlled by the
ratio between the forcing frequency \(\omega\) and the natural frequency of the
relevant modes of the interface \(\omega_0\), and by the external forcing term
\( \paramF = a \omega^2/g \) (where \(g\) is the gravitational acceleration).
For this system, linear stability analysis reveals unstable regions where the
response frequency is either subharmonic, which is the most amplified (in the
regime of interest here), or harmonic~\citep{Benjamin1954, Kumar_1994,
Ward2019}. In this work, we vary the forcing amplitude \(a\) while keeping the
frequency \(\omega\) fixed, so that the surface waves are always subharmonic and
have the same wavelength. This allows us to explore different kinds of surface
motion, from wavy, unbroken free surfaces to breaking waves. The breaking waves
can also lead to splashing droplets and jets from the collapse of large
cavities, which inject lighter fluid and turbulence into the stratified layer
(see, for instance, Figure \ref{Figure2:snapshot} and a video presented at the
APS/DFD Gallery of Fluid Motion \citep{CastilloCastellanos2023}). We also vary
the initial depth of the stratification and follow each experiment for several
hundred forcing periods to capture the long-term evolution of the system. To
access quantities that are difficult to measure in the laboratory, we complement
the experiments with direct numerical simulations (DNS).

We report two distinct mixing mechanisms arising from the interaction between
Faraday surface waves and the internal stratification. Turbulence injected from
the free surface drives interface displacement and entrainment, while secondary
parametric instabilities thicken the mixed layer. The oscillating strain
induced by the surface waves drives a secondary parametric instability of the
internal stratification, even at depths of order one wavelength below the
surface. The interface breaks down and re-stratifies into a thicker layer. Over
short timescales, especially when stratification is shallow, these instabilities
combine with near-surface turbulence and surface motion, leading to intense
mixing bursts: successive breaking of the interface enhances entrainment of
heavier fluid upward, while specific events such as wave splash-down and cavity
collapse inject lighter fluid downward. On longer timescales, the asymptotic
depth of the mixed layer varies non-monotonically with initial stratification
depth, reflecting the competition between energy decay and buoyancy barrier
reduction as entrainment proceeds. We capture the long-term behaviour
qualitatively using a one-dimensional $K$-$\varepsilon$ model.

The article is organised as follows. Section~\ref{section:setup} describes the
experimental setup and Section~\ref{section:governing} the governing equations
and direct numerical simulations, validated against linear theory during the
growth phase. Section~\ref{section:surface} characterises how the surface waves
inject energy and set the turbulent length scales. Sections
\ref{section:stratification:short} and \ref{section:stratification:long}
establish the two short-term mixing mechanisms and the long-term, non-monotonic
evolution of the mixed-layer depth. Section~\ref{sec:model} condenses these
ingredients into a one-dimensional $K$-$\varepsilon$ model that reproduces the
non-monotonic behaviour.

\section{Experimental Setup}
\label{section:setup}

We use a cuboidal tank with internal dimensions of \(94.6 \cm\) width, \(11
\cm\) depth, and \(67 \cm\) height, designed for free surface flow studies at
the Gaztransport and Technigaz\textregistered~(GTT) Motion Analysis and Testing
Laboratory. The tank is filled up to a height \(H\) with a mixture of salt water
and fresh water forming a two-layer system: a sharp miscible interface
separating fresh water (mixed with blue dye) and salt water, and a free surface
separating air and water (Figure~\ref{Figure1:setup}). We consider that
salt diffuses in water with molecular diffusivity \(\kappa_\mix\), while the free
surface is characterized by a surface tension coefficient
\(\sigma_\surf\). 

\subsection{Key Vocabulary and Interface Detection}
\label{section:setup:vocabulary}

We denote the properties of the fluid below (resp. above) the diffuse interface
by the subscript 1 (resp. 2), while the properties of the gas phase above the
free surface are denoted by the subscript 3. As shown in
figure~\ref{Figure1:setup}, it is convenient to distinguish between the
`mixed layer' --- the nearly homogeneous uppermost region formed by the history
of mixing --- and the `mixing layer' --- the zone in which mixing is currently
active~\citep{Brainerd1995}. The vertical positions of the interfaces separating
fluids 1 and 2 (resp. 2 and 3) are denoted by \(\zeta_\mix(x,y,t)\) (resp.
\(\zeta_\surf(x,y,t)\)). We also define the initial depth and density of the
mixed layer as \(h_\init\) and \(\rho_\init\), with their time-dependent
counterparts given by \(h_\mixed\) and \(\rho_\mixed\). 

In practice, the interfaces $\zeta_\mix$ and $\zeta_\surf$ are estimated from
recorded images by means of a segmentation algorithm discussed in
Appendix~\ref{section:appendix:experiments:camera}. From these, a mean depth is
estimated from
\begin{align}
  \label{eq2.1}
  h_\mixed(t) = \langle\zeta_\surf (x,t) - \zeta_\mix (x,t)\rangle_{x}
\end{align}
where \(\langle \cdot \rangle_{x}\) denotes the horizontal average, while
$\rho_\mixed$ is deduced from mass conservation.

The mixed-layer depth relaxes slowly from its initial value $h_\init$ toward an
asymptotic value $h_\infty$, which we characterize by the empirical form
\begin{align} 
  \label{eqn:2.7} 
  h_\mixed (t) \approx h_\infty - \Delta h_\mixed\exp\lbrace - \gamma_\mixed (t-t_\infty)\rbrace, 
\end{align} 
where \(t_\infty\) is a virtual origin, \(\Delta h_\mixed = h_\infty -
h_\init\), and \(\gamma_\mixed\) characterizes the entrainment process. We
further denote by $h_\final$ the depth measured at the end of a run, distinct
from the fitted asymptote $h_\infty$. This also implies that the entrainment
velocity, $U_e \equiv \frac{\mathrm{d}}{\mathrm{d}t} h_\mixed$, decreases
exponentially over time at the same rate.

\subsection{External Forcing}
\label{section:setup:forcing}

The tank is mounted on the platform of a hexapod capable of movement along six
axes (translation and rotation); however, in this study, it is used exclusively
to generate vertical sinusoidal oscillations $a\cos(\omega t)$. Additional
details on the application of the external forcing can be found on
\citet{Briard2020} and \citet{Cavelier_2022}. 

The forcing frequency \(\omega=20 \rad\) is selected to induce a sub-harmonic
resonance of the free surface, generating two-dimensional waves with a
longitudinal wavenumber \(k=3\pi/W=9.96\)~\unit{\per\meter} and frequency
\(\omega/2\). This wavenumber is the one closest to the resonance conditions, as
quantified by the detuning parameter \( \detune \equiv {g k} / {\omega^2} -
\frac{1}{4} \approx -0.006\). This choice of parameters gives the largest
possible sub-harmonic waves, while limiting contact with the container lid and
ensuring a good separation from the nearest unstable mode, even for intense
forcing. Each experiment typically runs for \(400\) to \(800\) periods in order
to reach the asymptotic regime. Additional details on the predicted natural
frequencies from inviscid theory are found in the
Appendix~\ref{section:appendix:stability:2}.

\subsection{Filling Process}
\label{section:setup:filling}
The filling process is controlled by an automatic pump system that injects fluid
through a bottom diffuser, ensuring a consistently thin, diffuse interface at
the desired position. We consider that fresh water has a density of
\(998~\kgm\), while the density difference between salt and fresh water ranges
from \(10\) to \(100~\kgm\) (see Table~\ref{Table1:fluids} in
Appendix~\ref{section:appendix:parameters} for fluid properties). We measure
vertical density profiles at the beginning and end of the experiment using a
fixed conductivity probe by raising and then lowering the tank at a small,
constant velocity. The initial stratification generally conforms to:
\begin{align}
  \label{eq2.2}
  \rho(\vec{x},0) = 
  \frac{\rho_1+\rho_2}{2} + \frac{\rho_1-\rho_2}{2} \tanh\left(\frac{z-h_\init}{L_\init/3}\right)
\end{align}
The initial interface depth \(h_\mathrm{init}\) ranges from \(2 \cm\) for
shallow interfaces to \(27 \cm\) for deep interfaces, while the initial
thickness \(L_\mathrm{init}\) varies between \(1 \cm\) and \(2 \cm\) (see
Appendix~\ref{section:appendix:parameters} for a complete
list of the experimental realizations). 

\subsection{Dimensionless parameters}
\label{section:setup:parameters}

The system dynamics depends on the following dimensionless numbers:

\renewcommand{\arraystretch}{1.15}
\begin{longtable}{@{}p{0.38\textwidth}ll@{\hspace{0.5em}}l@{}}
\toprule
\textbf{Parameter} & \textbf{Symbol} & \textbf{Definition} & \textbf{Value/Range} \\
\midrule
\multicolumn{4}{l}{\textbf{Control parameters (varied in this study)}} \\
\quad Forcing parameter & $\paramF$ & $a \omega^2 / g$ & $0.12$--$0.50$ \\
\quad Relative initial height & $\hstar$ & $h_\init / H$ & $0.07$--$0.67$ \\
\midrule
\multicolumn{4}{l}{\textbf{Other dimensionless numbers}} \\
\quad Vibrational Reynolds number & $\reynolds$ & $\rho_1 a^2 \omega / \mu_1$ & $190$--$3400$ \\
\quad Vibrational Weber number & $\weber$ & $(\rho_2-\rho_3) a^3 \omega^2 / \sigma_\surf$ & $0.16$--$9.5$ \\
\quad Schmidt number & $\schmidt$ & $\mu_1 / (\rho_1 \kappa_\mix)$ & $ 700$ \\
\bottomrule
\end{longtable}


In addition, the system depends on a density and viscosity ratio for each
interface, geometric aspect ratios, a dimensionless amplitude and an initial
interface width. Any other possible dimensionless groups can be expressed in
terms of these parameters. For
clarity, the density ratio \(\rho_2 / \rho_1\) is customarily
replaced by the nominal Atwood number:
\begin{align}
  \label{eqn:2.3}
  \atwood_\init = \frac{\rho_1 - \rho_2}{\rho_1 + \rho_2} \approx 0.047.
\end{align}

We organize the experiments into three forcing series (low, mid, and high
\(\paramF\)) corresponding to progressively larger surface waves ranging from
unbroken to breaking. Within each series, we systematically vary the
stratification depth \(\hstar\) from 6\% (shallow) to 70\% (deep). The remaining
dimensionless groups are either held constant or determined by other control
parameters.

\subsection{Experimental observations}
\label{section:surface:experiments}

Since Faraday waves are the driving mechanism for all subsequent mixing, a basic
characterisation of the surface dynamics is necessary before analysing the
mixing process itself. In this section, we describe: (i) the temporal evolution
of a typical experimental run (\S\ref{section:surface:overview}); (ii) a modal
decomposition of both interfaces that reveals how the surface wave forces the
miscible interface and how this forcing generates travelling waves that will
contribute to the horizontal homogenization of the mixing layer
(\S\ref{section:surface:POD}); and (iii) the asymptotic mixed-layer depth
observed after hundreds of forcing periods (\S\ref{section:surface:asymptotic}).

\subsubsection{Overview of an experimental run}
\label{section:surface:overview}
\begin{figure}   
	\hspace{8.5em}
	Transient \hspace{7.5em}
	Stationary \hspace{6.5em}
	Attenuation 	

	\centering
	\includegraphics[width=0.9\linewidth]{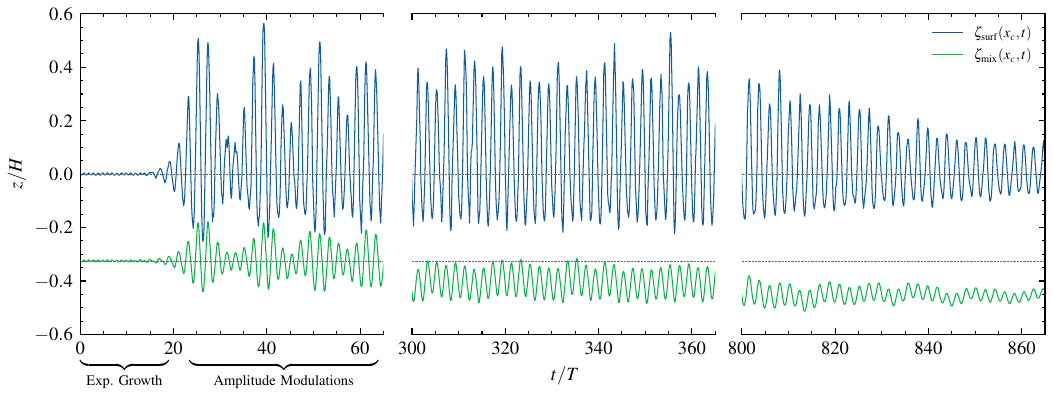}
	\caption{
		Time evolution of the free surface (upper curve, in blue) and miscible
		interface (lower curve, in green) evaluated at one of the wave
		anti-nodes, shown for the case with $\paramF=0.49$ and 30\%
		depth (EXP-F49-H30). Three phases are visible: a \emph{transient} phase,
		including exponential growth and amplitude modulation; a
		\emph{stationary} phase of roughly constant wave amplitude; and an
		\emph{attenuation} phase after forcing is disabled. The dashed
		horizontal lines mark the initial position of each interface and serve
		as a fixed reference against which the slow deepening of the miscible
		interface is visible.
	}
	\label{Figure3:Hovmoller}
\end{figure}

A typical example of the early stages of an experimental run is shown in the
sequence of snapshots in Figure~\ref{Figure2:snapshot}, while
Figure~\ref{Figure3:Hovmoller} shows the time evolution of $\zeta_\surf$ and
$\zeta_\mix$ taken at one of the wave anti-nodes for the case EXP-F49-H30,
illustrating the three phases common to all experimental realizations:
transient, stationary, and attenuation. The transient phase begins with the
linear regime of exponential growth. As the amplitude grows, nonlinear effects
lead to saturation, and the primary wave exhibits persistent, slowly modulated
amplitudes over hundreds of forcing periods \(T=2\pi/\omega\). This behaviour is
characterized by an initial overshoot before oscillating and settling into a
statistically stationary state with roughly constant amplitude (stationary
phase). After \(800T\), the external forcing is disabled and the wave decays
gradually until the experiment concludes (attenuation phase). 

For each experiment, we measure the growth and damping rates, \(\lambda_\surf\)
and \(\gamma_\surf\), by tracking the peaks in \(\zeta_\surf\) and fitting them
to an exponential function. For a given forcing amplitude \(\paramF\), the
internal stratification does not significantly affect the surface dynamics (see
Table~\ref{Table2:experiments}). The value of \(\lambda_\surf\) ranges between
\(0.5\) and \(1.3~\unit{\per\second}\), which aligns well with predictions from
linear stability analysis for the excited mode (a quantitative comparison is
postponed until \S\ref{section:validation:growth}). Meanwhile, the damping
rates \(\gamma_\surf\) vary between \(0.04\) and \(0.06\unit{\per\second}\),
consistent with observations from \cite{Jiang_1996} and \cite{Kalinichenko2020}
in similar-sized systems. Similarly, tracking the peaks in $\zeta_\mix$ yields
$\lambda_\mix \approx \lambda_\surf$, suggesting that this interface is
passively advected by the surface. 

\subsubsection{A simplified view of the surface and interface dynamics}
\label{section:surface:POD}

\begin{figure}
  \centering
  \begin{minipage}[t]{0.9\linewidth}
    (a) \nth{1} surface mode \hfill (b) \nth{2} surface mode \hfill (c) \nth{3} surface mode \hfill~\\
    \includegraphics[width=0.33\linewidth]{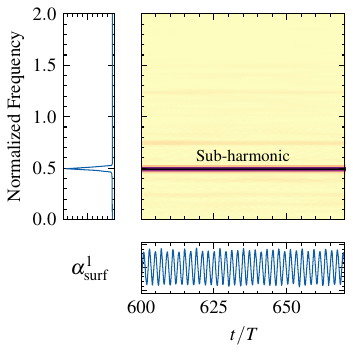}
    \includegraphics[width=0.33\linewidth]{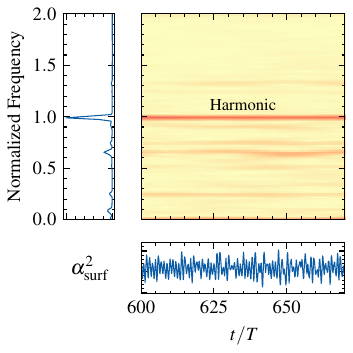}
    \includegraphics[width=0.33\linewidth]{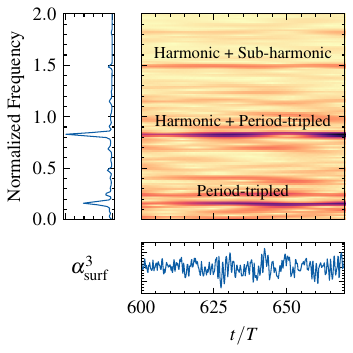}

    (d) \nth{1} interface mode \hfill (e) \nth{2} interface mode \hfill (f) \nth{4} interface mode \hfill~\\
    \includegraphics[width=0.33\linewidth]{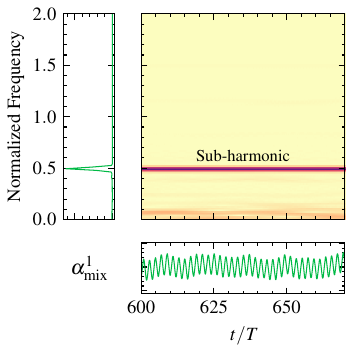}
    \includegraphics[width=0.33\linewidth]{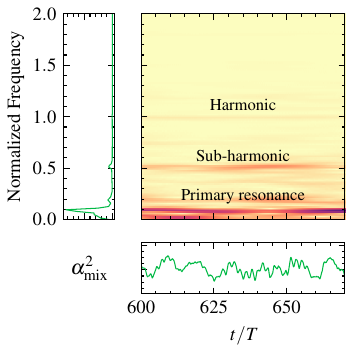}
    \includegraphics[width=0.33\linewidth]{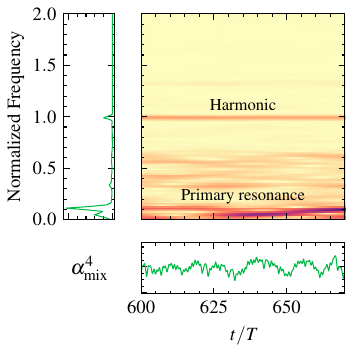}
  \end{minipage}
  \caption{
    Time-frequency representation of leading POD modes of the case EXP-F49-H30
    showing their time evolution alongside their spectral content and
    spectrogram analysis. Top row: free surface modes, with dominant frequencies
    $\omega/2$ (subharmonic), $\omega$ (harmonic), and $\omega/6$
    (period-tripling). Bottom row: miscible interface modes, with dominant
    frequencies $\omega/2$ (surface forcing), $\omega$ (harmonic), and the
    natural frequency of the miscible interface.
  }
  \label{Figure4:POD:coefficients}
\end{figure}


A way to highlight the coupling between the two interfaces, and more
specifically, how the surface wave forces the miscible interface, is to expand
both interfaces in a set of orthogonal modes, 
\begin{align}
\zeta_\surf(x,t) = \sum_{n=1}^\infty \alpha^{n}_\surf(t) ~\phi^{n}_\surf(x), \quad 
\zeta_\mix(x,t) = h_\mixed + \sum_{n=1}^\infty \alpha^{n}_\mix(t) ~\phi^{n}_\mix(x).
\end{align}
A natural choice for the orthogonal basis would be Fourier modes, but we retain
instead a POD basis computed from the experimental image sequences, as it
concentrates energy into fewer modes and better captures the dominant wave
patterns. A detailed description of the decomposition procedure is given in
Appendix~\ref{section:appendix:POD}.

Figure~\ref{Figure4:POD:coefficients} shows the modal coefficients and their
frequency spectra for the leading modes of each interface, illustrating how each
mode is associated with a distinct set of harmonics. The primary surface mode
takes the form 
\begin{align}
  \label{eq:pod:surface1}
  \phi^1_\surf = A_1 \sin(kx) + A_3 \sin(3kx) + A_5 \sin(5kx) + \cdots
\end{align}
where $\sin(kx)$ represents the primary wave excited by the linear instability,
while the odd harmonics represent the (nonlinear) steepening of the saturated
waves. The modal coefficient $\alpha^1_\surf$ displays a dominant frequency
\(\omega/2\) and closely resembles $\zeta_\surf$ measured at the anti-nodes, as
previously shown in Figure~\ref{Figure3:Hovmoller}. The secondary surface mode
takes the form 
\begin{align}
  \label{eq:pod:surface2}
  \phi^2_\surf = B_2 \cos(2kx) + B_4 \cos(4kx) + \cdots
\end{align}
which represents the spatial harmonic of $\phi^1_\surf$, while the modal
coefficient $\alpha^2_\surf$ displays a dominant frequency \(\omega\), making it
a temporal harmonic as well. This mode breaks the top/bottom symmetry, giving
the primary wave its characteristic cnoidal profile. Higher-order modes capture
progressively finer spatial features of the wave profile, concentrated near the
anti-nodes, and their modal coefficients display richer frequency contents. For
instance, the third modal coefficient is noisy, and its frequency content
includes \(\omega/6\), which may be linked  
to features associated with period-tripled breaking, such as peaks, depressions,
and jets~\citep{Jiang1998}. This low-frequency response emerges only at large
forcing amplitudes. Additionally, the frequency content includes \(5\omega/6\)
and \(3\omega/2\), which are consistent with interactions involving harmonic,
sub-harmonic, and period-tripled modes
(Figures~\ref{Figure4:POD:coefficients}c).

The primary and secondary interface modes have the same general form as
\eqref{eq:pod:surface1}, but the degree of steepening is less pronounced as the
interface depth increases, reflecting the progressive decoupling from the
surface modes. The primary and secondary modes are shifted by half a period and
their combined action creates a travelling wave propagating horizontally. The
modal coefficient $\alpha^1_\mix$ is well correlated to $\alpha^1_\surf$,
representing the advection of the miscible interface by the free surface motion.
This behaviour is consistent with a barotropic ---or zigzag --- mode obtained
from linear stability analysis~\citep{Bestehorn2016}. The secondary mode tells
a different story: $\alpha^2_\mix$ displays low-frequency, sub-harmonic, and (to
a smaller degree) harmonic content (Figure~\ref{Figure4:POD:coefficients}e),
where the lower frequency is close to the natural frequency of the miscible
interface for the primary wave's wavelength $(g k \atwood)^{1/2}$. This is
consistent with the miscible interface behaving as a forced harmonic oscillator
driven by the surface wave, with part of the energy radiated as horizontally
propagating waves that contribute to horizontal homogenization. The same
forced-oscillator picture extends to higher-order modes: modes 3 and 4 form a
travelling-wave pair at wavenumber $2k$, driven by the secondary surface mode at
frequency $\omega$, with a natural frequency contribution also visible in the
spectrum (Figure~\ref{Figure4:POD:coefficients}f).

\subsubsection{Asymptotic mixed-layer depth}
\label{section:surface:asymptotic}

\begin{figure}
	\centering
	\begin{minipage}[t]{0.8\linewidth}
	~\hspace{4em} EXP-F49-H07 \hfill EXP-F49-H30 \hfill EXP-F49-H67 \hfill~ \\
	\includegraphics[width=\linewidth, trim=-70 0 -5 0, clip]{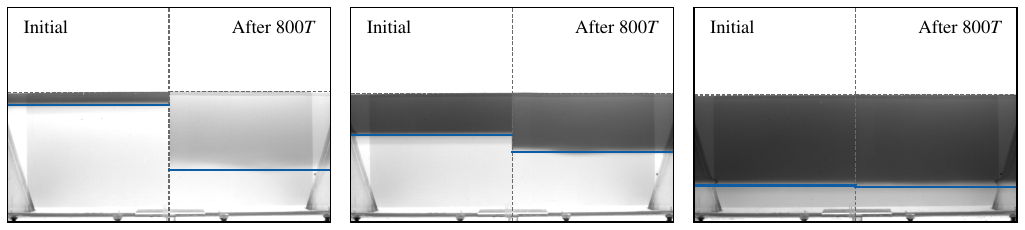}
	
	\centering
	\hspace{1em}\includegraphics[width=0.625\linewidth, trim=0 0 0 0, clip]{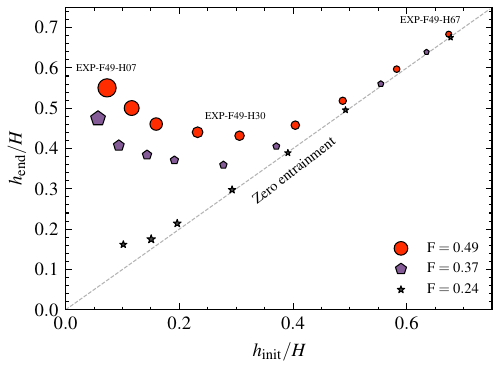}
	\end{minipage}
	\caption{
		Top panel: images taken before (left half) and after (right half) $800T$
		of forcing at $\paramF=0.489$ for increasingly deep initial interfaces
		(from left to right). The mixed layer thickens and deepens in all three
		cases, but the final depth varies non-monotonically on the initial
		depth. See also the time-lapse in the Supplementary Material. Bottom
		panel: final interface depth $h_\final$ as a function of the initial
		interface depth $h_\init$ for all experiments; marker size is
		proportional to the final interface thickness $L_\final$.
	}
	\label{Figure5:SideToSide}
\end{figure}


As shown in Figure~\ref{Figure3:Hovmoller}, $\zeta_\mix$ is initially well
correlated with $\zeta_\surf$, but the two interfaces gradually decouple as
entrainment drives the miscible interface deeper, slowly approaching an
asymptotic value at long times. Comparing the measured $h_\final$ with the
fitted asymptote $h_\infty$ from~\eqref{eqn:2.7}, we confirm that the
experiments ran long enough to reach the asymptotic regime: $h_\final$ is within
7\% of $h_\infty$ for most runs. The larger deviations occur for the shallowest,
most strongly entraining cases, where the large $\Delta h_\mixed$
leaves a bigger residual. The same fits give an entrainment rate
\(\gamma_\mixed\) between \(0.005\) and \(0.017~\unit{\per\second}\), with
larger values usually corresponding to shallower configurations (see
Table~\ref{Table2:experiments}). 

Because this evolution is slow, we contrast the initial and final states directly
in Figure~\ref{Figure5:SideToSide}. The three images correspond to
representative shallow, intermediate, and deep cases at equal forcing, while the
bottom panel plots the final depth $h_\final$ against the initial depth
$h_\init$ for all experiments. The outcome is counterintuitive: one might expect
the final depth to increase monotonically with the initial depth; instead,
$h_\final$ is non-monotonic in $h_\init$: it is largest at both ends of the
range studied, but for different reasons. At the shallow end this reflects
strong entrainment despite the small initial depth; at the deep end,
entrainment is instead negligible ($\Delta h_\mixed\approx0$) and $h_\final$
stays large simply because $h_\init$ already was. This non-monotonic response
is characteristic of strong forcing; under weak forcing the interface barely
moves, and the initial and final states remain roughly the same.

This dependence reflects a competition between the turbulent energy reaching the
interface, set by the forcing $\paramF$ and by the initial depth $h_\init$, and
the energy required to erode the density gradient, which is mostly set by the
initial depth $h_\init$. Their balance governs both the transition from a weak
to a strong response and the non-monotonic dependence of the final depth on
$h_\init$ --- one of the main findings of the paper, examined in detail in
\S\ref{section:stratification:long}. However, characterizing this process
requires quantities beyond the reach of the current experimental setup, and we
therefore complement our observations with numerical simulations.

\section{Numerical Framework}
\label{section:governing}

%
%

\subsection{Governing Equations}
\label{section:governing:equations}

\subsubsection{One-fluid formulation using a volume-of-fluid approach}

Equations of motion are based on the incompressible Navier-Stokes equations
for two-phase fluids, supplemented by a transport equation for the binary
mixture. In this approach, the governing equations are written in terms of
a volume fraction \(f(\vec{x},t) \in [0,1]\):
\begin{subequations}
  \label{eqn:3.1}
  \begin{align}
    \label{eqn:3.1a}
    \del\cdot\vec{u} & = 0
    \\
    \label{eqn:3.1b}
    \frac{\partial\vec{u}}{\partial t}
    + \vec{u}\cdot(\del\vec{u})
    & = \frac{1}{\rho} \left[-\del p + \del\cdot (2\mu\mathsfbi{S}) + \sigma_\surf\kappa_\surf\vec{n}\delta_\surf\right] + \vec{g}(t)
    \\
    \label{eqn:3.1c}
    \frac{\partial f}{\partial t} + \vec{u} \cdot(\del f) & = 0
  \end{align}
\end{subequations}
where \(\vec{u}=(u,v,w)\) is the velocity field, $\rho$ the density field,
\(\vec{g} = -(g + a\omega^2\cos(\omega t))\vec{e_z}\) the modulated gravity,
\(p\) the pressure field, \(\mu\) the dynamic viscosity,
\(\mathsfbi{S}=(\del\vec{u}+\del\vec{u}^T)/2\) the strain rate tensor,
\(\kappa_\surf\) the local curvature, \(\delta_\surf\) a surface Dirac delta
function, and \(\vec{n}\) is the unit normal vector of the free
surface~\citep{Tryggvason2011}. 

\subsubsection{Transport equation for the concentration field}

Stratification results from differences in density due to salt concentration in
water. This is included by means of a (volumetric) concentration field
\(c(\vec{x},t) \in [0,1]\), which is governed by the transport equation:
\begin{align}
  \label{eqn:3.2}
  \frac{\partial c}{\partial t}
  + \vec{u}\cdot(\del c)
  &= \del\cdot \left(\kappa_\mix \del c - \beta_\mix \delta_s c \right)
\end{align}
where \(\kappa_\mix\) is the diffusion coefficient of \(c\) and \(\beta_\mix\) is a
parameter that controls diffusion across the free surface (see, for instance,
\cite{Haroun2010,Marschall_2012}). We set \(c=0\) in the gas phase, and choose
\(\beta_\mix\) to restrict diffusion to the liquid phase.

\subsubsection{Coupling stratification with the momentum equation}

The equations above are coupled through the fluid properties:
\begin{subequations}
\begin{align}
  \label{eqn:3.3}
  \mu(\vec{x},t) &= f_1(\vec{x},t) \mu_1 + f_2(\vec{x},t) \mu_2 + f_3(\vec{x},t) \mu_3
  \\
  \rho(\vec{x},t) &= \underbrace{f_1(\vec{x},t) \rho_1}_\mathrm{salt-water}
  + \underbrace{f_2(\vec{x},t) \rho_2}_\mathrm{fresh-water}
  + \underbrace{f_3(\vec{x},t) \rho_3}_\mathrm{air}
\end{align}
\end{subequations}
where \(f_i\) is the volume fraction of the component \(i\) and the liquid phase
is treated as a binary mixture, where fluid properties have a linear dependence
on \(c(\vec{x},t)\). In practice, we take \(f_1 = f c\), \(f_2 = f (1-c)\), and
\(f_3 = (1-f)\), which satisfies~\eqref{eqn:3.1a}-\eqref{eqn:3.1c}, as the
solute is not allowed into the gas phase.  For the simulations, the viscosity of
the liquid phase is simplified by assuming constant, equal pure-component
dynamic viscosities, \(\mu_1=\mu_2\).

%
%

\subsection{Instantaneous density profiles, mixed layer and mixing layer width}
\label{section:governing:definitions}

In the experiments, density profiles were only measured at the beginning and at
the end of each experiment where the fluid is at rest. In the simulations, the
instantaneous density field $\rho(\vec{x},t)$ is available at all times, but is
spatially distorted by surface-wave advection. In order to filter wave-induced
distortions, we use the reference state of minimum potential
energy~\citep{Winters_1995}, obtained by sorting fluid parcels by density,
equivalent to an adiabatic restratification. By definition, this reference state
is characterized by a monotonic density profile $\bar{\rho}(z,t)$, which has the
same probability distribution as the instantaneous density field
$\rho(\vec{x},t)$ and can therefore be computed from its cumulative distribution
function~\citep{Tseng2001}. Additional details on this computation can be found
in \cite{Castillo2016, Briard2019}.

Changes in $\bar{\rho}$ (and therefore in the background potential energy)
reflect only irreversible mixing. From these profiles, we also identify the
instantaneous depth of the mixed layer from the peak in
\((\bar{\rho}-\bar{\rho}_\mathrm{min})(\bar{\rho}_\mathrm{max}-\bar{\rho})\),
while the interface width \(L(t)\) is calculated using the classical expression
for a mixing layer width from \cite{Andrews1990}:
\begin{align}
  \label{eqn:3.4}
  L(t) = 6 \int_{z_{\min}}^{z_{\max}} \frac{(\bar{\rho}-\bar{\rho}_\mathrm{min})(\bar{\rho}_\mathrm{max}-\bar{\rho})}{(\bar{\rho}_\mathrm{max}-\bar{\rho}_\mathrm{min})^2}~\mathrm{d}z.
\end{align}
This definition ensures $L(t=0) = L_\init$, consistent with the initial
condition \eqref{eq2.2}.


\subsection{Numerical Approach}
\label{section:governing:approach}

We performed numerical simulations using a finite-volume solver implemented in
the open-source code Basilisk~\citep{Popinet2009,Popinet2015}. Boundary
conditions include a static contact angle of 90$^\circ$, a no-slip condition,
and impermeability imposed on all boundaries. By using a face-centered velocity
for volume fraction advection --- which is not strictly zero even when a zero
velocity is imposed at the solid boundary --- the contact line can move due to
what is known as "numerical slip". This numerical slip is linked to the grid
size and tends toward the no-slip limit as the size of the mesh is
decreased~\citep{Tavares2024}. 

The solver has been validated for a variety of complex interfacial
flows~\citep{Hooft2018, Berny2020, Riviere2021, Mostert2022}.  In this work, the
governing equations are discretized on a regular Cartesian grid, where
\(\Delta_x=\Delta_y=\Delta_z\), using a second-order scheme in time and space.
Viscous terms are treated implicitly, while a projection method is used to deal
with the velocity-pressure coupling. A compromise in spatial resolution is
required to perform simulations for several hundred periods at a reasonable
computational cost. Here, we use 1024 \(\times\) 768 \(\times\) 128 grid points
such that \(\Delta_x\) is roughly one third of the size of the capillary length.

A conventional grid-convergence study is delicate for this system, because
different observables converge at very different resolutions. We verified that
the linear dynamics are well converged at the present resolution, an assessment
reinforced by the validation against linear theory and experiments in
\S\ref{section:validation}. The fragmentation scales, in contrast, are only
marginally resolved: with the capillary length spanning roughly three cells,
droplets are expected to be artificially large compared to the experiments. A
direct experimental verification is not available either, as individual droplets
are difficult to measure with our imaging setup owing to their weak luminosity
contrast. Terms such as \emph{under-resolved DNS} or \emph{implicit LES} would
arguably describe these fragmentation and mixing scales more precisely. However,
we retain \emph{DNS} throughout, since the linear and transient phases are direct
simulations at this resolution and switching labels partway through the same
runs would complicate the presentation without changing the substance of the
assessment above.

Initial conditions assume the fluid is at rest, with the density profile
following \eqref{eq2.2}. The resulting system is solved using an iterative
multigrid cycle with a small relative tolerance to ensure that \(c(\vec{x},t)\)
does not diffuse into the gas phase. Additional details can be found in
\cite{Farsoiya_2021}. The source code developed in this work, as well as the
simulation setups are available on the Basilisk website~\citep{Castillo2025}.

The numerical investigation comprises three series of three-dimensional
simulations: 
\begin{itemize}
  \item[---] Series A: varying the stratification depth. 
  \item[---] Series B: varying the forcing term.
  \item[---] Series C: run for several hundred periods to comment on the
  asymptotic regime.
\end{itemize}
These simulations aim to closely reproduce the phenomenology observed in
experiments; detailed parameters for the three simulation series are listed in
Table~\ref{Table3:simulations} (Appendix~\ref{section:appendix:experiments}).
However, spatial resolution constraints require setting the Schmidt number to
1---instead of 700, as for salt water. This implies that mixing will happen
faster in simulations than in experiments, but the underlying mechanisms are
still expected to hold.

Several considerations motivate the use of three-dimensional simulations rather
than a two-dimensional setup. First, the droplet and bubble fragmentation
observed at the wave anti-nodes is inherently three-dimensional. Second,
two-dimensional turbulence redistributes this energy across scales differently
from its three-dimensional counterpart. Third, although the experimental
geometry is designed to constrain the dominant Faraday mode to the longitudinal
direction, residual transverse motion is observed. Fourth, we shall see that the
dominant contribution to the surface-wave damping coefficient comes from
friction along the front and back walls; this contribution is absent in two
dimensions and would change the saturation amplitudes of both interfaces.

                             
\subsection{Comparison between Experimental Observations and DNS}
\label{section:validation}
\begin{figure}   
	\begin{center}
	\begin{minipage}{0.9\linewidth}
	\color{white}		
		\includegraphics[width=0.5\linewidth, trim=300 0 250 25, clip]
		{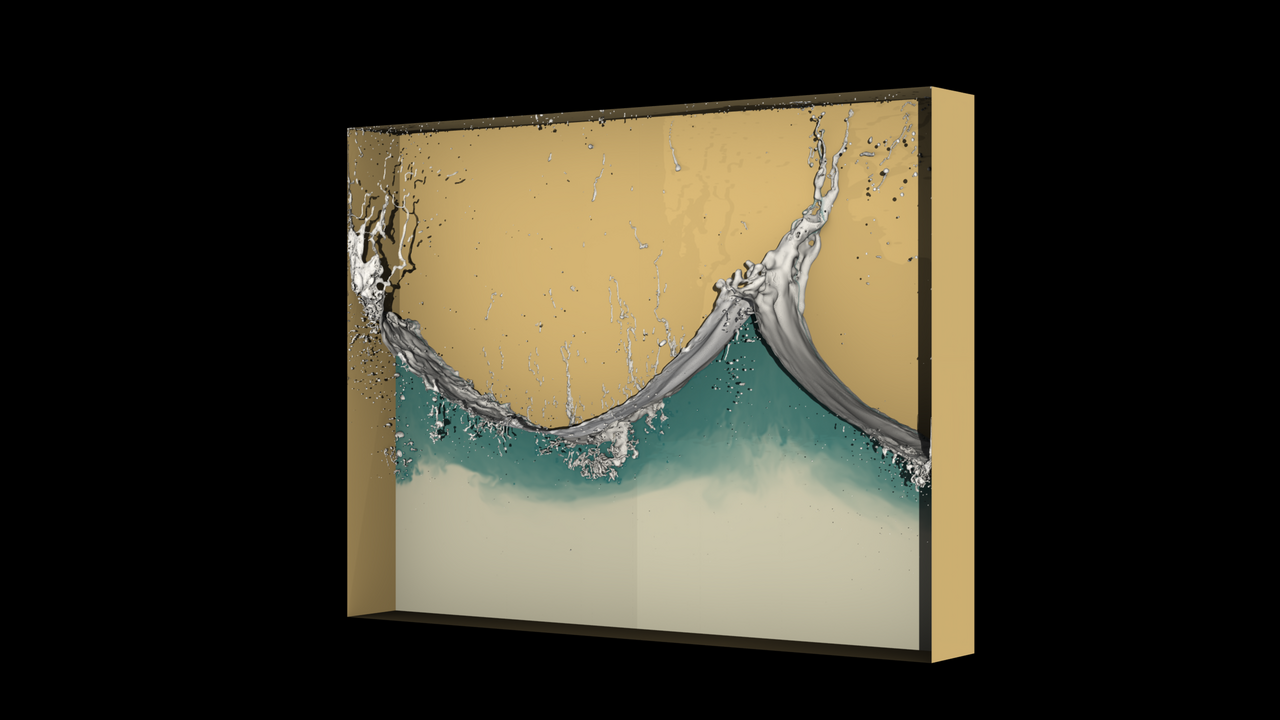}\hspace{-0.1cm}
		\includegraphics[width=0.5\linewidth, trim=300 0 250 25, clip]
		{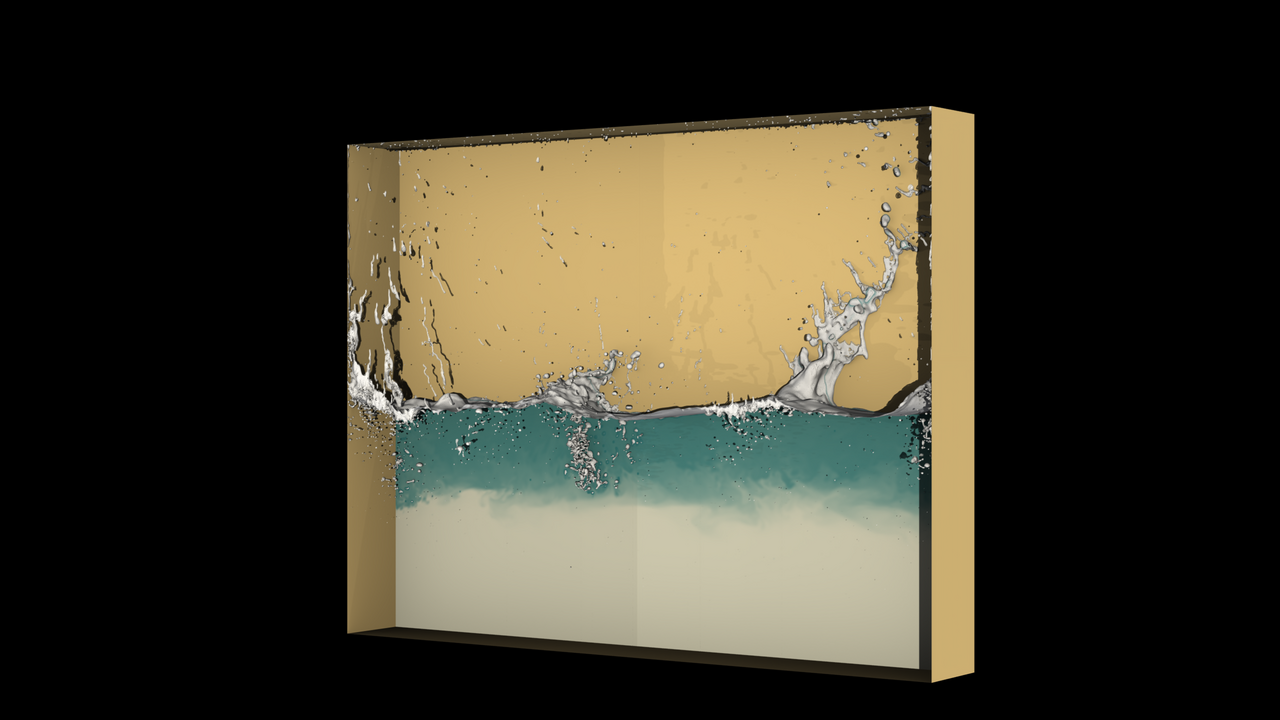}

		\vspace{-5.70cm}~~ (a) \hfill (b) \hfill ~ \\ \vspace{4.75cm}

		\includegraphics[width=0.5\linewidth, trim=300 0 250 25, clip]
		{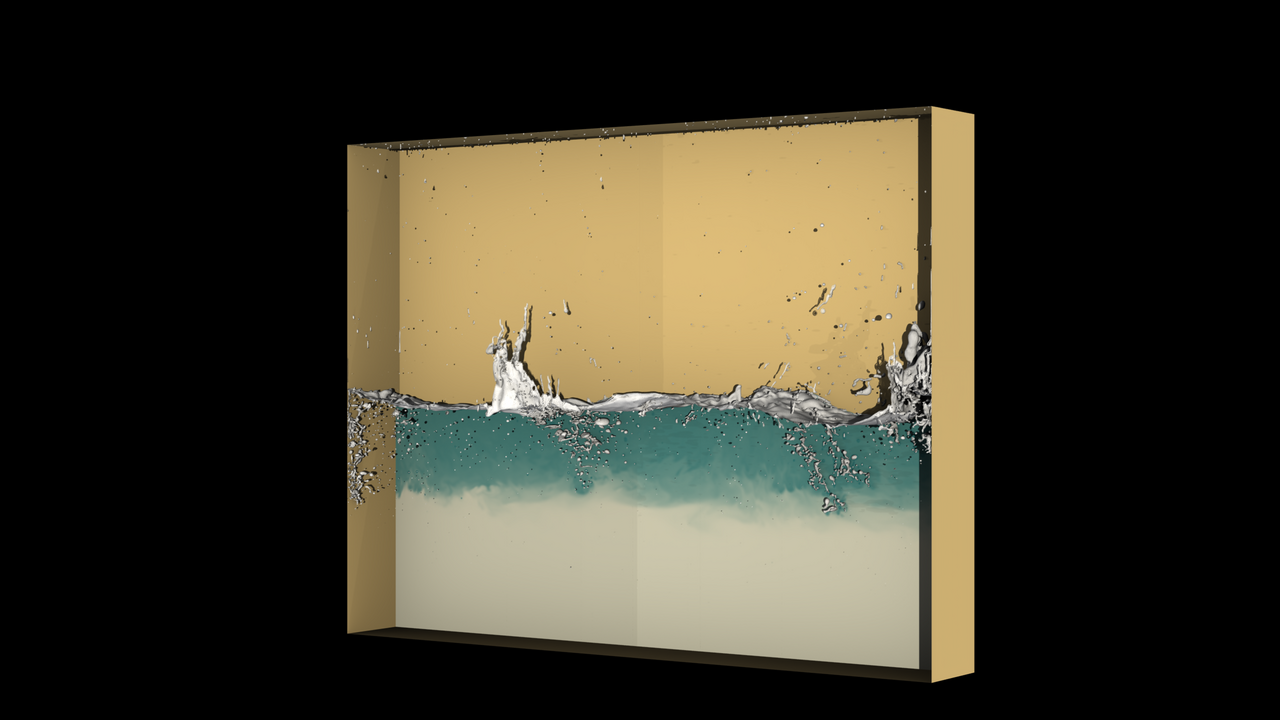}\hspace{-0.1cm}
		\includegraphics[width=0.5\linewidth, trim=300 0 250 25, clip]
		{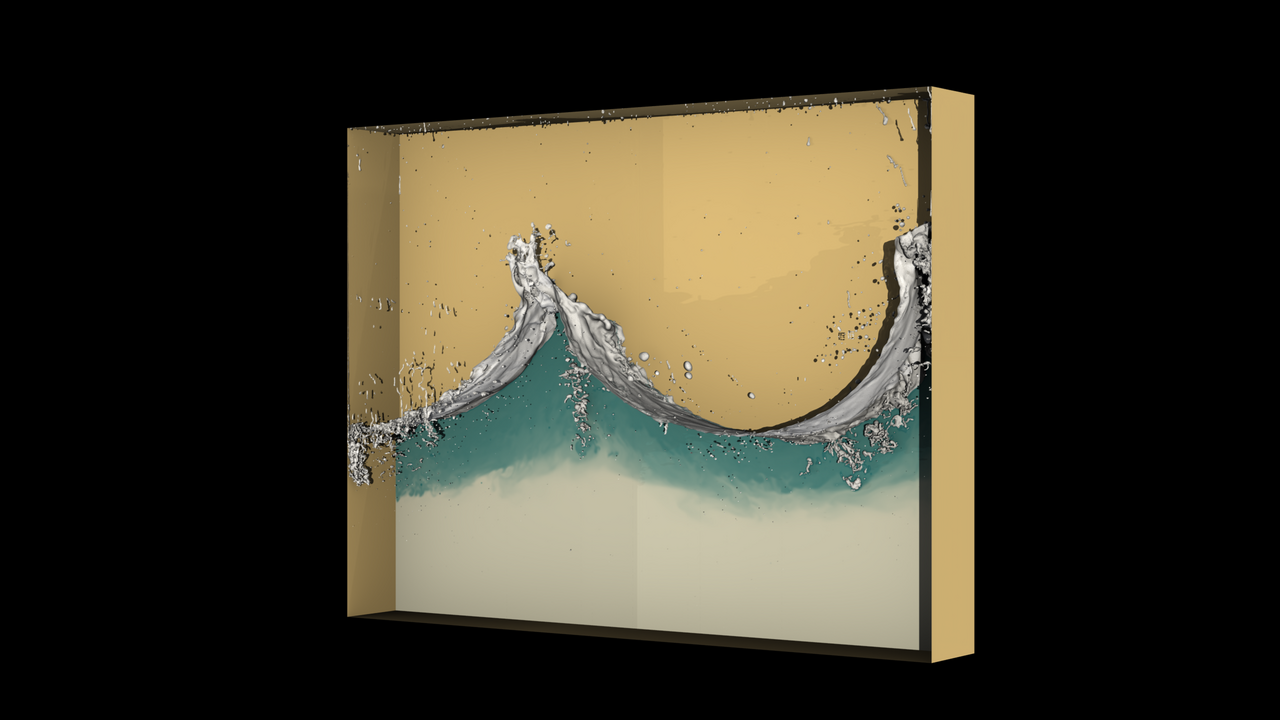}
				
		\vspace{-5.70cm}~~ (d) \hfill (c) \hfill ~ \\ \vspace{4.70cm}
	\end{minipage}
	\end{center}  
  \caption{    
	Simulation results for the case DNS-F49-H30. Here, the free surface is shown
	in white, while the concentration field at the median plane is depicted in
	colour. Snapshots are taken at regular intervals every $T/2$. See also
	animation in SM.
  }
  \label{Figure6:snapshots}
\end{figure}

Figure~\ref{Figure6:snapshots} (and the accompanying supplementary video) shows
a sequence of snapshots from DNS-F49-H30 taken in the stationary regime, where
surface waves are fully developed. This case, with strong forcing and
intermediate stratification, is the numerical counterpart of the experimental
run EXP-F49-H30. 

The sequence illustrates many of the features already observed in the
experiments, pointing to a good qualitative agreement. We identify the primary
surface wave with its subharmonic response; a transverse sloshing mode in which
a thin fluid layer slides along the front and back walls at a dominant frequency
near \(\omega/4\); and a significant amount of splashing droplets and bubble
columns, typically localized around the wave anti-nodes. All of these features
are observed in the experiments, although individual droplets and thin films are
difficult to detect because their luminosity contrast with the surrounding fluid
is weak. Droplet and bubble sizes, in particular, are more prominent in the
simulations; their precise distribution, however, depends on the grid size since
the breakup of thin films is mainly caused by numerical aspects rather than by
physical ones~\citep{Chirco2022}. We do not claim to accurately resolve all the
relevant scales of droplets and bubbles; their presence, however, even at sizes
larger than physical, is recognized as an important mechanism of injection of
turbulent energy into the mixed layer.

A quantitative comparison is more delicate due to the regimes considered. The
system passes through a linear, exponential-growth phase (deterministic and
well-characterized), before settling into an irregular, nonlinear saturated
state where comparisons can only be done in a statistical sense. We propose the
following diagnostics to validate our numerical approach. First, in
\S\ref{section:validation:growth}, we compare measured growth rates against
linear stability theory and experiments. Second, in
\S\ref{section:validation:amplitudes}, we compare the shape of both interfaces
during the transient phase, from the linear growth through the first couple of
amplitude overshoots. Third, in \S\ref{section:validation:amplitudes2}, we
extend this comparison to the stationary regime, where the surface wave shape
and amplitude are characterised statistically. Finally, in
\S\ref{section:validation:damping}, we measure damping rates to verify that the
DNS dissipates energy at the correct rate.

\subsubsection{Growth rates during the transient phase}
\label{section:validation:growth}
\begin{figure}   
	\centering	
	\begin{minipage}{0.9\linewidth}
	(a) \hfill (b) \hfill~ \\
	\includegraphics[width=0.5\linewidth]{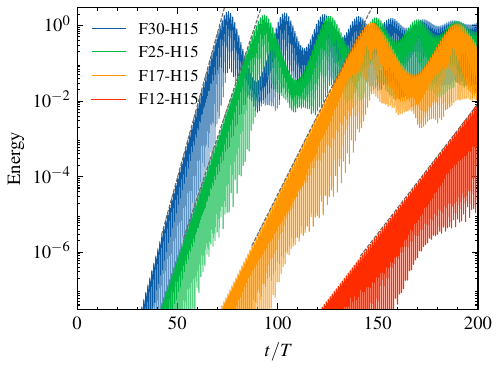}
	\includegraphics[width=0.5\linewidth]{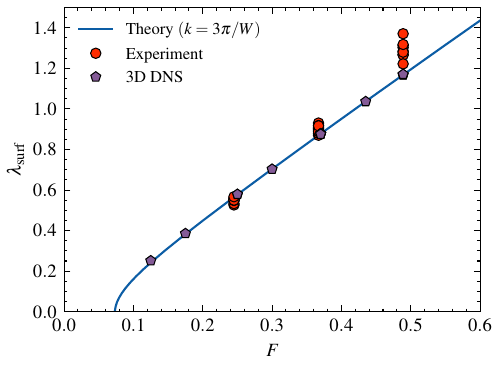}
	\end{minipage}
	\caption{
		(a) Time evolution of the kinetic energy during the linear phase, for
		the Series B simulations. (b) Exponential growth rate \(\lambda_\surf\)
		measured in experiments and simulations compared to linear theory.
	}
	\label{Figure7:Growth}
\end{figure}

As described in \S\ref{section:surface:experiments}, the experimental growth
rate $\lambda_\surf$ was obtained by tracking the peak wave amplitude during the
linear phase and fitting it to an exponential. In the simulations we corroborate
this rate via the time evolution of the kinetic energy, which grows with twice
the rate and, as a global quantity, is less sensitive to errors in interface
reconstruction. The procedure used to estimate the energy from the experiment is
described in \S\ref{section:appendix:energies}. Figure~\ref{Figure7:Growth}a
shows the time evolution of the kinetic energy for the Series B simulations,
which share the same initial stratification but different forcing. Since the
interface is initially flat, perturbations develop from numerical noise and
several decades of exponential growth are observed. For small Atwood numbers,
the stratification depth has no effect on the growth rates, which depend only on
\(F\). These measurements align closely with predictions from linear stability
theory for the wavenumber in question (see Figure~\ref{Figure7:Growth}b). In
comparison, the experiments are more scattered, which could be attributed to
measurement errors. Details on the predictions from inviscid linear theory are
found in the Appendix~\ref{section:appendix:stability:2}.

\subsubsection{Comparison with linear theory during the growth phase}
\label{section:validation:amplitudes}
\begin{figure}   
	\begin{center}
		\begin{minipage}{0.9\linewidth}
		\scalebox{-1}[1]{\includegraphics[width=0.33\linewidth, trim=2 0 245 0, clip]{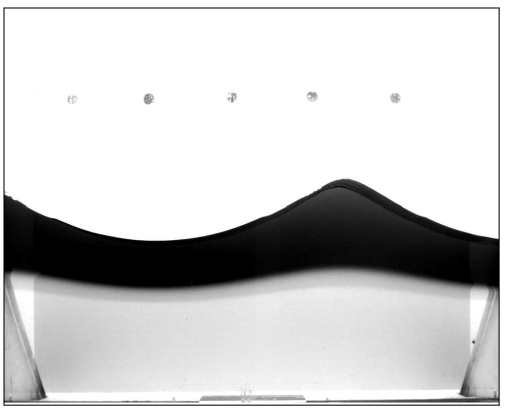}}\hfill
		\scalebox{-1}[1]{\includegraphics[width=0.33\linewidth, trim=2 0 245 0, clip]{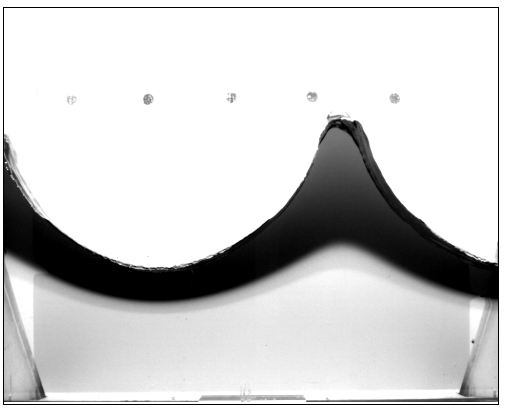}}\hfill
		\scalebox{-1}[1]{\includegraphics[width=0.33\linewidth, trim=2 0 245 0, clip]{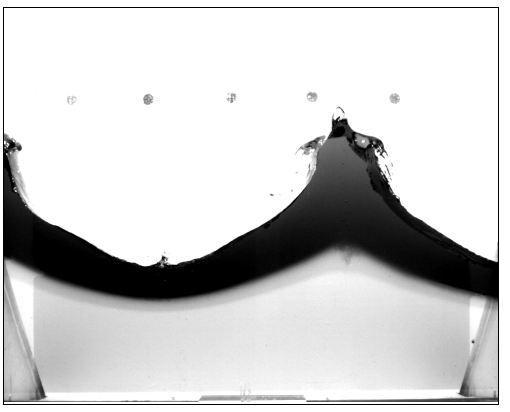}}	

		\vspace{-3.0cm} \quad(a) \hfill \quad(c) \hfill\quad (e) \hfill~\\ \vspace{2.20cm}

		\scalebox{-1}[1]{\includegraphics[width=0.33\linewidth, trim=248 0 0 0, clip]{./figures_R1/figure8a_alt-compressed.pdf}}\hfill
		\scalebox{-1}[1]{\includegraphics[width=0.33\linewidth, trim=248 0 0 0, clip]{./figures_R1/figure8b-compressed.pdf}}\hfill
		\scalebox{-1}[1]{\includegraphics[width=0.33\linewidth, trim=248 0 0 0, clip]{./figures_R1/figure8c-compressed.pdf}}

		\vspace{-3.0cm} \quad(b) \hfill \quad(d) \hfill\quad (f) \hfill~\\ \vspace{2.20cm}
		\end{minipage}
	\end{center}  
  \caption{    
	Experimental and numerical snapshots for EXP-F49-H30 (top, panels a,c,e) and
	DNS-F49-H30 (bottom, panels b,d,f). From left to right: (a-b) during the
	exponential growth phase, approximately four periods before saturation;
	(c-d) at the peak of the first overshoot; (e-f) at the peak of the second
	overshoot. 
  }
  \label{Figure8:snapshots}
\end{figure}


Figure \ref{Figure8:snapshots} compares experimental snapshots and DNS results
at three representative instants of the transient, chosen to match the same
physical stage in both datasets rather than the same absolute time (see
below). At these early times, the miscible interface has not yet been displaced from its
initial position, but oscillates vertically in response to the surface wave
motion. The sequence also illustrates how both interfaces become increasingly
disorganized as the transient phase progresses and the surface wave undergoes
breaking and successive splash-up cycles. The DNS results are in good
qualitative agreement with the experiments, with differences at small scales
attributable to transverse integration in the experimental images, which tends
to obscure fine structure.

In panels (a,b), taken during the exponential growth phase, the surface wave
already exhibits a cnoidal shape while the miscible interface retains a nearly
sinusoidal profile. The cnoidal shape arises from spatial and temporal
harmonics, which break the top/bottom symmetry of nonlinear waves. The asymmetry
between the two interfaces is explained by their relative amplitudes, set by the
linear eigenmode. During the linear phase, the ratio between the amplitudes,
$\vert A_\mix \vert /\vert A_\surf \vert $, is fixed and scales with
$\exp(-kh_\init)$ (see Appendix~\ref{section:appendix:stability} and
Figure~\ref{fig:appendix:B:2}b). For both EXP-F49-H30 and DNS-F49-H30, the
amplitude of the miscible interface is 40\% of that of the surface wave. Its
harmonic contributions, which scale with the square of the amplitude, therefore
remain negligible, and its shape stays nearly sinusoidal. At the larger
amplitudes of the first and second overshoots (panels c-f), both interfaces
display a cnoidal shape.

Several wave models capture this qualitatively. For instance, Rayleigh's
classical nonlinear standing-wave solution already predicts sharp crests and
flat troughs consistent with the cnoidal shape. However, for the larger
amplitudes observed here, a more accurate description is obtained using
Lagrangian models, which express the position \((X(a),Z(a))\) in terms of a
Lagrangian coordinate \(a \in [-W/2,W/2]\). For instance, in the deep water
approximation, the model used by \cite{Kalinichenko2019} reads
\begin{subequations}
\begin{align}
  \label{eqn:3.5a}
  X - a \approx& -\frac{1}{2}\vert A \vert \sin(ka)\cos\psi
  \\
  \label{eqn:3.5b}
  Z \approx& ~~~~\frac{1}{2} \vert A \vert \cos(ka)\cos\psi 
  + \frac{k}{16} \vert A \vert ^2 (1 + \cos(2\psi))
\end{align}
\end{subequations}
where $\psi = \omega t/2 + \theta$, and $\theta$ is a phase shift that remains
small in the limit of weak damping. Applying
\eqref{eqn:3.5a}--\eqref{eqn:3.5b} to both interfaces, using $A_\surf$ for the
surface wave and $A_\mix \approx 0.4\,A_\surf$ as predicted by linear theory,
yields good agreement with the DNS shown in panel (b).

A note on the comparison strategy: in Figure~\ref{Figure8:snapshots} we compare
DNS and experiments at the same physical stage of the transient --- the first
and second amplitude overshoots --- rather than at identical time instants. The
snapshot in the linear phase (panels a,b) was likewise selected relative to the
first overshoot, taken approximately four periods before it in each dataset.
This choice is deliberate: the time between overshoots, and more generally the
pace at which the wave reaches the stationary state, is faster in the
experiments than in the DNS. The source of this discrepancy is difficult to
pinpoint; multiple factors are plausible candidates, including differences in
initial conditions, experimental uncertainties, and the simplifications used 
in the DNS. In this respect, the agreement between DNS and experiments remains
qualitative rather than quantitative.

\subsubsection{Surface wave shape and amplitude during the stationary phase}
\label{section:validation:amplitudes2}
\begin{figure}
  \centering
  \begin{minipage}{0.9\linewidth}   
  \includegraphics[width=\linewidth, trim=0 0 0 0, clip]{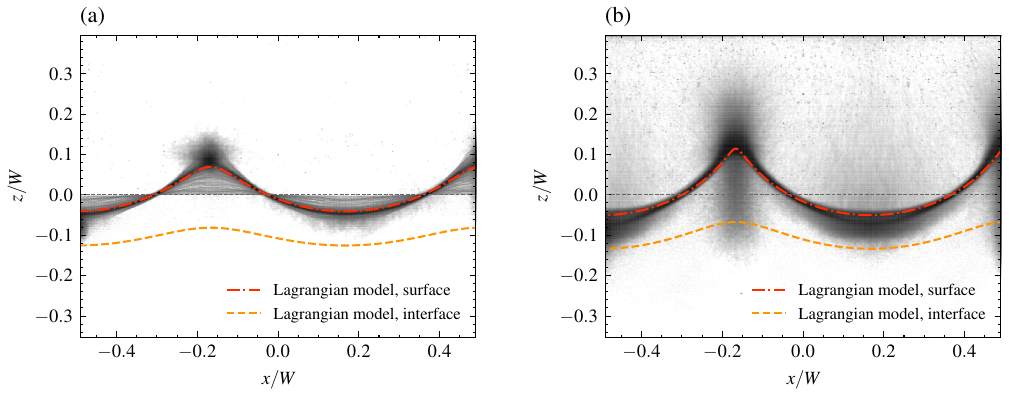}
  \end{minipage}
  \caption{    
    Conditionally averaged histograms $P(x_\text{bin}, z_\text{bin} \mid
    \varphi=0)$ representing the spatial distribution of $\zeta_\text{surf}$ for
    cases (a) DNS-F25-H30 and (b) DNS-F49-H30. Darker colours indicate more
    likely positions of the free-surface. The red dash-dotted curve shows the
    predicted free-surface shape from the model
    \eqref{eqn:3.5a}-\eqref{eqn:3.5b}, using $A_\steady$ from \eqref{eqn:3.8}.
    For reference, we also display the miscible interface.
  }
  \label{Figure9:histograms}
\end{figure}


Following the transient phase, the amplitude overshoots progressively decay and
the surface wave settles into a stationary regime in which it oscillates with a
nearly constant mean amplitude $A_\steady$. The system is nonetheless chaotic and
dynamically rich: the instantaneous wave shape fluctuates irregularly from cycle
to cycle, driven by splashing droplets, bubble columns, and transverse sloshing,
so that meaningful comparisons with theory can only be made in a statistical
sense. To this end, we characterise the wave shape through conditionally
averaged histograms: at each phase \(\varphi = (\omega t \mod 2\pi)\), we
accumulate the \((x, z)\) positions of all points on the free surface across
many cycles, so that the most probable positions trace out the mean wave shape.
Formally, $P(x_\bin, z_\bin \vert \varphi)$ is defined as
\begin{align}
  \label{eqn:3.6}
  P(x_\bin, z_\bin \vert\varphi) = 
  \frac{1}{N}\sum_{\zeta_\surf(x, y, z, t)} \delta(x \in x_\bin, z \in z_\bin, \varphi)
\end{align}
where \((x_\bin, z_\bin)\) are the discretized values of \((x, z)\), \(N\) is
the total number of data points, and \(\sum_{\zeta_\surf(x, y, z, t)}\)
indicates a sum over every point on the free surface.

Figure~\ref{Figure9:histograms} displays $P(x_\bin, z_\bin \vert \varphi)$ at
$\varphi = 0$, corresponding to the crest of the wave cycle, for two forcing
amplitudes representative of the experiments, $\paramF = 0.25$ and $\paramF =
0.49$. In addition to observing the typical wave amplitude, we can also identify
traces of droplet ejection near the wave crests and penetrating bubbles formed
by collapsing cavities near the wave troughs and along the sidewalls. As the
forcing amplitude $\paramF$ increases, the saturation amplitude $A_\steady$
increases but appears to plateau beyond a certain threshold. What changes most
markedly is the intensity of spray and bubble columns: for low forcing, the
surface retains sufficient structure that both are negligible. As $\paramF$
increases, spray ejected at the wave crests and bubble columns descending from
collapsing troughs become progressively more intense. For the strong forcing and
intermediate stratification, as in DNS-F49-H30, the bubble columns may penetrate
deep into the stratified layer, providing a direct pathway for mass exchange
between the surface and the interior.

The saturation amplitude $A_\steady$ can be predicted following the approach of
\cite{Rajchenbach2015}. A nonlinear dispersion relation of the form
\begin{align}
\label{eqn:3.7}
\omega_0' = \omega_0\left(1 + \tfrac{1}{2}K(k\alpha)^2\right)
\end{align}
is introduced into a damped Mathieu equation. Here, $\omega_0$ is the natural
frequency of the surface wave predicted by linear theory
(Appendix~\ref{section:appendix:stability}), $\omega_0'$ its amplitude-dependent
nonlinear counterpart, and $K$ a nonlinear detuning parameter that depends on
the fluid depth, obtained from a perturbation expansion to third order in the
wave steepness~\citep{Penney1952, Tadjbakhsh1960}. In the deep water
approximation, this gives $K = -1/8$.
At saturation, the amplitude-dependent frequency shift brings the oscillator
into resonance balance with the forcing, yielding the implicit condition
\begin{align}
  \label{eqn:3.8}
  3 K k^2 \vert A_\steady \vert^2 = 
  \left(\frac{1}{4}\frac{\omega^2}{\omega_0^2}-1\right) 
  \mp \frac{1}{2} \left[F^2 - \left(\frac{\omega\gamma}{\omega_0^2}\right)^2\right]^{1/2}
\end{align}
where the $\mp$ sign corresponds to the two stable solution branches. Details of
the derivation are given in Appendix~\ref{section:appendix:amplitude}. As shown
in Figure~\ref{Figure9:histograms}, the interface shape predicted by
\eqref{eqn:3.5a}--\eqref{eqn:3.5b} using the saturation amplitude $A_\steady$
from \eqref{eqn:3.8} is in good agreement with the DNS histograms for the range
of parameters considered.

\subsubsection{Damping rates during the attenuation phase}
\label{section:validation:damping}
\begin{figure}
  \centering
  \begin{minipage}{0.9\linewidth}      	
  (a) \hfill (b) \hfill~ \\
  \includegraphics[width=0.5\linewidth]{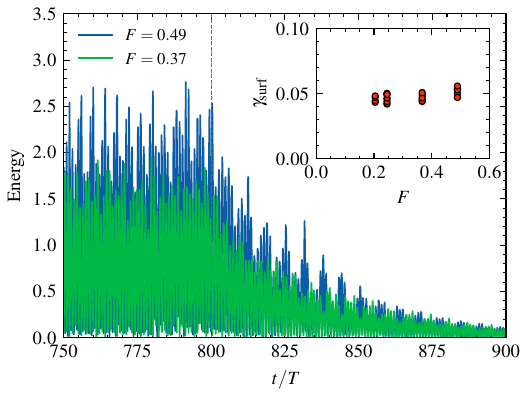}
  \includegraphics[width=0.5\linewidth]{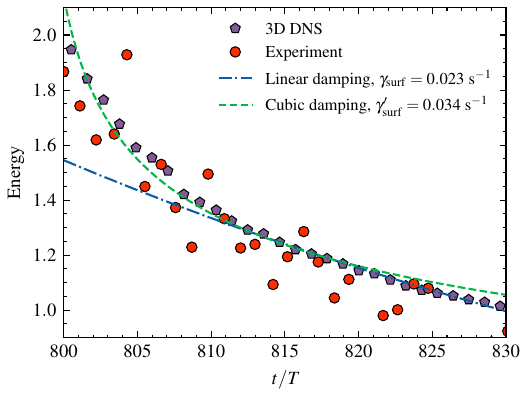}				
  \end{minipage}
  \caption{
    Mechanical energy as a function of time during the attenuation phase, which
    is used to measure the damping rate \(\gamma_\surf\). (a) Measurements from
    the experiments EXP-F49-H16 and EXP-F37-H14; Inset: average
    \(\gamma_\surf\) over the attenuation phase as a function of \(F\).
    (b) Comparison between the DNS-F49-H30 and EXP-F49-H30.
    For small amplitudes, measurements approximate a linear damping behaviour
    with \(\gamma_\surf=0.02\)~\unit{\per\second} (shown in
    {\color{blue}\dashdotted}), while for large amplitudes, they approximate a
    Duffing oscillator (shown in {\color{green}\dashed}).
  }
  \label{Figure10:Decay}
\end{figure}

As mentioned in \S\ref{section:surface:experiments}, in the experiments the
linear damping rates, \(\gamma_\surf\), were measured by tracking wave amplitude
peaks during the attenuation phase and fitting them to an exponential function.
As done in \S\ref{section:validation:growth} for the growth rates, measured
damping rates were corroborated from the evolution of mechanical energy, which
decays at twice the rate \(2\gamma_\surf\). 

Figure~\ref{Figure10:Decay}a shows the average damping rate over the attenuation
phase, \(\gamma_\surf \approx 0.05~\unit{\per\second}\), nearly independent of
the forcing. This average over-estimates the linear damping, however, because
the decay proceeds in two distinct regimes (Figure~\ref{Figure10:Decay}b). At
large amplitudes, typical of early attenuation, the decay follows a Duffing
free-decay envelope (see Appendix~\ref{appendixB:mathieu:cubic2}):
\begin{align}
  \label{eqn:3.9}
  A(t) &= A_\steady \left(1 + \beta t \right)^{-1/2}, 
  \quad
  \text{with}
  \quad
  \beta=\gamma'_\surf k^2 A^2_\steady,
\end{align}
where \(A_\steady\) is the wave amplitude at the onset of attenuation, \(\beta\)
the effective nonlinear damping rate, and $\gamma'_\surf$ the cubic damping
coefficient. This behaviour is expected given the strongly nonlinear state of the
wave at the onset of attenuation, and is confirmed by both experiments and DNS;
DNS results are less scattered owing to the absence of mechanical noise. At
smaller amplitudes, the decay transitions to an exponential with rate
\(\gamma_\surf = 0.023~\unit{\per\second}\), consistent with a linear damping
term. The larger value in panel~(a) is thus an averaging artifact, not a
discrepancy between the experiments and the DNS.

For small-amplitude waves, the classical analysis of \cite{Keulegan1959} and
\cite{Miles1967}, as detailed in \cite{Cavelier_2022}, identifies several sources
of damping: viscous losses in the bulk, losses near the solid walls, contact-line
dynamics, and damping at the liquid--liquid interface. For this configuration,
viscous dissipation at the front and back walls dominates, with all remaining
contributions at least one order of magnitude smaller,
\begin{align}
  \label{eqn:3.10}
  \gamma_\surf \approx \frac{\nu}{D\delta_w} = \frac{\sqrt{\nu \omega}}{\sqrt{2}D},
\end{align}
where \(\delta_w = (2\nu/\omega)^{1/2}\) is the boundary layer width. Equation
\eqref{eqn:3.10} yields \(\gamma_\surf = 0.027~\unit{\per\second}\), in good
agreement with the measured value of \(0.023~\unit{\per\second}\) in the linear
regime. This suggests that the DNS correctly captures the energy dissipation
rate of surface waves.

\subsubsection{Density profiles during the long-term evolution}
\label{section:validation:longterm}
\begin{figure}   
	\begin{center}
		(a) \hfill (b) \hfill (c) \hfill~ \\
		\includegraphics[width=0.32\linewidth]{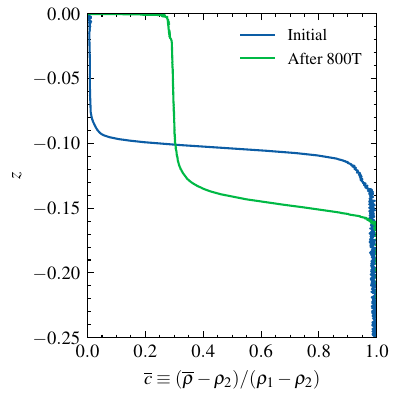}		
		\includegraphics[width=0.32\linewidth]{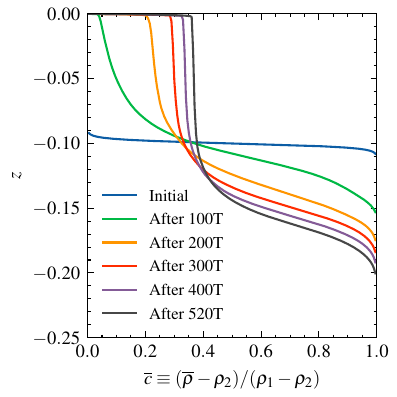}		
		\includegraphics[width=0.32\linewidth]{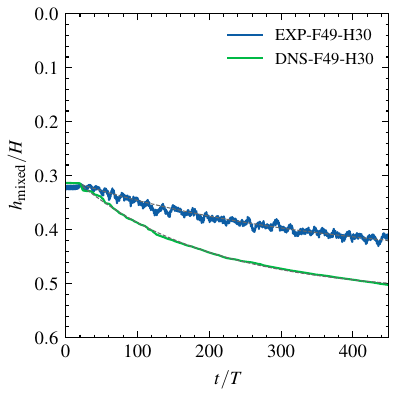}		
	\end{center}  
  \caption{
	(a) Normalized density profiles obtained from DNS-F49-H30 with
	\(\schmidt=1\) using the spatial CDF of \(\rho\). (b) Corresponding
	measurements from EXP-F49-H30. (c) Mixed-layer height, \(h_\mixed\), as a
	function of time, comparing DNS and experiment.
  }
  \label{Figure11:profiles}
\end{figure}

Figure~\ref{Figure8:snapshots} showed that the interfacial wave is well captured
in the DNS, despite some differences observed at small scales, due in part to
the transverse integration of the experimental imaging. The cumulative effect of
these differences is most apparent in the normalized density profiles presented
in Figure~\ref{Figure11:profiles}. As \(\bar{\rho}(z,t)\) evolves, the upper
layer gradually becomes a homogeneous mixture, while the lower layer remains
largely unmixed, reducing the concentration gradient and pushing the interface
downwards (panel a). This is in good qualitative agreement with
conductivity-probe measurements taken at the end of the experiment (panel b),
although the mixed-layer deepens faster in the DNS (panel c). Both the DNS and
the experiment are well described by the exponential evolution~\eqref{eqn:2.7},
but the entrainment rate $\gamma_\mixed$ --- the relaxation rate of $h_\mixed$
toward $h_\infty$ in \eqref{eqn:2.7} --- in the DNS is more than three times
that of the experiment.

We attribute this difference to a combination of physical and numerical factors.
The main physical factor is the difference in molecular diffusivity
(\(\schmidt=700\) in the experiments, \(\schmidt=1\) in the DNS).  Numerically,
both the dissipation accumulating over the long simulations and the finite
spatial resolution, which may bias the drop-size distribution toward
artificially large droplets, can contribute to the faster mixing in the DNS.
Overall, the DNS reproduces the wave amplitudes, growth rates, and damping
rates quantitatively (\S\ref{section:validation:growth}--\ref{section:validation:damping}).
Long-term mixing is reproduced only qualitatively: the essential features of
the density evolution match, but the DNS entrainment rate is up to three times
faster than the experiment. We rely on the DNS in what follows to access
quantities beyond the reach of the measurements.
%
%
\section{Energy injection by the surface waves}
\label{section:surface}

\begin{figure}   
  \begin{center}
    \begin{minipage}{0.9\linewidth}
    \begin{tabular}{lll}
      (a) KE from leading POD modes, 
      \(\Kwave=\frac{1}{2}\vec{\tilde{u}}^2\) 
      & 
      (b) KE from the remaining POD modes, 
      \(\Kturb=\frac{1}{2}{\vec{u}'}^2\)
      \\
      \includegraphics[width=0.5\linewidth]
      {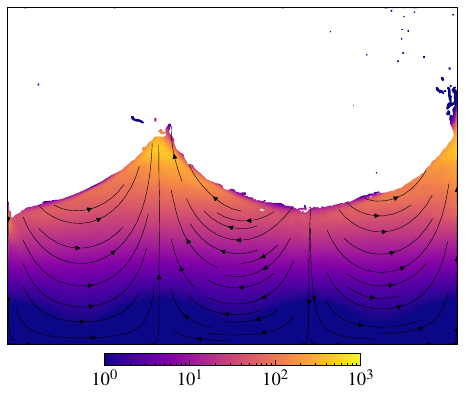}
      &
      \includegraphics[width=0.5\linewidth]
      {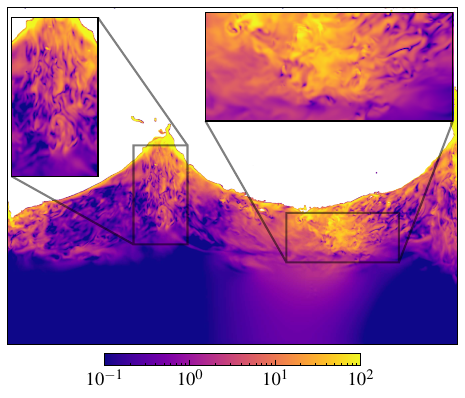}
      \\
      (c) Viscous dissipation, \( \epsilon = 2\nu(\mathsfbi{S}:\mathsfbi{S}) \) 
      &
      (d) Anisotropy coefficient, $b_{zz} = \Kturb_v/\Kturb$
      \\
      \includegraphics[width=0.5\linewidth]
      {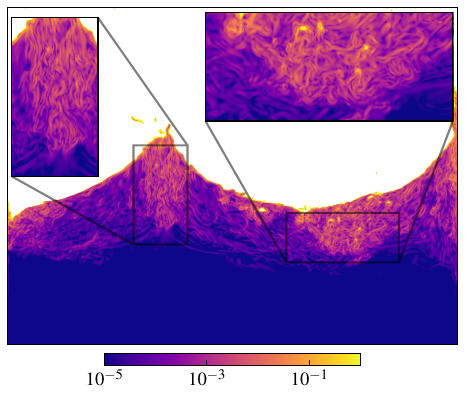}
      &
      \includegraphics[width=0.5\linewidth]
      {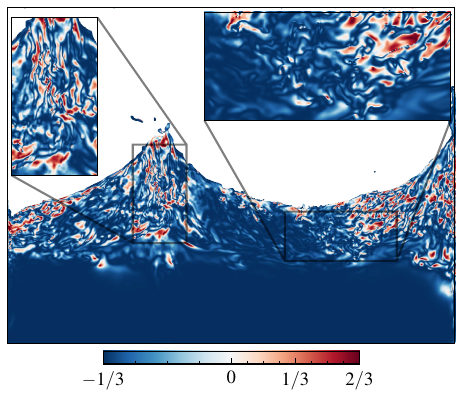}      
    \end{tabular}
    \end{minipage}
  \end{center}

  \caption{
    Spatial distribution of different quantities measured from DNS-F49-H30 taken
    at the transversal midplane \(y=0\) and \(t=200T\). Subfigure (a) shows the
    kinetic energy contained in the first 10 POD modes (large-scale features
    associated to the oscillating motion) with superimposed streamlines, while
    subfigure (b) shows the remainder (small-scale features associated to
    turbulence). Subfigure (c) shows the viscous dissipation rate, while
    subfigure (d) shows the vertical anisotropy coefficient $b_{zz} =
    \Kturb_v/\Kturb-1/3$. 
  }
  \label{Figure12:scales}
\end{figure}

Having established in \S\ref{section:validation} that the DNS reproduces the
correct wave amplitudes, growth rates, and dissipation rates, we now use the
numerical simulations to investigate how surface waves inject energy into the
fluid. In this section, we characterise the spatial distribution of kinetic
energy, dissipation, and turbulent transport as functions of depth, and show how
these quantities set the length scales relevant to mixing.

\subsection{Separating wave motion from turbulence}
\label{section:surface:validation:energy}

To separate oscillating wave motion from turbulent fluctuations, we apply a
POD-based scale decomposition (see Appendix~\ref{section:appendix:decomposition}
for details). We use separate POD bases for the velocity and concentration
fields. A joint decomposition, used in studies of stratified turbulence and
convection \citep{podvin2015, Castillo2019}, would directly connect flow
structures to concentration transport, but requires fixing a scaling parameter
that relates the two field norms. The overall conclusions of the analysis are
unchanged by this choice.

For the velocity field, the large-scale component
$\vec{\widetilde{u}}(\vec{x},t)$, reconstructed from the first 10 POD modes,
captures the oscillating wave motion, while the remainder $\vec{u}'(\vec{x},t)$
is associated with small-scale turbulence. For the concentration field, the
decomposition involves three contributions: a short-time average
$\overline{c}(\vec{x},t)$, which captures the slow progression of the mixing
layer; an oscillating component $\widetilde{c}(\vec{x},t)$, obtained from the
POD modes of the fluctuation; and a turbulent remainder $c'(\vec{x},t)$. 
From these decompositions, we evaluate the following : the large-scale kinetic
energy, $\Kwave = \tfrac{1}{2}|\vec{\widetilde{u}}|^2$; the horizontal
fluctuations, $\Kturb_h = \tfrac{1}{2}({u'_x}^2 + {u'_y}^2)$, and vertical
fluctuations, $\Kturb_v = \tfrac{1}{2}{u'_z}^2$ (with
$\Kturb=\Kturb_h+\Kturb_v$); the square of the vertical shear, $\Shear^2 = (\partial
u_x/\partial z)^2 + (\partial u_y/\partial z)^2$ (as with the velocity field,
$\Shear=\ShearWave+\ShearTurb$ decomposes linearly into a coherent
contribution and a turbulent remainder $\ShearTurb$); the viscous energy dissipation
rate, $\epsilon = 2\nu \mathsfbi{S}:\mathsfbi{S}$; the vertical anisotropy
coefficient, $b_{zz} = \Kturb_v / \Kturb-1/3$ ; and the turbulent concentration
flux, $q_z = u_z' c'$.

\subsection{Spatial organisation of energy injection, dissipation and anisotropy}

Figure~\ref{Figure12:scales} shows the spatial distribution of $\Kwave$,
$\Kturb$, $\epsilon$, and $b_{zz}$ at the transversal midplane and $t=200T$, as
the surface wave approaches its peak amplitude. The streamlines of
$\vec{\widetilde{u}}$ (panel a) reveal the characteristic orbital motion of a
standing wave with wavenumber $k$. The associated large-scale kinetic energy
$\Kwave$ decreases monotonically from the free surface to the bottom wall,
consistent with the potential-flow solution for standing gravity waves. In
contrast, the small-scale structures --- reflected in $\Kturb$ and $\epsilon$
(panels b-c) --- are strongly localized below the wave anti-nodes, where energy
injection is most intense. These structures are generally confined to the mixed
layer, but may locally penetrate into the stratified region below (see
Supplementary Material). Near the anti-nodes, the turbulent structures are
elongated in the vertical direction, with vertical velocity fluctuations
dominating over horizontal ones, as evidenced by the elevated values of $b_{zz}$
in panel (d).

\subsection{Depth dependence in the presence of stratification}

\begin{figure}   
  \begin{center} 
    \begin{minipage}{0.9\linewidth}     
    \hspace{2em}(a) \hfill (b) \hfill (c) \hfill (d) \hfill~\\
    \includegraphics[width=\linewidth, trim=0 180 0 0, clip]{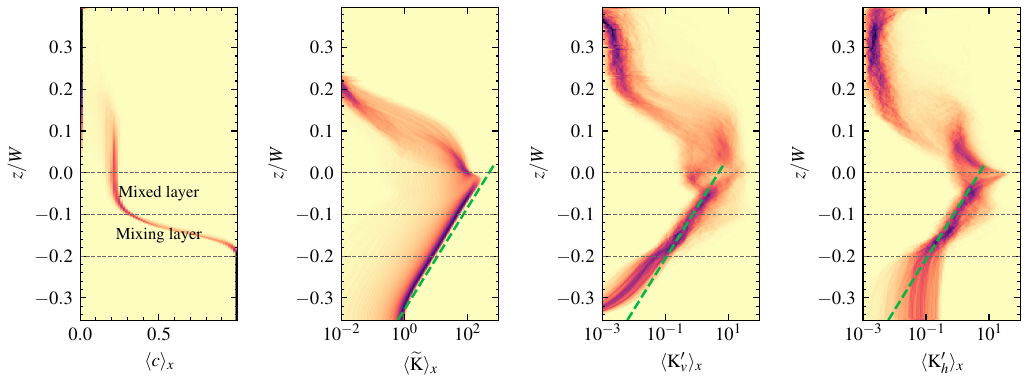}

    \hspace{2em}(e) \hfill (f) \hfill (g) \hfill (h) \hfill~\\
    \includegraphics[width=\linewidth, trim=0 0 0 180, clip]{./figures_R1/figure13_alt-compressed.pdf}
    \end{minipage}
  \end{center}

  \caption{
    Vertical profiles from DNS-F49-H30 for $t \in [150T, 210T]$,
    represented as 2D histograms with darker colours indicating the most likely
    profiles:
    (a) concentration field;
    (b) large-scale kinetic energy;
    (c) vertical fluctuations;
    (d) horizontal fluctuations;
    (e) anisotropy coefficient;    
    (f) viscous dissipation rate;    
    (g) turbulent concentration flux;
    (h) turbulent length scale.
    Lines with $e^{2kz}$ ({\color{gray}\dashed}) and
    $e^{3kz}$ ({\color{gray}\dashdotted}) are shown for reference; these
    exponents follow from the wave-velocity decay $e^{kz}$ through the TKE
    budget and the depth-independence of $l_\turb$ (\S\ref{section:surface}).
  }
  \label{Figure13:scales}
\end{figure}

Figure~\ref{Figure13:scales} complements this picture with time-averaged
vertical profiles of the same quantities, quantifying their depth dependence
over the stationary phase. Two key observations can be made here. 
First, regarding the spatial distribution of energy: large-scale flow features
contain most of the kinetic energy. This energy typically peaks at the surface
level and decreases with $e^{2kz}$, consistent with the characteristics of
the potential solution (panels a-b). Notably, these large
scales showed little change with the initial stratification; most differences in
energy distribution are concentrated in the small scales. 
In contrast, contributions from the small scales also decrease with
$e^{2kz}$ within the mixed layer. Because stratification acts as a barrier
at the interface, fluctuations are confined to the mixed layer, producing a
visible kink in the profiles (panels c-d). In this picture, the energy injected
at the surface would be strongly anisotropic, dominated by vertical oscillations
(near the surface, panel c). However, within the mixed layer, the horizontal and
vertical fluctuations become comparable, suggesting that turbulent processes
redistribute the anisotropic input evenly across directions. Below the mixed
layer, stratification prevents vertical fluctuations from penetrating further;
only horizontal motion persists and $b_{zz}$ drops.
The same barrier effect confines other quantities to the mixed layer:
$\epsilon$, $\Shear^2$, and $q_z$ all show sharp transitions at the stratified
interface (panels f-g).  

Second, within the mixed layer, contributions to $\epsilon$ decay more steeply
than $e^{2kz}$, with profiles broadly consistent with $e^{3kz}$ (panel f). This
dependence can be understood by examining the balance of terms in the
phase-averaged transport equation for the turbulent kinetic energy, derived
from the triple decomposition of \citet{Reynolds1972}:
\begin{align}
  \label{eqn:4.1}
  \frac{\partial }{\partial t}  \tfrac{1}{2} \overline{u'_i u'_i} 
  + \overline{
    \widetilde{u}_j \frac{\partial}{\partial x_j} \langle \tfrac{1}{2} u'_i u'_i \rangle_{\varphi}
  }
  =&
  - \frac{\partial}{\partial x_j}\left[
      \frac{1}{\rho} \overline{p'u'_j} 
      + \tfrac{1}{2} \overline{u'_iu'_iu'_j} 
      - 2\nu \overline{u'_i S_{ij}'}
    \right]
  \\\nonumber&
  - \overline{\langle u'_i u'_j \rangle_{\varphi} \frac{\partial \widetilde{u}_i}{\partial x_j}}
  - \frac{g \overline{u'_z \rho'}}{\rho_0}
  - 2\nu \overline{S'_{ij}S'_{ij}}
\end{align}
where the flow has been decomposed as $\vec{u} = \vec{\widetilde{u}} + \vec{u}'$
with no mean component, $\langle \cdot \rangle_{\varphi}$ denotes an average over repeated
wave cycles at fixed phase, and $\rho_0=(\rho_1+\rho_2)/2$. Since both
$\vec{\widetilde{u}}$ and $\vec{u}'$ decay as $e^{kz}$, the advection of energy
by the oscillating field and the turbulent transport flux divergence in
\eqref{eqn:4.1} would both scale as $e^{3kz}$. If transport is the dominant
balance against dissipation, the dissipation would follow a similar scaling
$\epsilon \propto e^{3kz}$, consistent with the observed profile (panel f).

Although a regime with $\epsilon \propto e^{2kz}$ is also compatible with the
TKE equation, two independent measurements allow us to discriminate between the
two. If we consider that both quantities decrease exponentially with depth, 
\begin{align}
  \Kturb(\vec{x},t) = C_E(x,y,t) e^{mkz}, ~~
  \epsilon(\vec{x},t) = C_\epsilon(x,y,t) e^{nkz},
\end{align}
then, according to Kolmogorov the turbulent length scale scales as 
\begin{align}
  l_\turb \propto \frac{{\Kturb}^{3/2}}{\epsilon} = \frac{C_E^{3/2}}{C_\epsilon} e^{(\frac{3}{2}m-n)kz}
\end{align}
Our measurements show that $l_\text{turb}$ is independent of $z$ within the
mixed layer (panel h), which requires $\tfrac{3}{2}m - n = 0$. An independent
test is to consider the scaling of the fluctuating vertical shear. If we approximate
$\ShearTurb^2$ as 
\begin{align}
  \ShearTurb^2(\vec{x},t) \approx (u_\turb/l_\turb)^2 = C_S(x,y,t) e^{pkz},
\end{align}
then $p=n-\frac{1}{2}m$. With $m=2$ measured directly from the $\Kturb$-profile,
both constraints give $n=3$ and $p=2$. We will revisit these quantities in
\S\ref{sec:model} in the context of a turbulent diffusion model.

\subsection{Depth dependence in the absence of stratification}

\begin{figure}   
	\begin{center}
    \begin{minipage}{\linewidth}
      (a) \hspace{0.26\linewidth} (b) \hspace{0.26\linewidth} (c) \hfill~ \\
      \begin{minipage}{0.58\linewidth}
        \begin{tabular}{cc}
          \scriptsize
          Fresh-water and salt-water & 
          \scriptsize
          Fresh-water and passive tracer \\[0.5em]
          \includegraphics[width=0.49\linewidth]{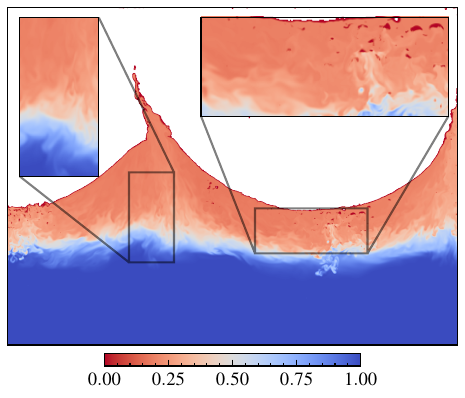} &		
          \includegraphics[width=0.49\linewidth]{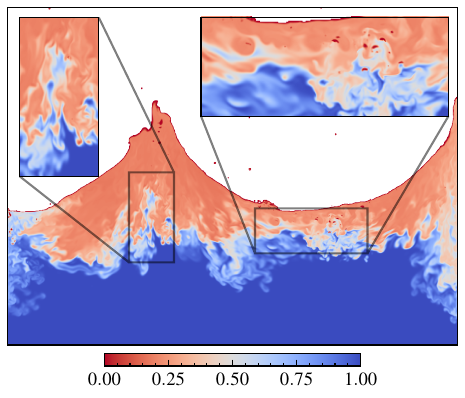}
        \end{tabular}
      \end{minipage}\hfill
      \begin{minipage}{0.41\linewidth}
        \includegraphics[scale=0.6, trim= 0 0 397 0, clip]{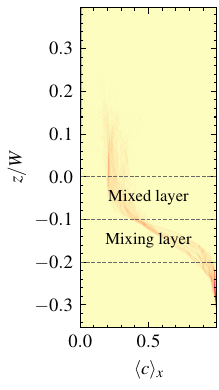}
        \includegraphics[scale=0.6, trim= 268 0 161 0, clip]{./figures_R1/figure14c.pdf}
        \includegraphics[scale=0.6, trim= 426 0 0 0, clip]{./figures_R1/figure14c.pdf}
      \end{minipage}
    \end{minipage}
	\end{center}  
  \caption{
    Comparison of the concentration field at \(t=175T\) for DNS-F49-H30, for (a)
    the fresh-water and salt-water case and (b) the fresh-water and passive
    tracer case at the same instant. Without the buoyancy barrier (b), turbulent
    structures penetrate markedly deeper into the lower layer and the interface
    does not re-stratify, in contrast to the stratified case (a). Panel (c)
    shows vertical profiles of the horizontally-averaged concentration, viscous
    dissipation, and turbulent length-scale for the passive-tracer case,
    presented in a similar manner as in Figure~\ref{Figure13:scales}; in the
    absence of a barrier the dissipation follows a clear \(e^{3kz}\) decay.
    Lines with \(e^{2kz}\) ({\color{gray}\dashed}) and \(e^{3kz}\)
    ({\color{gray}\dashdotted}) are shown for reference, see text.
  }
  \label{Figure14:tracer}
\end{figure}

To isolate the role of the stratification, we now consider the limiting case of
a passive scalar. This is a dedicated diagnostic rather than one of the
production simulations. For this comparison, we took a snapshot of a weakly
stratified simulation once the Faraday instability had fully developed and used
its concentration field to initialise a passive tracer, which is equivalent to
taking a stratified simulation and lowering the Atwood number artificially to
zero. The surface forcing is left unchanged, and the flow is allowed to evolve
for about $30T$ so that the results are representative of a passive scalar.

Without the buoyancy barrier the flow develops a markedly different structure
(Figures~\ref{Figure14:tracer}a-b). Turbulent eddies penetrate far deeper into
the lower layer, the mixing region acquires sharp discontinuities at its edges,
and the horizontal homogenization and re-stratification that characterise the
stratified case are absent, so fresh fluid parcels injected by breaking waves
tend to remain where the turbulence left them. Despite this different structure,
the horizontally-averaged dissipation follows the same $e^{3kz}$ decay as in
the stratified case, now extending well into the bottom layer
(Figure~\ref{Figure14:tracer}c). Conversely, the turbulent length-scale remains
more or less constant throughout the mixing region, rising by about an order of
magnitude only as it approaches the bottom wall. This suggests the
$e^{3kz}$ decay is intrinsic to the Faraday-wave forcing --- set by the
production terms of \eqref{eqn:4.1} --- and not a consequence of the
stratification. Instead, the stratification fixes the penetration depth at which
the turbulence is arrested, without changing the $e^{3kz}$ dependence in the
upper layer. 
%
%
\section{Short-term evolution of the stratification}
\label{section:stratification:short}

In \S\ref{section:surface} we characterised the turbulence injected into the
mixed layer by the surface waves. Turbulent entrainment, as described by
\cite{Linden_1973}, requires turbulent eddies to reach the interface, work
against the density gradient, and lift heavy fluid from the stratified layer
into the mixed layer. While the interface remains organised, two mechanisms
enhance this entrainment by breaking its resistance to the forcing. The first,
resonant and long-range, drives a secondary parametric instability even at
depths of order one wavelength; the second, confined to shallow interfaces,
couples this instability with near-surface turbulence and breaking. Both
dominate the early-stage dynamics and gradually fade as mixing proceeds---the
parametric instability once the interface no longer resonates with the primary
wave, the coupling once the renewed interface lies too deep to reach the
surface---leaving turbulent entrainment to act alone over the long term
(presented in \S\ref{section:stratification:long}). 

\subsection{Long-range mechanism: secondary parametric instability}
\label{section:stratification:secondary}

\begin{figure}   
	\centering
	\def\markfrac{0.5}
	\begin{tikzpicture}
		\node[anchor=south west, inner sep=0] (specb) at (0,0)
			{\includegraphics[scale=0.7, trim=0 0 0 0, clip]{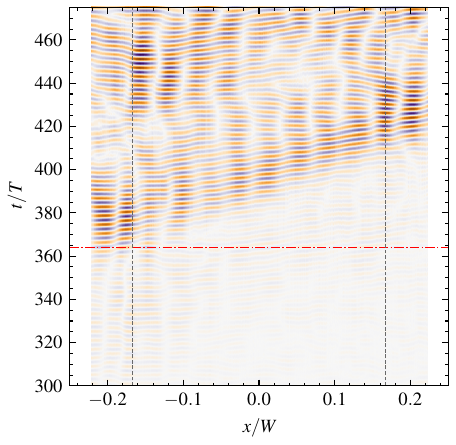}};
		
		\coordinate (mark1) at (0.85,2.34);
		\coordinate (mark2) at (5.3,2.34);
		\draw[black, thick] (mark1) -- (mark2);
		\draw[black, thick] (mark2) -- (6.5,5.2);
		\draw[black, thick] (mark2) -- (6.5,0.8);

		\node[anchor=south west, inner sep=0] (snapa) at ($(specb.south east)+(1cm,0.75cm)$)
			{\includegraphics[scale=0.73, trim=0 0 0 0, clip]{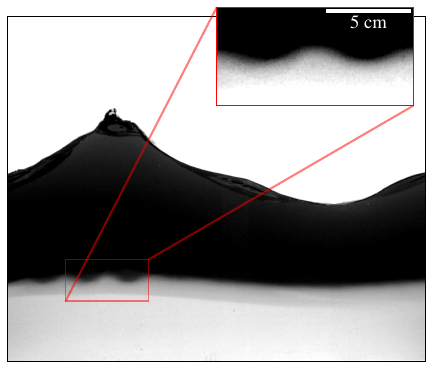}};
	\end{tikzpicture}
	\caption{
	(a) Spatiotemporal diagram of $\zeta_\mix(x,t)$ filtered at frequency
	\(\omega/4\), showing the horizontal propagation of the wave; vertical lines
	mark the positions of the wave anti-nodes, and the horizontal line indicates
	the instant at which the snapshot in panel~(b) was taken. (b) Image taken
	during EXP-F24-H49 at that instant, showing an interfacial wave developing
	via a secondary Faraday instability. The observed spatial frequency is
	consistent with linear
	stability predictions for forcing at \(\omega/2\) (see
	Appendix~\ref{section:appendix:stability}).
	}
	\label{Figure15:Secondary}
\end{figure}

As the free surface oscillates, it advects the miscible interface below,
subjecting it to an advection-induced acceleration in addition to gravity and
the external forcing. This oscillating motion stretches and compresses the
interface at the wave anti-nodes while inducing an oscillating shear at the
nodes, destabilising it even at depths of order one wavelength below the
surface. We identify this as a secondary parametric resonance, of the kind
previously reported by~\cite{Cavelier_2022} and~\cite{Liu2022}. Two distinct
forms are observed, depending on the amplitude of the induced acceleration, and
are described in the following.

The first form is a secondary Faraday instability of the miscible interface.
Figure~\ref{Figure15:Secondary} shows an example of an interfacial Faraday wave
with frequency $\omega/4$, which forms near one of the anti-nodes of the primary wave
before spreading horizontally to the rest of the domain; this propagation is
clearly visible in the spatiotemporal diagram of panel~(a).  The spatial
frequency $k \approx 19\pi/W$, measured from panel~(b) and confirmed by the
filtered signal in panel~(a), is consistent with linear stability predictions
for a forcing at $\omega/2$ (see Appendix~\ref{section:appendix:stability}). The
interface grows until it no longer resonates with the primary wave. We estimate
this effect using the saturation width expression from \cite{Grea2018}, taken
here as \(L_\mathrm{sat}=32\atwood g/\omega^2\), since the acceleration induced
by the primary wave is vanishingly small with depth and the forcing frequency is
\(\omega/2\). The resulting estimate is of the order of \(2-3~\cm\), which
aligns well with the experimental measurements.

\begin{figure}
	\centering
	\begin{minipage}{0.9\linewidth}
	\includegraphics[width=0.33\linewidth, trim=150 65 167.5 52.5, clip]{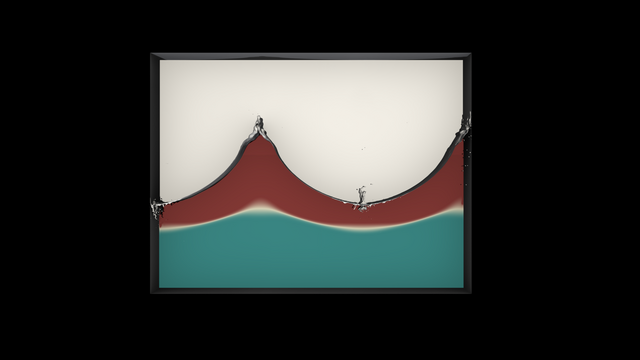}\hfill
	\includegraphics[width=0.33\linewidth, trim=150 65 167.5 52.5, clip]{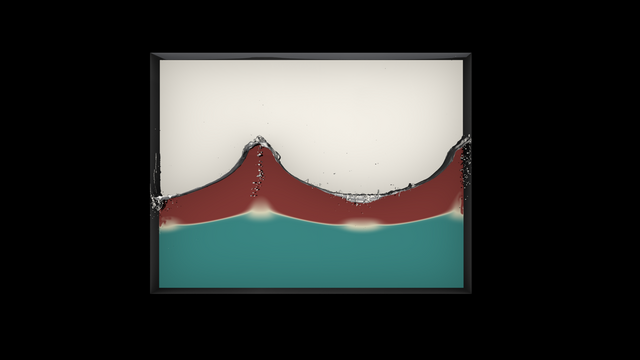}\hfill
	\includegraphics[width=0.33\linewidth, trim=150 65 167.5 52.5, clip]{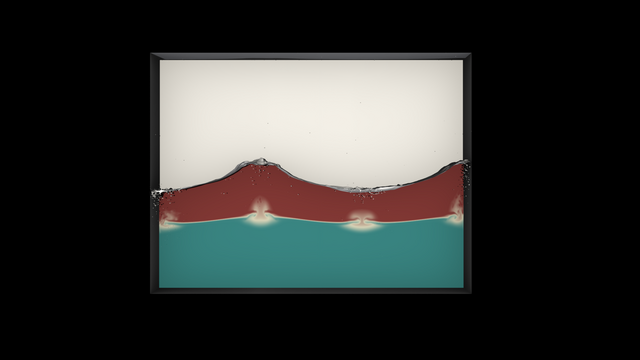}

	\vspace{-2.9cm} \quad(a) \hfill \quad(b) \hfill\quad (c) \hfill~\\ \vspace{2.20cm}

	\includegraphics[width=0.33\linewidth, trim=150 65 167.5 52.5, clip]{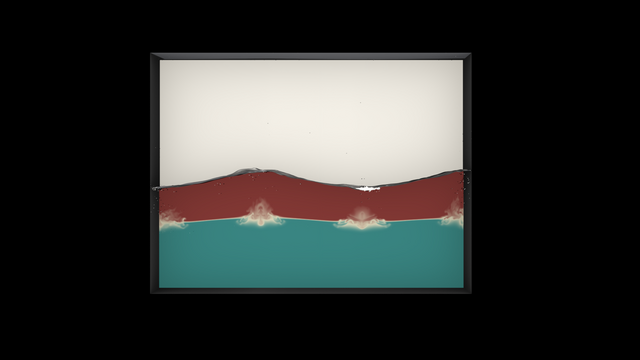}\hfill
	\includegraphics[width=0.33\linewidth, trim=150 65 167.5 52.5, clip]{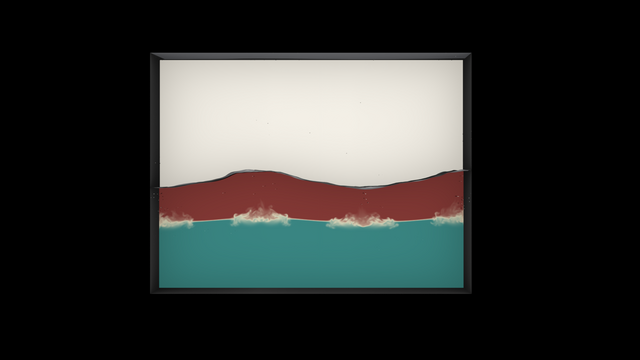}\hfill
	\includegraphics[width=0.33\linewidth, trim=150 65 167.5 52.5, clip]{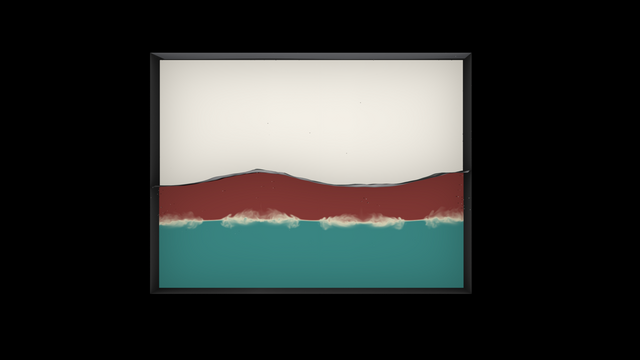}

	\vspace{-2.9cm} \quad(d) \hfill \quad(e) \hfill\quad (f) \hfill~\\ \vspace{2.20cm}
	\end{minipage}
	\caption{
	Sequence of snapshots taken every \(2T\) for DNS-F37-H37, showing the
	destabilization of the internal stratification below the wave anti-nodes,
	followed by horizontal homogenization and re-stratification into a thicker
	mixed layer.
	}
	\label{Figure16:deep}
\end{figure}


The second form occurs when the induced acceleration is strong. As shown in
Figure~\ref{Figure16:deep}, the interface below the anti-nodes is destabilized,
forming mushroom-like structures that break down and re-stratify into a thicker
interface after horizontal homogenization. Although the mushroom-like structures
resemble Rayleigh-Taylor instability, local acceleration in the reference frame
attached to the interface does not reverse sign, indicating the mechanism is
parametric rather than convective. This second form is observed in all DNS runs
at some stage of the evolution. For shallow stratifications, however, this
instability couples with near-surface turbulence and surface breaking, enhancing
mixing as described in the following section.

\subsection{Short-range mechanism: coupling with near-surface turbulence}
\label{section:stratification:coupling}

\begin{figure}
	\centering
	\begin{minipage}{0.9\linewidth}
	\includegraphics[width=0.33\linewidth, trim=150 65 167.5 52.5, clip]{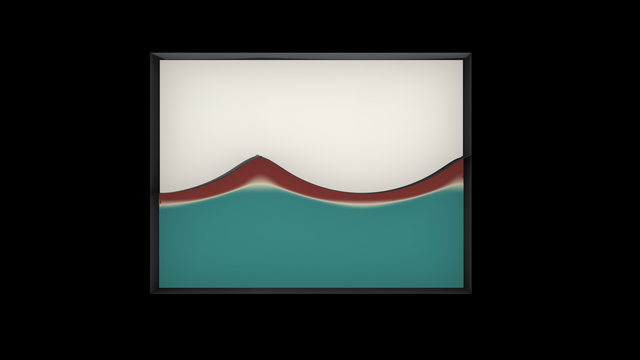}\hfill
	\includegraphics[width=0.33\linewidth, trim=150 65 167.5 52.5, clip]{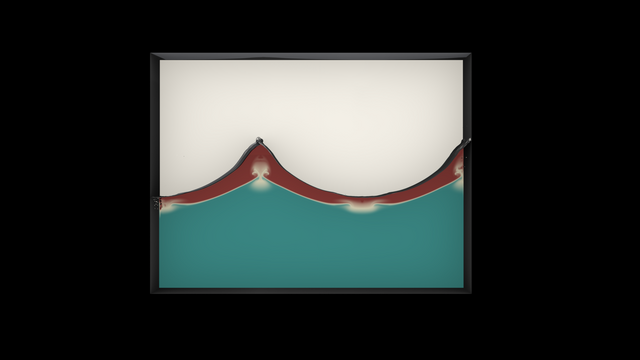}\hfill
	\includegraphics[width=0.33\linewidth, trim=150 65 167.5 52.5, clip]{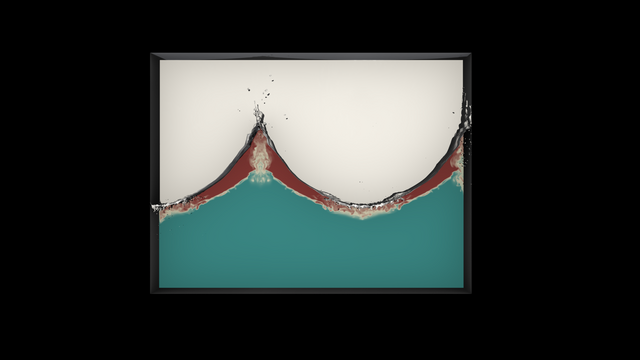}

	\vspace{-2.9cm} \quad(a) \hfill \quad(b) \hfill\quad (c) \hfill~\\ \vspace{2.20cm}

	\includegraphics[width=0.33\linewidth, trim=150 65 167.5 52.5, clip]{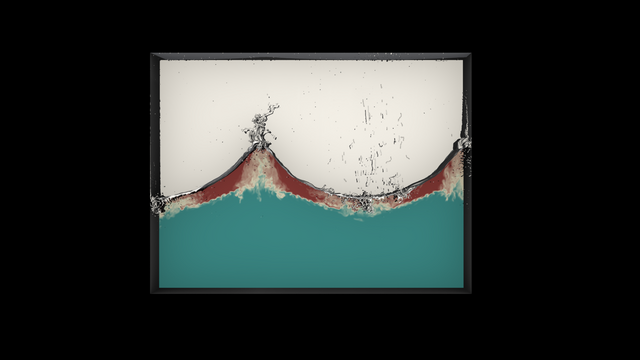}\hfill
	\includegraphics[width=0.33\linewidth, trim=150 65 167.5 52.5, clip]{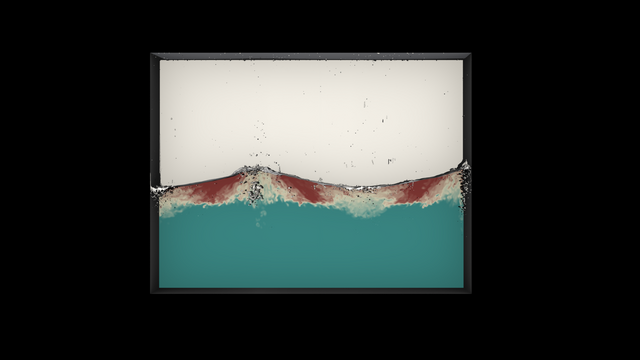}\hfill
	\includegraphics[width=0.33\linewidth, trim=150 65 167.5 52.5, clip]{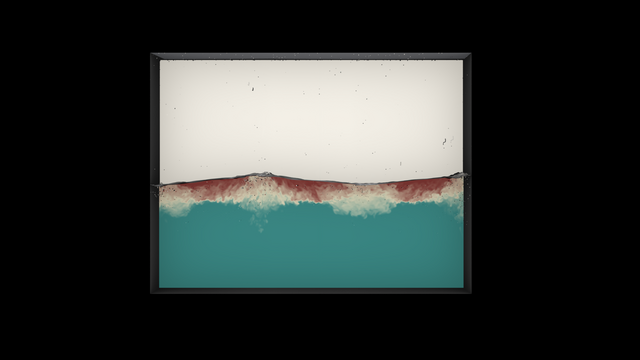}

	\vspace{-2.9cm} \quad(d) \hfill \quad(e) \hfill\quad (f) \hfill~\\ \vspace{2.20cm}
	\end{minipage}
	\caption{
	Sequence of snapshots taken every \(2T\) for DNS-F37-H15, showing fluid
	ejected from the miscible interface being lifted by the surface wave and
	redistributed into the mixed layer, followed by horizontal homogenization.
	}
	\label{Figure17:shallow}
\end{figure}


For shallow stratifications, two effects combine to enhance turbulent
entrainment. The parametric instabilities described in
\S\ref{section:stratification:secondary} lower the energy barrier by breaking
down the miscible interface, making turbulent lifting more efficient. Combined
with this, proximity to the free surface allows the ejected fluid to be further
lifted by the motion of the surface wave and redistributed across the mixed
layer.

Figure~\ref{Figure17:shallow} shows a sequence of six snapshots taken every $2T$
for DNS-F37-H15. In this sequence, the miscible interface is destabilized below
the wave anti-nodes, forming the characteristic mushroom-like structures
localised at the anti-nodes, while leaving the interface between nodes
relatively undisturbed (panels a-b). The ejected fluid is then lifted by the
orbital motion of the surface wave toward the free surface, while surface
breaking injects bubbles and entrains fresh fluid into the stratification
(panels c-d). Finally, the vertically transported fluid spreads horizontally,
homogenising the mixed layer and leaving a new, thicker interface in its place
(panels e-f). 

Unlike the long-range mechanism described in
\S\ref{section:stratification:secondary}, which increases $L$ while leaving
$h_\mixed$ essentially unchanged, the short-range coupling increases both: the
interface becomes thicker and deeper after each mixing burst. If the newly
formed interface remains within coupling range of the surface, the process may
repeat, producing a staircase-like succession of mixing events. As we shall see
in \S\ref{section:stratification:forcing}, the conditions for this coupling
depend on both the forcing amplitude and the initial stratification depth.

\subsection{Influence of the stratification depth and the forcing amplitude}
\label{section:stratification:forcing}     

\begin{figure}
	\centering
	\begin{minipage}{0.9\linewidth}   
	(a) \hfill (b) \hfill~\\
	\includegraphics[width=\linewidth]{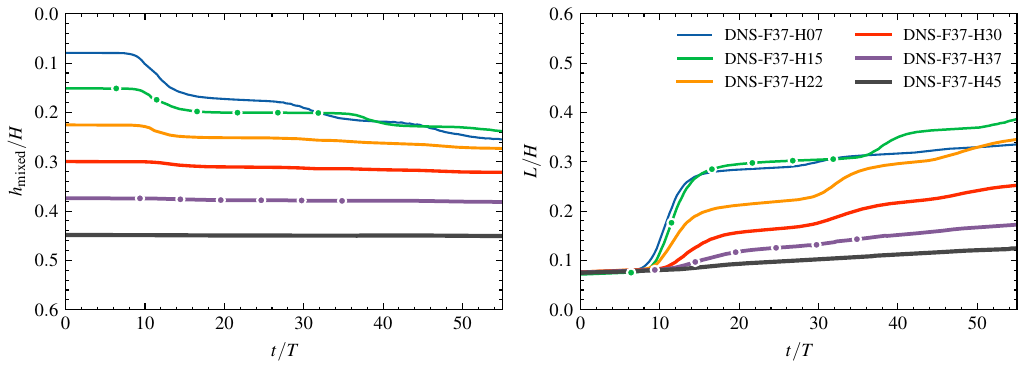}		

	(c) \hfill (d) \hfill~\\
	\includegraphics[width=\linewidth]{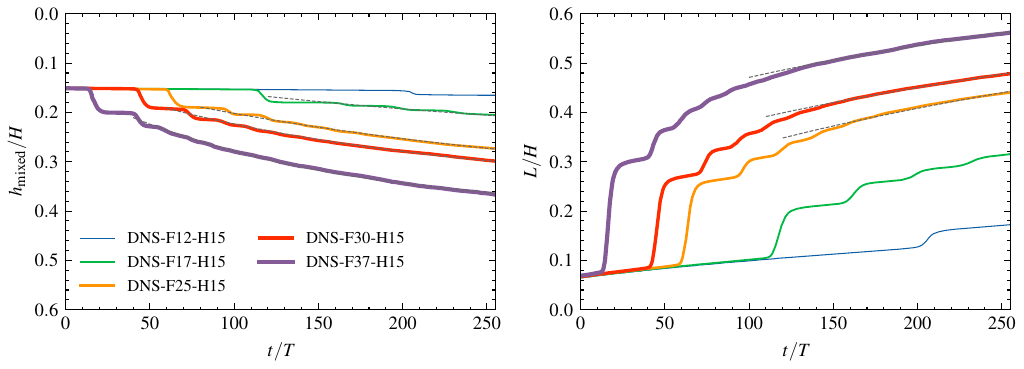}		
	\end{minipage}
	\caption{
		Time evolution of the size of the mixed layer \(h_\mixed\) (left panels)
		and the thickness of the mixing layer \(L\) (right panels). Top
		panels~(a)-(b): DNS series A at constant forcing \(\paramF = 0.37\),
		with varying initial stratification depth \(h_\init\). Bottom
		panels~(c)-(d): DNS series B at constant \(h_\init\), with varying
		forcing amplitude \(\paramF\). Markers in panels~(a)-(b) indicate the
		time instants of the snapshots presented in Figures~\ref{Figure16:deep}
		and~\ref{Figure17:shallow}.
	}
	\label{Figure18:same_forcing}
\end{figure}

Naturally, the coupling between the interface destabilization and surface motion
is expected to depend on the initial stratification depth and on the forcing
intensity. Figure~\ref{Figure18:same_forcing} presents two DNS series: series A,
at constant forcing $\paramF=0.37$ with varying initial depth $h_\text{init}$
(panels a-b), and series B, at constant $h_\text{init}$ with varying forcing
$\paramF$ (panels c-d). Together they provide a qualitative criterion to
distinguish the two mixing mechanisms.

When only the long-range mechanism is present, the breakdown and subsequent
re-stratification of the miscible interface are reflected as jumps in $L$, with
$h_\text{mixed}$ remaining essentially unchanged. As the initial stratification
depth $h_\text{init}$ decreases and the short-range coupling becomes
increasingly prominent, the mixing bursts are reflected in simultaneous jumps in
both $L$ and $h_\text{mixed}$ (series A, panels a-b). The magnitude of each jump
grows proportionally with $\paramF$, as larger forcing results in more energetic
eddies reaching the interface, lifting proportionally more fluid into the mixed
layer, resulting in more intense mixing bursts (series B, panels c-d). Note that
\(L\) initially follows a square-root behaviour with an effective molecular
diffusion coefficient, grows rapidly during the mixing bursts, before
re-stratification returns the system to a new, thicker interface, producing the
staircase pattern visible across all panels. This effective diffusion
coefficient, inferred from the growth of \(L\) in panels (c-d), is between 80
and 120 times
larger than the molecular diffusion coefficient \(\kappa_\text{mix}\). As the system
evolves, the intense mixing events reflected in these jumps become progressively
less prominent, and the dynamics settle into an asymptotic regime dominated by
turbulent entrainment.
\section{Long-term evolution of the stratification}
\label{section:stratification:long}

The resonant mechanisms of \S\ref{section:stratification:short} are transient by
nature : the secondary parametric instability subsides once the interface no
longer resonates with the primary wave, and the short-range coupling weakens as
the renewed interface sinks out of reach of the surface. What remains over the
long term is turbulent entrainment, slowly deepening the mixed layer until it
settles toward the apparent final depth reported in
\S\ref{section:surface:asymptotic}. This section presents two different aspects
of this long-term behaviour. We first consider the energetic cost of mixing
(\S\ref{section:stratification:long:energetics}), which explains why
intermediate interfaces mix the least and so accounts for the non-monotonic
dependence of the asymptotic depth on $h_\init$. We then turn to the dynamic
barrier that restricts the entrainment
(\S\ref{section:stratification:long:barrier}), set by the competition between
the stratification and the turbulence reaching the interface, and show how it
strengthens as the layer deepens, driving the mixing rate toward a small
residual value.

\subsection{Energetics of mixing}
\label{section:stratification:long:energetics}

\subsubsection{Energetic cost of mixing}

One way to understand the non-monotonic behaviour of the asymptotic depth,
previously described in \S\ref{section:surface:asymptotic}, is by considering
the energetics of mixing. In the framework of \cite{Winters_1995}, a fraction of
the injected kinetic energy is irreversibly converted into background potential
energy due to diapycnal mixing,
\begin{align} 
  \label{eqn:6:1}
  \Phi_d(t) &\equiv - \kappa_\mix ~ g \int_V \frac{\mathrm{d}z_*}{\mathrm{d}\rho} \lVert \del \rho \rVert^2~dV
\end{align} 
where $\kappa_\mix$ is the molecular diffusion coefficient. Here,
$z_*(\rho)$ is the height in the adiabatically sorted reference state, obtained
from the spatial CDF of the density field (the normalized profiles of
Figure~\ref{Figure11:profiles}), and its derivative is evaluated at the local
value density. For a closed system, in which no buoyancy flux crosses the
boundaries, $\Phi_d(t)$ is equal to the rate of change of the background
potential energy, 
\begin{align}
  \label{eqn:6:2}
  \frac{\mathrm{d}\BPE}{\mathrm{d}t} = \Phi_d(t).
\end{align}
As shown in the Appendix~\ref{section:appendix:energies}, by assuming a
piecewise density profile, we may write the amount of energy required for the
mixed layer to transition from an initial depth \(h_\init\) to a depth
\(h_\mixed\) due to irreversible mixing as 
\begin{align}
  \label{eqn:6:3}
  \Delta\BPE = \int_0^t\Phi_d(u)~du \approx \frac{g(\rho_1 - \rho_2)}{2} ( h_\mixed - h_\init )~(h_\init W D).
\end{align} 
The quadratic dependence on $h_\init$ in this equation indicates that mixing in
either very shallow or very deep interfaces requires less energy than mixing an
intermediate one. For a comparable energy input and comparable mixing
efficiency, these interfaces therefore mix further, reaching the greater final
depths reported in \S\ref{section:surface:asymptotic}.

\subsubsection{Cumulative mixing efficiency}
\begin{figure}   
  \centering
	\begin{minipage}{0.9\linewidth}   
		(a) \hfill (b) \hfill~ \\
		\includegraphics[width=0.495\linewidth]{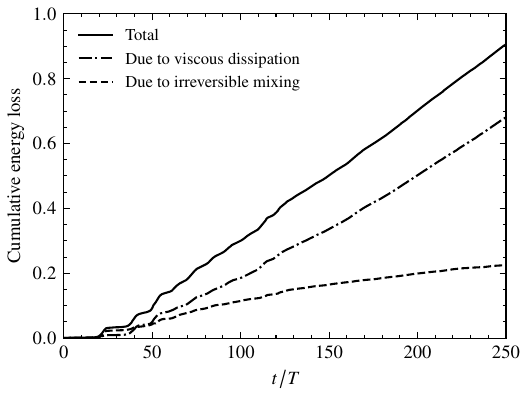}		
		\includegraphics[width=0.495\linewidth]{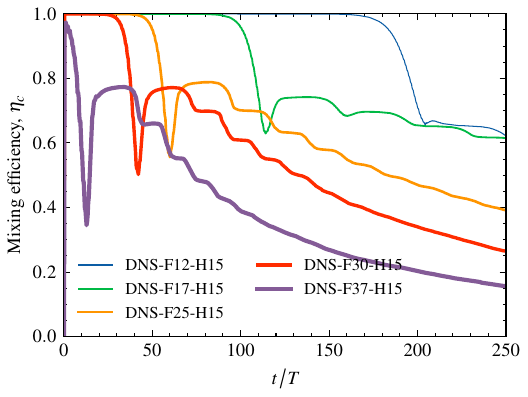}		
	\end{minipage}  
  \caption{
    (a) Cumulative energy losses as function of time obtained from DNS-F49-H30.
    (b) Cumulative mixing efficiency $\eta_c$ as function of time for DNS
    series B, previously shown in Figure \ref{Figure18:same_forcing}.
  }
  \label{Figure19:efficiency}
\end{figure}

Mixing efficiency can be defined as the fraction of energy lost to irreversible
mixing rather than to viscous dissipation~\citep{Peltier2003, Caulfield_2021}.
In the context of Faraday waves it is convenient to reason in terms of a
cumulative mixing efficiency rather than an instantaneous one, as done for
instance
by~\cite{Singh2024},
\begin{align}
  \label{eqn:6:4}
  \eta_c = \frac{\int_0^t\Phi_d(u) du}{\int_0^t\Phi_d(u) du + \int_0^t \mathcal{E}(u) du },
\end{align}
with 
\begin{align}
  \label{eqn:6:5}
  \mathcal{E}(t) \equiv \int_V \epsilon(\vec{x},t)~dV
\end{align}
where the time interval may be selected as to isolate the manner in which
mixing efficiency evolves in time.

Figure~\ref{Figure19:efficiency}a shows the cumulative losses due to mixing and
viscous dissipation, as a function of time for one of the simulations. At the
beginning of each experiment, the fluid is at rest and viscous losses are
negligible, while mixing is primarily driven by molecular diffusion giving a
high initial mixing efficiency. As the instability develops, overdriven waves
lead to enhanced mixing during the mixing bursts and viscous dissipation due to
wave breaking. In our case, energy injection is driven by the surface dynamics,
which depends very little on the stratification. In the asymptotic regime, the
viscous losses continue to grow steadily, while the entrainment slows to a low
residual rate without stopping entirely. The cumulative mixing efficiency
therefore starts around 1.0, drops sharply with each mixing burst, and continues
to decrease in the asymptotic regime (Figure~\ref{Figure19:efficiency}b). Mixing
never strictly stops, but the residual entrainment is too slow to deepen the
layer appreciably over the duration of our experiments, which therefore reach an
apparent final depth.

\subsection{Dynamic evolution of the energetic barrier}
\label{section:stratification:long:barrier}
\begin{figure}   
  \centering
	\begin{minipage}{0.9\linewidth}   
		\hspace{1.5em} (a) \hfill (b) \hfill (c) \hfill (d) \hfill~ \\
		\includegraphics[width=\linewidth]{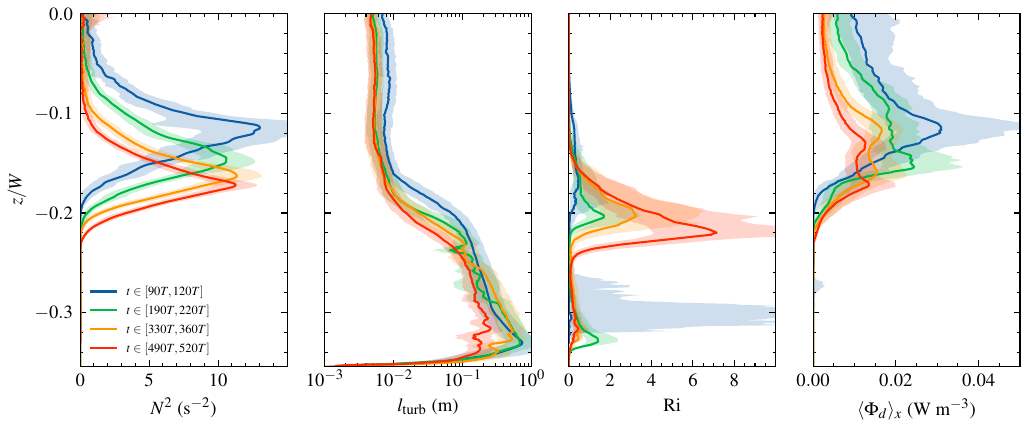}			
	\end{minipage}  
  \caption{
    Vertical profiles from DNS-F49-H30: (a) squared Brunt--Väisälä frequency
    $N^2$, (b) turbulent length-scale $l_\turb$, (c) bulk Richardson number
    $\richardson$, and (d) average diapycnal mixing rate $\Phi_d$. In
    each panel the four lines correspond to profiles averaged over the time
    windows $t \in [90T, 120T]$, $[190T, 220T]$, $[330T, 360T]$ and $[490T,
    520T]$. Here, solid lines indicate the mean value, while shaded regions
    indicate the standard deviation.
  }
  \label{Figure20:richardson}
\end{figure}

The energetic barrier reflects the competition between the stratification, which
opposes vertical motion, and the turbulence impinging on the interface, which
acts to erode it. This balance is expressed by the bulk Richardson number
\begin{align}
  \label{eqn:6:6}
  \richardson \equiv \dfrac{N^2}{(u_\turb/l_\turb)^2},
\end{align}
where $N$ is the Brunt--Väisälä frequency and $u_\turb$, $l_\turb$ are a
characteristic velocity and integral length scale of the eddies reaching the
interface (both defined in \S\ref{section:surface}). When $\richardson$
exceeds a critical value $\richardson_c$, impinging eddies cannot
penetrate any further. 

Consider first the numerator. From the initial stratification \eqref{eq2.2}, the
squared buoyancy frequency $N^2=-(g/\rho_0)(\partial\overline\rho/\partial z)$
follows a $\operatorname{sech}^2$ profile, concentrated in the mixing layer of
thickness $L$ and essentially zero elsewhere. Evaluated at the centre of the
interface, its peak value is $N^2_{\max}\approx 100~\mathrm{s^{-2}}$ at the
beginning of each simulation, but decreases rapidly during the transient phase,
due to the thickening of the interface and mixing in the upper layer. In the
stationary state the density gradient continues to weaken, but only very
gradually. Overall, the shape of the $N^2$ profile evolves slowly, and simply
follows the progression of the mixed layer as it deepens
(Figure~\ref{Figure20:richardson}a).

The denominator $(u_\turb/l_\turb)^2$ follows the behaviour established in
\S\ref{section:surface}. Because the energy is injected at the free surface and
the wave velocity field decays with depth as the potential solution
$\propto\exp(kz)$, the turbulent kinetic energy decays as $\exp(2kz)$ through the
mixed layer, while $l_\turb$ remains more or less constant
(Figure~\ref{Figure20:richardson}b). This attenuation is set by the surface
forcing alone and would be present even in the absence of stratification. On
crossing the mixing layer, however, the eddies work against the buoyancy of the
stable stratification and lose energy, so that $u_\turb$ drops while $l_\turb$
increases. Together these cause the denominator to fall sharply, halting the
eddies at the interface.

Combining the two quantities shapes the spatial profile of $\richardson$
(Figure~\ref{Figure20:richardson}c). In the mixed layer the density gradients are
small, so the numerator is negligible and $\richardson\approx0$. On entering
the mixing layer $N^2$ rises steeply towards $N^2_{\max}$ while the denominator
decays with depth, so $\richardson$ increases rapidly, peaks within the
interface, and sharply drops to a residual value in the bottom layer. The
resulting profile is asymmetric: because the turbulent denominator decays with
depth, $\richardson$ peaks below the centre of the mixing layer, where $N^2$
is maximal. The upper part of the layer therefore sits at lower $\richardson$
and is eroded most readily, consistent with entrainment proceeding from the
mixed layer downward. The large variance of $\richardson$ near
$z\approx-0.3$ in the earliest window ($t\in[90T,120T]$) despite its mean
remaining close to zero there reflects the transient phase of the wave
amplitude: the overshoot drives intense splash-down events that reach into
the bottom layer and produce short, intermittent mixing bursts.

This interpretation is confirmed by the diapycnal mixing rate $\Phi_d$
(Figure~\ref{Figure20:richardson}d). The mixing rate peaks on the upper flank,
where eddies first engage the interface, and decreases with depth as they lose
the energy needed to penetrate further ($\richardson$ eventually exceeding
$\richardson_c\sim O(1)$). This upper-flank concentration is
consistent with the turbulent entrainment mechanism, where heavy fluid is
continuously pulled upward into the mixed layer and homogenized there.

The barrier also evolves in time, through two competing effects. First, although
the total energy injected at the surface remains constant, the energy reaching
the interface decreases exponentially as the mixed layer deepens. Second, as the
layer deepens it weakens the interfacial buoyancy gradient, so that $N^2_{\max}$
decreases, allowing eddies to engulf fluid more efficiently. Overall, since the
first effect is dominant, the energetic barrier strengthens as the layer
approaches its apparent final depth. The same upward-entrainment mechanism
explains how $\Phi_d$ evolves (Figure~\ref{Figure20:richardson}d): the active
mixing zone spreads over a wider range even as its overall intensity weakens,
approaching a small residual value that mirrors the slow residual entrainment of
\S\ref{section:stratification:long:energetics} as the system settles toward its
apparent final depth.
\section{One-dimensional model for the stratification}
\label{sec:model}

Having identified the surface-injected turbulence (\S\ref{section:surface}) and
the energetic barrier that arrests it
(\S\ref{section:stratification:long:barrier}), we now ask whether these two
ingredients alone reproduce the non-monotonic dependence of the asymptotic
mixed-layer depth $h_\infty$ on the initial depth $h_\init$
(\S\ref{section:surface:asymptotic}). We reduce the dynamics to a single
vertical coordinate forced by a prescribed Faraday wave, in the spirit of
upper-ocean wave-mixing models \citep{Qiao2004, Ghantous2014}. The concentration
equation is closed by gradient diffusion, with the eddy diffusivity supplied by
a two-equation $K$-$\varepsilon$ model that solves transport equations for the
turbulent kinetic energy and its dissipation~\citep{Burchard1995,
BurchardBolding2001}. We first derive the one-dimensional transport equations
(\S\ref{section:model:averaging}), then specify and calibrate the closure
(\S\ref{section:model:keps}), and finally compare its predictions against the
full set of experiments (\S\ref{section:model:results}).

\subsection{Horizontal averaging}
\label{section:model:averaging}

Every field is split into a horizontal mean and horizontal fluctuations, 
\begin{align}
  \label{eqn:7.1}
  X = \langle X \rangle_{xy} + X',
  \qquad
  \langle X \rangle_{xy}
  = A^{-1}\!\int_A X~\mathrm{d}x\,\mathrm{d}y,
\end{align}
with $A$ the horizontal cross-section of the tank. The mean buoyancy equation
illustrates how this works. Since the density is linear in $c$, the buoyancy $b
= -g(\rho-\rho_0)/\rho_0$, with $\rho_0=(\rho_1+\rho_2)/2$ the reference
density, satisfies the same transport equation \eqref{eqn:3.2} as the
concentration. Averaging it horizontally eliminates the horizontal flux
divergences and the mean horizontal advection, leaving only the vertical fluxes,
\begin{align}
  \label{eqn:7.2}
  \frac{\partial}{\partial t}\langle b \rangle_{xy}
  = \frac{\partial}{\partial z}\left[
    \kappa_\mix \frac{\partial}{\partial z}\langle b \rangle_{xy}
    - \langle u_z' b' \rangle_{xy}
  \right].
\end{align}

We close the turbulent flux with a gradient-diffusion hypothesis, which turns
\eqref{eqn:7.2} into a vertical diffusion equation, retaining the molecular
diffusivity $\kappa_\mix$ alongside the eddy diffusivity $\kappa_\turb$. The
latter is obtained from a two-equation $K$-$\varepsilon$ closure
\citep{Burchard1995, UmlaufBurchard2003}, in which the turbulent energy
evolves dynamically. In the remainder of this section, all variables are
understood as horizontally averaged in this sense, and we drop the decoration
$\langle\cdot\rangle_{xy}$ to lighten the notation, while primes retain their
meaning throughout.

\subsection{Two-equation, $K$-$\varepsilon$ closure}
\label{section:model:keps}
\renewcommand{\Kturb}{\mathrm{K}}

Closing the turbulent flux of \eqref{eqn:7.2} with the diffusivity supplied by
the closure, and adding dynamic equations for the turbulent kinetic energy
$\Kturb$\footnote{From here on we write the turbulent kinetic energy as
$\Kturb$ rather than $K'$ (\S\ref{section:surface}), matching the standard
$K$-$\varepsilon$ notation} and its dissipation $\varepsilon$, gives the closed
system,
\begin{subequations}
  \begin{align}
    \label{eqn:7.4a}
    \frac{\partial\Kturb}{\partial t} &=
    \frac{\partial}{\partial z}\left[ \nu_k \frac{\partial\Kturb}{\partial z} \right]
    - B - \varepsilon,
    \\
    \label{eqn:7.4b}
    \frac{\partial\varepsilon}{\partial t} &=
    \frac{\partial}{\partial z}\left[ \nu_\varepsilon \frac{\partial\varepsilon}{\partial z} \right]
    - \frac{\varepsilon}{\Kturb} \left[
      C_{\varepsilon 3} B + C_{\varepsilon 2} \varepsilon
    \right],
    \\
    \label{eqn:7.4c}
    \frac{\partial b}{\partial t} &=
    \frac{\partial B}{\partial z}
  \end{align}
\end{subequations}
where, on the right-hand side, the remaining terms correspond to transport,
buoyancy production $B \equiv \kappa_\turb N^2$, and dissipation $\varepsilon$.
Molecular diffusivity is neglected in \eqref{eqn:7.4c}, since $\kappa_\mix$ is
orders of magnitude smaller than $\kappa_\turb$. A shear-production term is
absent by construction, since with $\langle u \rangle_{xy}=0$. Each of these
unknown turbulent fluxes is closed in terms of the mean fields through a single
turbulent viscosity,
\begin{align}
  \label{eqn:7.5}
  \nu_t = C_\mu\,\frac{\Kturb^2}{\varepsilon},
  \quad
  \nu_k = \frac{\nu_t}{\sigma_k},
  \quad
  \nu_\varepsilon = \frac{\nu_t}{\sigma_\varepsilon},
  \quad
  \kappa_\turb = \frac{\nu_t}{\sigma_b},
\end{align}
with the standard values $C_\mu=0.09$, $\sigma_k=1.0$ and
$\sigma_\varepsilon=1.3$. The dissipation equation also carries the empirical
constants $C_{\varepsilon2}$ and $C_{\varepsilon3}$, weighting destruction and
buoyancy respectively. Here, we adopt the standard value
$C_{\varepsilon2}=1.92$. We shall see that $\sigma_b$ and $C_{\varepsilon3}$ are
not independent parameters but depend on the definition of a critical Richardson
number~\citep{BurchardBolding2001}.

\subsubsection{Initial conditions and boundary conditions}                                          

The closed system \eqref{eqn:7.4a}--\eqref{eqn:7.4c} requires initial and
boundary conditions. We initialise the integration with the concentration
profile of the corresponding experiment (or DNS) at $t=0$, and with a small,
uniform seed for $\Kturb$ and $\varepsilon$, to which the results are
insensitive. At the boundaries we impose homogeneous Neumann conditions on the
concentration $c$, so that no mass crosses the surface or the bottom. The
turbulent quantities $\Kturb$ and $\varepsilon$ require special care, since the
injected turbulence is the main driver of this system. 

We consider that wave breaking and splashdown deliver a fraction $I_k$ of the
wave kinetic energy to turbulence---the injection efficiency---so that
$\Kturb_\surf = I_k \tfrac{1}{2}U_0^2$. As shown in
Appendix~\ref{section:appendix:energies}, we may estimate the wave energy from
the wave amplitude as $U_0^2 \simeq g' k \vert A_\surf \vert^2$, with $g' =
g(\rho_1-\rho_3)/\rho_0$ the reduced gravity. The surface amplitude $A_\surf =
R\,e^{i\theta}$ can be either measured directly from the recorded free surface
or estimated from the weakly nonlinear wave model of
Appendix~\ref{appendixB:mathieu:cubic1}, whose amplitude
equations~\eqref{eqn:B2.12a}--\eqref{eqn:B2.12b}, integrated numerically,
reproduce its linear growth, saturation, and exponential decay once the forcing
stops.

With $\Kturb_\surf$ fixed, the surface dissipation follows from the standard
length-scale relation
\begin{align}
  \label{eqn:7.6}
  \varepsilon_\surf = C_\mu^{3/4}\,\frac{\Kturb_\surf^{3/2}}{l_k},
\end{align}
where $l_k$ is a turbulent length scale of the order of a centimetre, consistent
with our observations of the surface breaking. In practice we treat the
injection efficiency $I_k$ and the length scale $l_k$ as the two adjustable
parameters that set the surface forcing.

\subsubsection{Calibrated parameters}
\label{section:model:calibration}

\begin{table}
  \centering
  \begin{tabular}{ccccccc}
    \toprule
    $C_\mu$ & $\sigma_k$ & $\sigma_\varepsilon$ & $\sigma_b$ & $C_{\varepsilon2}$ & $C_{\varepsilon3}$ \\
    \midrule
    0.09 & 1.0 & 1.30 & 7.95 & 1.92 & 0.01 \\
    \bottomrule
    \\
    \\
  \end{tabular}
  \quad
  \begin{tabular}{ccc}
    \toprule
    $F$ & $I_k$ & $l_k~(\unit{\centi\meter})$ \\
    \midrule
    0.24 & 0.003 & 2.2 \\
    0.37 & 0.019 & 2.0 \\
    0.49 & 0.027 & 1.9 \\
    \bottomrule
  \end{tabular}
  \caption{Calibrated shared parameter values used in the $K$-$\varepsilon$ turbulence model.}
  \label{Table4:parameters}
\end{table}

We calibrated the model in stages against the full set of 24 experimental probe
casts, taking as objective function the root-mean-square error between modelled
and measured concentration profiles, averaged over the casts. We first fitted
the reference experiment EXP-F49-H30 by a coarse parameter scan followed by a
Nelder--Mead polish, and verified that the resulting parameters, held fixed,
predict the remaining $F = 0.49$ experiments without further adjustment. The two
lower-forcing series were then fitted through their per-series $I_k$ and $l_k$
alone, before a final joint optimisation over all 24 experiments in which the
shared parameters were released. The calibration yielded $I_k = 0.003$, $0.019$
and $0.027$ and $l_k = 2.2$, $2.0$ and $1.9~\unit{\centi\meter}$ for $F = 0.24$,
$0.37$ and $0.49$ respectively, with shared $\sigma_b = 7.95$ and
$C_{\varepsilon3} = 0.01$. The increase of $I_k$ with the forcing is consistent
with the state of the surface. At low $F$ the wave field remains coherent and
injects little turbulence, while stronger forcing brings increasingly frequent
breaking and splashdown events, which convert a growing fraction of the wave
energy. The complete set of parameters is summarized in
table~\ref{Table4:parameters}.

\subsection{Results}
\label{section:model:results}

\begin{figure}   
  \centering
	\begin{minipage}{0.9\linewidth}   
		(a) \hfill (b) \hfill (c) \hfill~ \\
		\includegraphics[width=\linewidth]{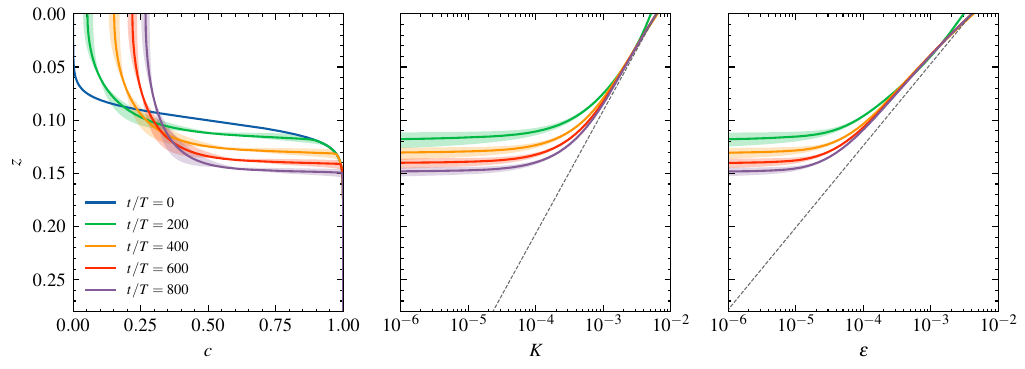}			
	\end{minipage}  
  \caption{
    Vertical profiles predicted by the $K$-$\varepsilon$ closure for parameters
    matching EXP-F49-H30: (a) concentration $c$, (b) turbulent kinetic energy
    $\Kturb$, and (c) dissipation rate $\varepsilon$. For reference, panels (b) and
    (c) also show dashed lines with $\exp{(2kz)}$ and $\exp{(3kz)}$,
    respectively. Solid lines correspond to the fitted value $\sigma_c=7.95$,
    while the shaded bands span $\pm 50\%$ to illustrate the sensitivity to this
    parameter.
  }
  \label{Figure21:model}
\end{figure}
\begin{figure}   
  \centering
	\begin{minipage}{0.9\linewidth}   
		(a) \hfill\hspace{1.5cm} (b) \hfill~ \\
		\includegraphics[width=\linewidth]{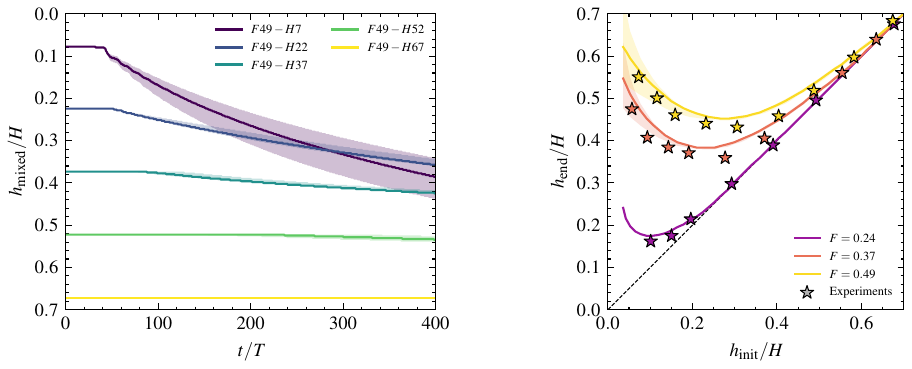}			
	\end{minipage}  
  \caption{
    Comparison between the calibrated $K$-$\varepsilon$ closure (solid lines)
    and the experiments (stars). (a) Mixed-layer depth $h_\mixed$ vs.\ time
    for a fixed forcing $\paramF=0.489$ and varying initial stratification
    depth. (b) Final interface depth $h_\final$ (after $800T$) vs.\ the
    initial depth $h_\init$ for the three forcing amplitudes, displayed as in
    Figure~\ref{Figure5:SideToSide}. The model reproduces the non-monotonic
    dependence of $h_\final$ on $h_\init$; solid lines use the fitted
    $\sigma_c=7.95$, and shaded bands span $\pm50\%$ to show the sensitivity
    to this parameter.
  }
  \label{Figure22:model}
\end{figure}

\subsubsection{Predicted profiles and mixed-layer depth}
\label{section:model:results:comparison}

Figure \ref{Figure21:model} shows the numerical integration of the
$K$-$\varepsilon$ closure for the reference experiment EXP-F49-H30, using the
parameters calibrated above. As expected, the model reproduces a gradual
homogenization of the upper layer, while the bottom layer remains unmixed. This
process reduces the concentration gradient and pushes the interface downwards
(Figure \ref{Figure21:model}a). Figures \ref{Figure21:model}b and
\ref{Figure21:model}c display the turbulent kinetic energy and the dissipation,
respectively. Both develop exponential profiles $\Kturb \propto e^{2kz}$
and $\varepsilon \propto e^{3kz}$, and reveal the progression of the
energy and dissipation fronts as the energetic barrier is eroded. In each
panel, the shaded bands indicate the sensitivity to the transport constant
$\sigma_b$, which remains modest for this experiment.

Figure \ref{Figure22:model} compares the model against the full set of
experiments. Figure \ref{Figure22:model}a shows the mixed-layer depth $h_\mixed$
as a function of time for a fixed forcing amplitude and varying initial
stratification depth, while Figure \ref{Figure22:model}b shows the final
interface depth $h_\final$ (after $800T$) as a function of the initial depth
$h_\init$ for the three forcing amplitudes, in the same format as Figure
\ref{Figure5:SideToSide}. The calibrated model (solid lines) reproduces the
experimental data (stars) well, including the non-monotonic dependence of
$h_\final$ on $h_\init$. This agreement is obtained with only five adjusted
parameters across the 24 experiments, one injection efficiency $I_k$ per forcing
amplitude together with the shared $\sigma_b$ and $C_{\varepsilon3}$ (the
turbulent length scale having converged to $l_k \approx 2~\unit{\centi\meter}$).
Within each forcing series the parameters are fixed, so the non-monotonic
dependence on $h_\init$ is an output of the model, not a calibration target. As
before, the shaded bands indicate the sensitivity to $\sigma_b$. They widen only
for the shallowest initial stratifications, confirming that this parameter is
only relevant there.

\subsubsection{Comparison with predictions from assuming self-similar profiles}
\label{section:model:results:predictions}

The calibrated model develops turbulent profiles of the form $\Kturb \propto
e^{2kz}$ and $\varepsilon \propto e^{3kz}$ (Figure~\ref{Figure21:model}b,c),
broadly consistent with the shapes suggested by the simulations. While it is
unclear whether such a solution exists, we postulate one in the unstratified
limit and follow where the assumption leads.

Consider the unstratified limit, $B=0$, and seek a self-similar solution of the
form $\Kturb = \Kturb_\surf\,e^{2kz}$ and $\varepsilon =
\varepsilon_\surf\,e^{3kz}$. Substituting these profiles into the
$\Kturb$-transport equation yields an estimate for the turbulent length scale,
\begin{align}
  l_k k = C^{1/4}_\mu\left[\frac{\sigma_k}{6}\right]^{1/2}.
\end{align}
With the standard constants and the forcing wavenumber $k$, this gives $l_k
\approx 2~\unit{\centi\meter}$, in agreement with the value recovered by the
calibration (table~\ref{Table4:parameters}). Applying the same substitution to
the $\varepsilon$-transport equation gives the same relation, but with the
bracketed term replaced by $(\sigma_\varepsilon C_{\varepsilon2})/12$. Matching
the two expressions would then require that $\sigma_\varepsilon =
{2\sigma_k}/{C_{\varepsilon2}}$. With the standard values $\sigma_k = 1.0$ and
$C_{\varepsilon2} = 1.92$ this gives $\sigma_\varepsilon \approx 1.04$, not far
from the standard value $\sigma_\varepsilon = 1.3$ we retained. This suggests
the solution is weakly sensitive to $\sigma_\varepsilon$.

The self-similar profiles also offer a way to illustrate the barrier effect and
a critical Richardson number (\S\ref{section:stratification:long:barrier}).
Evaluated on the self-similar profiles, the transport term of \eqref{eqn:7.4a}
reduces to $T = 6 k^2 \nu_t \Kturb/\sigma_k$, and combining transport and
buoyancy gives
\begin{align}
  \label{eqn:7.9}
  T - B = \frac{6 k^2 \nu_t \Kturb}{\sigma_k}
  \left( 1 - \frac{\richardson}{\richardson_c} \right),
  \qquad
  \richardson_c \equiv \frac{\sqrt{C_\mu}\,\sigma_b}{2},
\end{align}
where $\richardson$ is the bulk Richardson number \eqref{eqn:6:6}, evaluated
with the closure's own scales, $u_\turb = (2\Kturb)^{1/2}$ and $l_\turb$. The
turbulence is thus sustained where $\richardson < \richardson_c$ and decays
where $\richardson > \richardson_c$. The same reasoning can be applied to the
$\varepsilon$-equation \eqref{eqn:7.4b}, whose transport term reduces to
$T_\varepsilon = 12 k^2 \nu_t
\varepsilon/\sigma_\varepsilon$ on the same profiles, giving
\begin{align}
  \label{eqn:7.10}
  T_\varepsilon - C_{\varepsilon3}\frac{\varepsilon}{\Kturb} B
  = \frac{12 k^2 \nu_t \varepsilon}{\sigma_\varepsilon}
  \left( 1 - \frac{\richardson}{\richardson_{c}^{\varepsilon}} \right),
  \qquad
  \richardson_{c}^{\varepsilon} \equiv
  \richardson_c \frac{C_{\varepsilon2}}{C_{\varepsilon3}},
\end{align}
which has the same form as \eqref{eqn:7.9}, with the ratio
$C_{\varepsilon2}/C_{\varepsilon3}$ determining which equation limits the
turbulence: for $C_{\varepsilon2}/C_{\varepsilon3}>1$, the $\Kturb$-equation
reaches its critical Richardson number first
($\richardson_{c}^{\varepsilon}>\richardson_c$); otherwise the
$\varepsilon$-equation does.

With the calibrated constants obtained below (\S\ref{section:model:calibration}
and table~\ref{Table4:parameters}), \eqref{eqn:7.9} gives $\richardson_c \approx
1.3$, of order unity as expected for the arrest of stratified turbulence. The
second critical number is far larger, $\richardson_{c}^{\varepsilon} =
\richardson_c\,C_{\varepsilon2}/C_{\varepsilon3} \approx 200$, which tells
us that the arrest is felt by the $\Kturb$-equation and not the
$\varepsilon$-equation. The buoyancy sink is effectively inactive in the latter,
so it is the change of sign of \eqref{eqn:7.9} that arrests the turbulence at the
interface. The large value should not be over-interpreted: since $\Kturb$ is
arrested before $\varepsilon$, the $\varepsilon$-equation never operates near its
own critical number.

This model captures the main features of the entrainment process observed in
both experiments and simulations. Crucially, it makes the physical origin of the
non-monotonicity explicit. The interface advances until the Richardson number
\eqref{eqn:7.9} reaches its critical value $\richardson_c$, at which point
turbulent transport can no longer outpace the buoyancy sink. The arrest depth
thus results from a balance between the exponentially decaying energy supply
$e^{2kz}$ and the potential energy barrier of the stratification. The barrier
itself is not fixed: it depends on the initial depth $h_\init$ through the
progressive dilution of the mixed layer as it entrains. The final depth
$h_\final$ therefore inherits a non-monotonic dependence on $h_\init$.
\section{Conclusions}
\label{section:conclusion}

We investigated how Faraday surface waves mix two miscible fluids with a small
density contrast. Large-amplitude standing waves were generated at the free
surface by a time-periodic vertical acceleration triggering the Faraday
instability. We followed the short- and long-term evolution of the
stratification through laboratory experiments and numerical simulations.

The sustained driver of mixing is turbulent entrainment. Wave breaking,
splash-down and the cavity collapse of large crests inject turbulent energy into
the fluid below, homogenizing the upper layer and progressively pushing the
interface downward. Two further mechanisms modify the mixing over short
timescales, a long-range mechanism and a short-range mechanism, both driven by
the oscillating strain of the surface wave.

The long-range mechanism originates in the advection of the interface by the
surface wave, whose oscillating strain drives a secondary parametric
instability. The interface breaks down and re-stratifies into a thicker layer,
even at depths of order one wavelength. For shallow stratification, this
destabilization couples with the near-surface turbulence, giving rise to the
short-range mechanism. Heavy fluid parcels released from the disrupted interface
are lifted into the mixed layer by the surface motion, while the same splash-down 
and cavity-collapse events inject bubbles and light fluid downward. Whereas the
long-range mechanism mainly thickens the interface, this short-range coupling
both thickens it and drives it deeper, in a succession of intense mixing bursts.

Over long timescales, the asymptotic depth of the mixed layer varies
non-monotonically with the initial stratification depth. This reflects the
competition between two effects. The energy barrier at the interface decreases
as the upper layer homogenizes, while the injected energy decays exponentially
with depth. Their balance eventually saturates the mixed-layer depth. As
entrainment proceeds, the interface progressively decouples from the surface
dynamics.

The DNS was validated systematically against the experiments, comparing the
linear growth rates, the surface shape, the saturation amplitudes and the
damping rates of the waves. The simulations show faster entrainment and
diffusion than the experiments, however. We attribute this mainly to the Schmidt
number, $\schmidt=1$ in the DNS against $\schmidt=700$ in the experiments, which
broadens the interface more rapidly in the simulations. A secondary contribution
could come from the spatial resolution, which affects the droplet-size
distribution and the scalar dissipation. These compromises between accuracy and
computational cost have effects that, while modest at any instant, may
accumulate over the long integration times considered here.

To interpret these observations, we condensed the dominant ingredients into a
one-dimensional model of the stratification. A one-equation closure with a
prescribed energy profile, in the spirit of K-profile
parametrizations~\citep{Large1994}, was deemed insufficient. In such a model the
stratification cannot feed back on the turbulence, and the stabilization
function is empirical by nature. We therefore turned to a two-equation
$K$-$\varepsilon$ closure, in which the turbulent energy evolves dynamically.
The surface waves enter solely through a boundary injection of turbulent
energy that is transported downward into the interior; with only a horizontal
mean flow, there is no shear-production term to consider. With this closure, the model solves for the
concentration, turbulent energy and dissipation, and reproduces the
non-monotonic dependence of the asymptotic depth on the initial stratification.

Such a closure necessarily carries several adjustable parameters. We retained
standard values for $C_\mu$, $\sigma_k$, $\sigma_\varepsilon$ and
$C_{\varepsilon2}$, leaving the injection efficiencies $I_k$, the turbulent
length scale $l_k$, the transport constant $\sigma_b$ and $C_{\varepsilon3}$ as
free parameters. The calibration returned a turbulent length scale
$l_k\approx2~\unit{\centi\meter}$ for all three forcing amplitudes,
$C_{\varepsilon3}\approx0$, and while $\sigma_b$ is larger than its standard
value, the associated critical Richardson number, $\richardson_c\approx1.3$,
remains within the typical range.

A self-similar solution accounts for some of these values. It predicts
a depth-independent length scale $l_k =
C_\mu^{1/4}(\sigma_k/6)^{1/2}/k\approx2~\unit{\centi\meter}$, in agreement with
the fitted value, and, by matching the $\Kturb$- and $\varepsilon$-equations, a
dissipation constant $\sigma_\varepsilon = 2\sigma_k/C_{\varepsilon2}$ close to
the standard value we retained. Evaluated on these profiles, the turbulence is
arrested when the Richardson number reaches $\richardson_c =
\sqrt{C_\mu}\,\sigma_b/2$, while the corresponding critical value for the
$\varepsilon$-equation is larger by a factor
$C_{\varepsilon2}/C_{\varepsilon3}$. A small $C_{\varepsilon3}$ thus pushes it
out of reach, so the arrest enters only through the $\Kturb$-transport equation,
consistent with the fitted $C_{\varepsilon3}\approx0$.

The closure represents the interface only through its bulk stratification and
does not resolve the processes acting there. In particular, the long-range and
short-range mechanisms described above lie beyond its reach, as do the
stretching and compression of the interface by the oscillating strain and the
Kelvin--Helmholtz instabilities driven by the wave-induced shear.

Several perspectives follow from this work. The one-dimensional model could be
refined through a more systematic, data-driven calibration~\citep{Thevenin2022,
Thevenin2025} and by incorporating the interface instabilities discussed above.
Our present calibration is naive in this respect. Because $I_k$, $l_k$,
$\sigma_b$ and $C_{\varepsilon3}$ are fitted jointly, part of what is actually a
defect in the injection efficiencies could be absorbed into $\sigma_b$ and
$C_{\varepsilon3}$ instead. A more careful calibration could
yield better estimates of these constants and make the present closure,
calibrated here for a single configuration, genuinely predictive. Beyond the
model, an important question is how these results scale to larger systems and to
the fluid properties encountered in industrial settings, such as the storage and
transport of liquefied natural gas. More broadly, because this configuration
isolates wave-driven mixing in a controlled way, it offers a route to improving
the parametrizations used in geophysical models, even though Faraday and ocean
waves differ in their generation and characteristics.

\backsection[Supplementary data]{
  \label{SupMat}Supplementary material and movies will be available upon publication.
}

\backsection[Acknowledgements]{ 
  This work was performed at the Gaztransport and Technigaz\texttrademark (GTT)
  Motion Analysis and Testing Laboratory (LAMT) in St. Rémy-lès-Chevreuse,
  France. We thank GTT for providing the experimental facility and technical
  support. This work was granted access to the HPC resources of TGCC/CCRT. We
  also acknowledge the Mesocentre Ruche computing centre at Université
  Paris-Saclay, supported by CNRS and Région Île-de-France. The first author
  would also like to thank D. Peláez-Zapata and L. Robles-D\'iaz for valuable
  discussions.
}

\backsection[Funding]{
  No additional external funding was received for this study.
}

\backsection[Declaration of interests]{
  The authors report no conflict of interest.
}

\backsection[Data availability statement]{
  The data that support the findings of this study will be openly
  available in the French Fluid Dynamics Database
  \hyperlink{https://entrepot.recherche.data.gouv.fr/dataverse/f2d2}{https://entrepot.recherche.data.gouv.fr/dataverse/f2d2}.
}

\backsection[Author ORCIDs]{\\
\orcidlink{0000-0003-2175-324X} Andrés Castillo-Castellanos, \hyperlink{https://orcid.org/0000-0003-2175-324X}{https://orcid.org/0000-0003-2175-324X};\\
\orcidlink{0000-0001-7355-4790} Benoît-Joseph Gréa, \hyperlink{https://orcid.org/0000-0001-7355-4790}{https://orcid.org/0000-0001-7355-4790};\\
\orcidlink{0000-0002-3796-0588} Antoine Briard, \hyperlink{https://orcid.org/0000-0002-3796-0588}{https://orcid.org/0000-0002-3796-0588};\\
\orcidlink{0000-0001-9683-4461} Louis Gostiaux, \hyperlink{https://orcid.org/0000-0001-9683-4461}{https://orcid.org/0000-0001-9683-4461}
}

\clearpage
\appendix
\renewcommand{\thefigure}{\Alph{section}.\arabic{figure}}
\setcounter{section}{0}

\section{Details on the experimental and numerical realizations}
\label{section:appendix:experiments}
\setcounter{table}{0}
\setcounter{figure}{0}

\subsection{Camera setup and interface segmentation}
\label{section:appendix:experiments:camera}

\begin{figure}
  \begin{center}
    (a) \hfill (b) \hfill~\\
    \includegraphics[width=\linewidth]{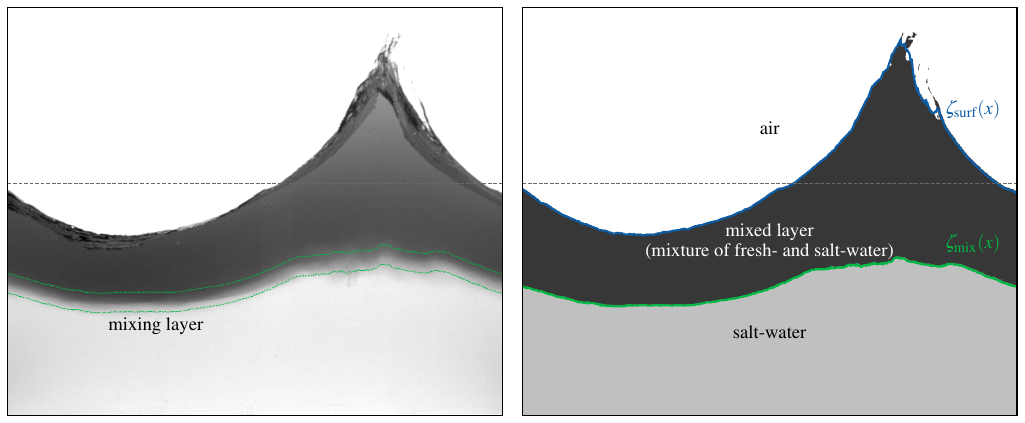}
  \end{center}
  \caption{
    (a) Typical greyscale image recorded during the experiments and (b) the
    corresponding segmentation using the procedure described in
    \S\ref{section:appendix:experiments:camera}. In subfigure (a),
    {\color{green!50!black}\dotted}  indicate the detected mixing layer, while
    in subfigure (b) {\color{blue!50!black}\full} (resp.
    {\color{green!50!black}\full}) indicates the detected free surface
    $\zeta_\surf$ (resp. middle of the mixing layer $\zeta_\mix$).
  }
  \label{FigureA1:Grayscale}
\end{figure}

\subsubsection{Image acquisition}

A monochromatic camera, operating at 26 frames per second with a 16-bit colour
depth, captured images of an LED-backlit white screen as seen through the tank
using a resolution of 1920 by 1080 pixels. The camera was mounted on the
oscillating platform via an extended arm, mechanically reinforced to minimize
vibrations. Residual effects of persistent small vibrations were removed during
post-processing. Additionally, images were cropped to span roughly one period of
the primary wave, as to reduce boundary-induced artefacts. A typical raw image
of the camera is seen in Figure \ref{FigureA1:Grayscale}a. The positions of both
interfaces were extracted from the recorded images using a segmentation process.

\subsubsection{Upper-interface segmentation}

For the upper interface, the image intensity was
transformed into a transversally averaged volume fraction field \(f(x,y,t)\),
which was set to 0 (gas phase) and 1 (liquid phase) using a binary threshold. As
mixing progresses, the intensity in the upper layer decreases over time. To
accommodate this, the threshold values were determined for each frame using the
triangle method. The interface separating the two phases was identified with a
contour-tracking algorithm, representing the interface as a one-dimensional
curve embedded in two-dimensional space. Alternatively, the free surface
position \(\zeta_\surf(x,t)\) was approximated by summing pixel intensities in
each column of \(f(x,y,t)\), yielding a monotonic function of \( x \). Both
methods produce equivalent results for non-breaking waves but may diverge in the
presence of breaking waves (see, for instance, the blue line in Figure
\ref{FigureA1:Grayscale}b).

\subsubsection{Lower-interface segmentation}

The lower interface position, \(\zeta_\mix(x,t)\), was estimated using a similar
approach. However, the binary threshold was based on the mean pixel
value~\citep{Glasbey_1993} within the liquid phase, while the gas phase was
ignored. 

\subsection{Estimation of potential and kinetic energies from the experiment}
\label{section:appendix:energies}

A practical way to validate the simulations is to consider global quantities
such as the potential and kinetic energies. In this section, we search for
reasonable estimates which can be obtained from the recorded images. By focusing
on the effects of the deepening of the interface and initially ignoring
molecular diffusion, the density field \(\rho(\vec{x},t)\) can be approximated
as a piecewise profile:
\begin{align}
  \label{eqn:A.1}
  \rho(\vec{x},t) =
  \begin{cases}
    \rho_1, & \text{for layer 1 } (z \leq \zeta_\mix(x,y,t)) \\
    \rho_\mixed = \rho_1 - (\rho_1-\rho_2)\frac{h_\init}{h_\mixed}, & \text{for layer 2 } (\zeta_\mix(x,y,t) < z \leq \zeta_\surf(x,y,t)) \\
    \rho_3, & \text{for layer 3 } (z \geq \zeta_\surf(x,y,t))
  \end{cases}
\end{align}
where the upper layer is taken as a homogeneous mixture of fresh and salt water,
while the lower layer remains salt water. 

\subsubsection{Evolution of the instantaneous Atwood number}

Replacing this profile in Eq. \eqref{eqn:2.3} gives the following approximation
for the instantaneous Atwood number:
\begin{align}
  \label{eqn:A.2}
  \atwood_\mixed \approx \frac{ \rho_1-\rho_2 }{\rho_1 (2\frac{h_\mixed}{h_\init} - 1) + \rho_2}
\end{align}
which is always lower than \(\atwood_\init\). In this sense, as the interface
height \(h_\mixed\) increases, less energy is required to mix both layers.

\subsubsection{Evolution of the change in the global potential energy}

Now consider the change in the global potential energy:
\begin{align}
  \label{eqn:A.3}
  \GPE \equiv& \iiint \rho(\vec{x},t) g z \, dx \, dy \, dz.
\end{align}
Replacing the piecewise profile into Eq. \eqref{eqn:A.3} gives the
following approximation for the change in \(\GPE\) between time \(t_0\) and time
\(t\):
\begin{align}
  \label{eqn:A.4}
  \Delta\GPE =&
  ~\frac{g(\rho_1 - \rho_3)}{2} \iint \zeta^2_\surf \, dx \, dy
  \\\nonumber
  &- \frac{g(\rho_1 - \rho_2)}{2} \frac{h_\init}{h_\mixed} \iint \left( \zeta^2_\surf - \zeta^2_\mix + h_\init h_\mixed \right) dx \, dy.
\end{align}
Note that the first term in equation \eqref{eqn:A.4} represents the
\(\Delta\GPE\) of an equivalent one-layer system, while the second term is a
small correction due to stratification which is proportional to the Atwood
number.

For small amplitudes, we may use a wave model to obtain an estimate for the
maximum potential energy of the equivalent one-layer system,
\begin{align}
  \label{eqn:A.5}
  \Delta\GPE_{\max} \approx \frac{g(\rho_1 - \rho_3)}{2} WD
  \left[
    \frac{1}{2} |A|^2 - \frac{k^2}{4} |A|^4
    \right]
\end{align}
or by averaging over a period, we obtain a mean potential energy
\begin{align}
  \label{eqn:A.6}
  \Delta\GPE_{\mathrm{avg}} \approx \frac{g(\rho_1 - \rho_3)}{2} WD
  \left[
    \frac{1}{4} |A|^2 - \frac{3k^2}{32} |A|^4
    \right]
\end{align}
where $A$ is the wave amplitude. In either case, the potential energy is
expected to scale with \(|A|^2\) and to have a correction term proportional to
\(|A|^4\).

\subsubsection{Evolution of the change in the background potential energy}

Another relevant quantity is the background potential energy (\(\BPE\)), defined
as the minimum potential energy attainable through adiabatic
motions~\citep{Winters_1995, Sutherland_2010}. This is equivalent to rearranging
fluid parcels into a statically stable system with the same density
distribution~\citep{Tseng2001}. The change in \(\BPE\) between time \(t_0\) and
time \(t\)---the energy lost to irreversible mixing---is:
\begin{align}
  \label{eqn:A.7}
  \Delta\BPE \approx \frac{g(\rho_1 - \rho_2)}{2} ( h_\mixed - h_\init ) ~(h_\init W D).
\end{align}
This term is also proportional to the Atwood number, the change in
stratification depth, and the initial volume in layer 2. Equation
\eqref{eqn:A.7} shows that a shallow interface requires less energy to
displace, since $\Delta\BPE\propto h_\init$. At the deep end of the range
studied, $\Delta\BPE$ is small for a different reason: the interface is
effectively shielded from the surface-injected turbulence, so little mixing
occurs ($\Delta h_\mixed\to0$) regardless of the energy nominally required
(\S\ref{section:surface:asymptotic}).

\subsubsection{A rough estimate of the total kinetic energy}

Locally, kinetic energy is injected through the surface motion via a resonant
instability. In this system, kinetic energy
\begin{align}
  \label{eqn:A.8}
  \KE \equiv& \iiint \frac{1}{2}\rho(\vec{x},t) \vec{u}^2 (\vec{x},t) \, dx \, dy \, dz
\end{align}
is at its minimum when potential energy is at its maximum, and the magnitudes of
both quantities are expected to be comparable. For a single wavenumber \(k\),
kinetic energy is expected to peak at the surface level and decrease with
\(\exp{(2kz)}\). We also consider a barrier at height \(H\) to be determined,
\begin{align}
  \label{eqn:A.9}
  \KE \propto \Delta\GPE \cdot
  \frac{2k (e^{2kz} - e^{-2kH})}{ 1 - e^{-2kH}(1 + 2kH)}
\end{align}
which, in the limit of $kH \gg 1$ tends to
\begin{align}
  \label{eqn:A.10}
  \KE \propto \Delta\GPE \cdot
  2ke^{2kz}.
\end{align}

\subsection{Experimental and Numerical Parameters}
\label{section:appendix:parameters}

Both the experimental and numerical investigations explore the same parameter
space: a range of forcing amplitudes and initial stratification depths. This
enables direct validation and comparison of experimental and numerical
phenomenology. Table \ref{Table1:fluids} shows the nominal fluid properties
considered.

\subsubsection{Experimental parameters}

\begin{table}
  \newcolumntype{R}[1]{>{\raggedleft\arraybackslash}p{#1}}
  \begin{center}
    \begin{tabular}{ll R{2cm} R{2cm}}
      && Density & Viscosity \\
      && [\unit{\kilogram\per\cubic\meter}] & [\unit{\milli\pascal\second}] \\
      \midrule
      Fluid 1 & Salt water  & $1010-1100$ & $0.97-1.09$ \\
      Fluid 2 & Fresh water & $998$       & $0.89-1.00$ \\
      Fluid 3 & Air         & $1.25$      & $0.0182$ \\
    \end{tabular}\hspace{0.25cm}
    \begin{tabular}{l R{2.0cm} R{2.0cm}}
      & Surface tension & Mol. diffusion \\
      & [\unit{\milli\newton\per\meter}] & [\unit{\meter\squared\per\second}] \\
      \midrule
      & $72.8$   & $1.3 \times 10^{-9}$ \\ \\ \\
    \end{tabular}
  \end{center}
  \caption{Nominal fluid properties considered in the experiments.}
  \label{Table1:fluids}
\end{table}
\begin{table}
	\centering
	\begin{tabular}{cc | cc | ccc | rrrrr}
		Case & \multicolumn{1}{c|}{Forcing parameter} & \multicolumn{2}{c|}{Initial strat.} & \multicolumn{3}{c|}{Final strat.} & \multicolumn{3}{c}{Exponential rates} \\
		& $a\omega^2/g$ & $L_\init$ & $h_\init$ & $L_\final$ & $h_\final$ & $h_{\infty}$ & $\lambda_\surf$ & $\gamma_\surf$ & $\gamma_\mixed$ \\
		&				& \cm & \cm & \cm & \cm & \cm & \unit{\per\second} & \unit{\per\second} & \unit{\per\second} \\
		\midrule
		EXP-F49-H67 & 0.49 & 2.3 & 22.6 &  2.2 & 22.9 & 23.0 & 1.15 & 0.06 & 0.007 \\
		EXP-F49-H58 &      & 2.0 & 19.5 &  2.5 & 20.0 & 20.3 & 1.20 & 0.05 & 0.005 \\
		EXP-F49-H49 &      & 3.0 & 16.3 &  3.1 & 17.4 & 17.9 & 1.21 & 0.05 & 0.006 \\
		EXP-F49-H40 &      & 2.9 & 13.5 &  4.2 & 15.3 & 16.1 & 1.17 & 0.05 & 0.007 \\
		EXP-F49-H30 &      & 3.4 & 10.2 &  5.2 & 14.5 & 14.8 & 1.16 & 0.05 & 0.009 \\
		EXP-F49-H23 &      & 3.3 &  7.8 &  6.7 & 14.7 & 15.4 & 1.19 & 0.05 & 0.011 \\
		EXP-F49-H16 &      & 3.5 &  5.3 &  8.9 & 15.4 & 16.1 & 1.16 & 0.05 & 0.012 \\
		EXP-F49-H12 &      & 3.9 &  3.9 & 12.7 & 16.8 & 17.2 & 1.30 & 0.05 & 0.013 \\
		EXP-F49-H07 &      & 3.8 &  2.4 & 19.1 & 18.4 & 19.5 & 1.22 & 0.06 & 0.013 \\ \midrule
		EXP-F37-H64 & 0.37 & 1.1 & 21.3 &  1.7 & 21.4 & 21.3 & 0.91 & 0.05 & 0.007 \\
		EXP-F37-H56 &      & 1.7 & 18.6 &  2.3 & 18.8 & 18.9 & 0.86 & 0.05 & 0.007 \\
		EXP-F37-H46 &      & 0.8 & 15.3 &  2.5 & 15.8 & 16.3 & 0.89 & 0.04 & 0.011 \\
		EXP-F37-H37 &      & 1.5 & 12.4 &  2.9 & 13.6 & 14.0 & 0.85 & 0.05 & 0.011 \\
		EXP-F37-H28 &      & 1.8 &  9.3 &  3.4 & 12.0 & 13.0 & 0.86 & 0.04 & 0.011 \\
		EXP-F37-H19 &      & 1.9 &  6.4 &  4.2 & 12.4 & 13.1 & 0.88 & 0.05 & 0.011 \\
		EXP-F37-H14 &      & 2.1 &  4.8 &  5.5 & 12.9 & 13.6 & 0.85 & 0.05 & 0.011 \\
		EXP-F37-H09 &      & 2.2 &  3.1 &  7.3 & 13.7 & 14.3 & 0.87 & 0.04 & 0.011 \\
		EXP-F37-H06 &      & 2.3 &  1.9 & 13.8 & 15.9 & 17.0 & 0.92 & 0.05 & 0.016 \\ \midrule
		EXP-F24-H68 & 0.24 & 2.1 & 22.7 &  2.1 & 22.6 & 22.4 & 0.55 & 0.05 & 0.006 \\
		EXP-F24-H58 &      & 1.6 & 19.4 &  1.8 &      & 19.5 & 0.54 & 0.05 & 0.005 \\
		EXP-F24-H49 &      & 2.4 & 16.5 &  2.5 & 16.6 & 16.6 & 0.54 & 0.04 & 0.010 \\
		EXP-F24-H39 &      & 2.5 & 13.1 &  2.6 & 13.0 & 13.4 & 0.52 & 0.04 & 0.011 \\
		EXP-F24-H29 &      & 2.8 &  9.8 &  3.6 & 10.0 & 10.6 & 0.55 & 0.04 & 0.011 \\
		EXP-F24-H20 &      & 2.6 &  6.6 &  3.8 &  7.2 &  7.8 & 0.54 & 0.04 & 0.017 \\
		EXP-F24-H15 &      & 2.2 &  5.0 &  4.6 &  5.9 &  6.7 & 0.53 & 0.05 & 0.014 \\
		EXP-F24-H10 &      & 3.0 &  3.4 &  3.2 &  5.4 &  5.8 & 0.52 & 0.04 & 0.013 \\
		EXP-F24-H05 &      & 1.7 &  1.7 &  3.3 &      &  6.0 & 0.54 & 0.05 & 0.010
	\end{tabular}
	\caption{
		List of parameters used in the experimental realizations: forcing
		parameter $\paramF$; initial thickness $L_\init$ and height $h_\init$;
		final thickness $L_\final$ and height $h_\final$ ; asymptotic height
		$h_\infty$; exponential growth and decay rates in $\zeta_\surf$. First
		two series ran for $800T$ before disabling forcing, while the third
		series only ran for $400T$. Cases are labelled by their parameters;
		e.g. EXP-F49-H30 has $\paramF=0.49$ and $h^*_\init=30\%$.
	}
	\label{Table2:experiments}
\end{table}


The experiments are organized into three series, each characterized by a
distinct forcing parameter: $\paramF = 0.24$, $0.37$, and $0.49$, corresponding
to progressively larger surface waves ranging from unbroken to breaking. Within
each series, the initial stratification depth is systematically varied from $2$
to $20 \cm$, allowing us to explore the interaction between surface dynamics and
stratification over a wide parameter space. Table~\ref{Table2:experiments}
provides a complete summary of all experimental realizations.

\subsubsection{Numerical simulation parameters}

\begin{table}
	\centering
	\begin{tabular}{ccc | cc | ccc | rrrrr}
		Series & Case & \multicolumn{1}{c|}{Forcing parameter} & \multicolumn{2}{c|}{Initial strat.} & \multicolumn{3}{c|}{Final strat.} & \multicolumn{2}{c}{Exp. rates} \\
		& & $a\omega^2/g$ & $L_\init$ & $h_\init$ & $L_\final$ & $h_\final$ & $h_{\infty}$ & $\lambda_\surf$ & $\gamma_\mixed$ \\
		& &				& \cm & \cm & \cm & \cm & \cm & \unit{\per\second} & \unit{\per\second} \\
		\midrule
		A & DNS-F37-H07 & 0.37 & 2.7 &~2.5 & 12.8 & 12.2 &      & 0.88 \\
      & DNS-F37-H15 &      &     &~5.0 & 18.9 & 12.3 &      & 0.88 \\
      & DNS-F37-H22 &      &     &~7.5 & 19.2 & 12.2 &      & 0.88 \\
      & DNS-F37-H30 &      &     &10.0 & 15.1 & 12.4 &      & 0.88 \\
      & DNS-F37-H37 &      &     &12.5 & 11.1 & 13.8 &      & 0.87 \\
      & DNS-F37-H45 &      &     &15.0 & ~7.3 & 15.5 &      & 0.87 \\
    \midrule

    B & DNS-F12-H15 & 0.12 & 2.7 &~5.0 & 6.0  &  5.5 &  --- & 0.25 \\
      & DNS-F17-H15 & 0.17 &     &     & 11.2 &  7.2 &  9.1 & 0.38 \\
      & DNS-F25-H15 & 0.25 &     &     & 14.9 &  9.2 & 12.0 & 0.58 \\
      & DNS-F30-H15 & 0.30 &     &     & 16.1 & 10.0 & 12.3 & 0.70 \\
      & DNS-F37-H15 & 0.37 &     &     & 18.9 & 12.3 & 13.6 & 0.88 \\
      & DNS-F43-H15 & 0.43 &     &     & 16.8 & 10.4 & 14.8 & 1.04 \\
    \midrule
    C & DNS-F49-H30 & 0.49 & 2.7 &10.0 & 24.7 & 16.1 & 16.6 & 1.17 & 0.017\\
	\end{tabular}
	\caption{
    List of parameters used in the numerical simulations. Quantities are defined
    as in table \ref{Table2:experiments}. Here, grid size \(\Delta_x\) is
    roughly 3 times smaller than the capillary length. Cases are labelled by
    their parameters; e.g. DNS-F49-H30 has $\paramF=0.49$ and $h^*_\init=30\%$.
	}
	\label{Table3:simulations}
\end{table}


The numerical investigation comprises three series of 3D simulations: Series A
explores the stratification depth at a fixed mid-range forcing ($\paramF =
0.37$); Series B varies the forcing parameter while keeping the stratification
depth fixed; and Series C provides extended-duration runs ($\sim 400$ periods)
to characterize the asymptotic regime. Together, these series span the same
parameter space as the experiments, enabling meaningful comparison of the two
approaches. Table~\ref{Table3:simulations} provides a complete summary of all
numerical simulations.

Series A and B are used to study the influence of the stratification depth and
the forcing parameter, in particular when exploring the secondary instabilities.
This sweep allows us to show that the secondary instability is quite general
(observed in every DNS listed here), even for small forcing. Varying the
stratification depth in Series A was important to highlight the coupling of the
instability with surface dynamics, while Series B was also used for the
comparison with linear theory. Series C consists of a single extended run and
was used to examine the long-term evolution.  

\clearpage
\section{Details on the linear and weakly nonlinear theory}
\label{section:appendix:theory}
\setcounter{figure}{0}

This appendix summarizes the theory used to validate the experiments and
simulations. In \S\ref{section:appendix:stability}, a linear three-layer model
provides the natural frequencies and the stability chart under periodic
forcing. In \S\ref{section:appendix:amplitude}, weakly nonlinear theory yields
the saturation amplitude and the decay laws --- exponential for linear damping,
algebraic for cubic damping --- used in \S\ref{section:validation}.

\subsection{Linear three-layer (two-interface) model}
\label{section:appendix:stability}

We present a simple model for three-layer (two-interface) miscible fluids based
on potential inviscid theory. Although large-scale experiments may lie outside
its stated range of validity, the model highlights key aspects of the system
response (see, for instance, \cite{Labrador2021,Huang_2024}), and within our
parameter range its results are qualitatively similar to viscous approaches
such as \cite{Pototsky2016, Ward2019}.

\subsubsection{Linear equations for three-layer (two-interface) fluids}
\label{section:appendix:stability:1}

Following \cite{Benjamin1954}, we consider each layer to be irrotational and
retain only linear terms. The velocity potentials \(\phi_j(\vec{x},t)\) and
interface displacements are expanded in a Fourier basis,
\begin{subequations}
  \begin{align}
    \label{eqn:B1.1a}
    \phi_j =& \sum_{m,n=1}^\infty [P_{jmn}\cosh(k z) + Q_{jmn}\sinh(k z)] \cos(\vec{k}\cdot\vec{x}) 
    \\
    \label{eqn:B1.1b}
    \zeta_j =& \sum_{m,n=0}^\infty \alpha_{jmn}\cos(\vec{k}\cdot\vec{x}) 
  \end{align}
\end{subequations}
where \(\vec{k} = (k_n \vec{\hat{e}}_x + k_m \vec{\hat{e}}_y)\) is the
horizontal wavenumber and \(k = \vert\vec{k}\vert\) its magnitude. The
boundary conditions give \(P_{jmn}\) and \(Q_{jmn}\) in terms of the modal
amplitudes \(\alpha_{jmn}\)~\citep{Veletsos1993}, so the interface motion
reduces to two coupled equations for each wavenumber \(\vec{k}\),
\begin{align}
  \label{eqn:B1.2}
  \mathsfbi{B} \vec{\ddot{\alpha}} + 2\omega\mathsfbi{D} \vec{\dot{\alpha}} + (1 + F\cos(\omega t)) \mathsfbi{M} \vec{\alpha} = 0,
\end{align}
where \(\vec{\alpha}(t) = [\alpha_{1}(t), \alpha_{2}(t)]\) and the indices
\(mn\) are omitted. The \(2 \times 2\) matrix \(\mathsfbi{B}\) contains the
finite-depth corrections and a coupling term:
\begin{subequations}
  \begin{align}
    \label{eqn:B1.3a}
    B_{11} &= \rho_1[\coth{(k(H-h_\init))}+\tfrac{1}{2}kL] + \rho_2[\coth{(kh_\init)}+\tfrac{1}{2}kL],
    \\
    \label{eqn:B1.3b}
    B_{22} &= \rho_2\coth{(kh_\init)} + \rho_3\coth{(kH)},
    \\
    \label{eqn:B1.3c}
    B_{12} &= B_{21} = -\rho_2\csch(kh_\init),
\end{align}
\end{subequations}
where \(h_1=H-h_\init\), $h_2=h_\init$, and \(h_3=H\) are the layer heights.
The term $\tfrac{1}{2}kL$ in $B_{11}$ corrects for the interface thickness
\(L\), valid for \(kL \ll 1\). The off-diagonal term $B_{12}$ decays
exponentially in $kh_\init$, so the two interfaces decouple as the
stratification deepens.

The diagonal matrices \(\mathsfbi{D}\) and \(\mathsfbi{M}\) contain the linear
damping and stiffness coefficients,
\begin{align}
    \label{eqn:B1.4}
    \mathsfbi{D} &= 
    \begin{bmatrix}
      B_{11} \gamma_1 & 0 \\ 0 & B_{22} \gamma_2
    \end{bmatrix}
    ,&
    \mathsfbi{M} &= \begin{bmatrix}
    k g (\rho_1-{\color{black}\rho_2})  & 0 \\ 0 & k g ({\color{black}\rho_2}-\rho_3) + \sigma k^3
  \end{bmatrix}
\end{align}
where the damping coefficients \(\gamma_1\) and \(\gamma_2\) may be estimated
from the viscosity following \cite{Kumar_1994}:
\begin{align}
  \label{eqn:B1.5}
  \gamma_j \approx 2k^2 (\mu_j\coth(kh_j) + \mu_{j+1}\coth(kh_{j+1})),
  ~~\text{for}~~j=1,2,
\end{align}
or fitted from the experimental data. 

\subsubsection{Influence of stratification depth on the natural frequencies}
\label{section:appendix:stability:2}

\begin{figure}
  \begin{center}
    (a) \hfill  (b) \hfill~\\
    \includegraphics[width=\linewidth]{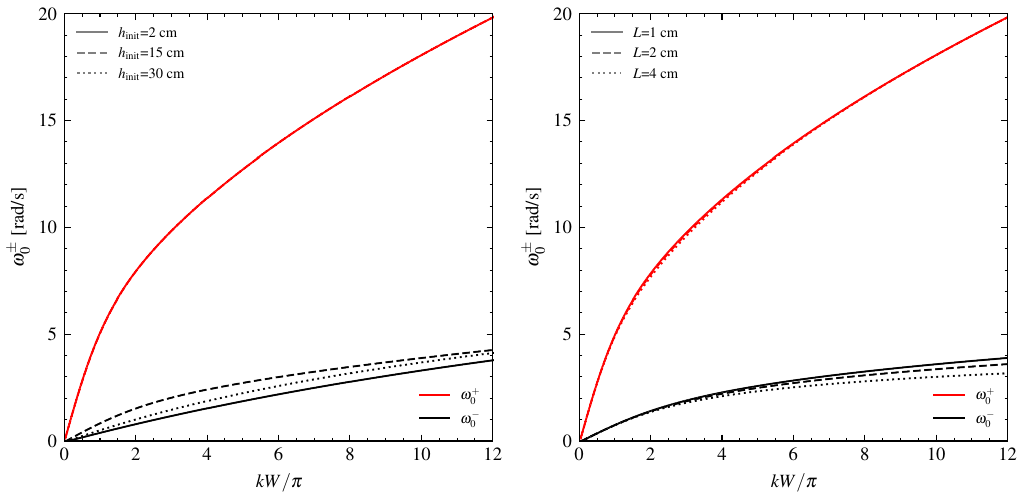}    
  \end{center}
  \caption{
    Roots of equation \eqref{eqn:B2.1} as function of $k W/\pi$ for different
    scenarios: (a) $L=0$, $k H \gg 1$ and varying $h_\init$; and (b) $k H \gg
    1$, $h_\init=10$ \unit{\centi\meter} and varying $L$.
  }
  \label{fig:appendix:B:1}
\end{figure}

The natural frequencies of \eqref{eqn:B1.2} are the roots of a quadratic
equation in $\omega_0^2$,
\begin{align}
  \label{eqn:B1.6}
  \underbrace{\left( k g (\rho_1 - \rho_2) - \omega_0^2 B_{11} \right)}_\text{miscible interface}
  \underbrace{\left( k g (\rho_2 - \rho_3) + \sigma k^3 - \omega_0^2 B_{22} \right)}_\text{free surface} - \omega_0^4 B_{12} B_{21} = 0.
\end{align}
namely
\begin{align}
  \label{eqn:B1.7}
  \omega_0^2 &= \frac{1}{2} \left[
    (\omega^2_\mix + \omega^2_\surf) \pm \sqrt{(\omega^2_\mix - \omega^2_\surf)^2 + 4\epsilon\omega^2_\mix\omega^2_\surf} 
  \right]
\end{align}
where $\omega_\mix$ and $\omega_\surf$ are the natural frequencies of each interface,
\begin{align}
  \label{eqn:B1.8}
  \omega^2_\mix = \left[\frac{B_{11}B_{22}}{\det{\mathsfbi{B}}}\right]\frac{kg(\rho_1-\rho_2)}{B_{11}}, && 
  \omega^2_\surf = \left[\frac{B_{11}B_{22}}{\det{\mathsfbi{B}}}\right]\frac{kg(\rho_2-\rho_3)+ \sigma k^3}{B_{22}},
\end{align}
and 
\begin{align}
  \label{eqn:B1.9}
  \epsilon = \frac{B_{12}B_{21}}{B_{11}B_{22}}
\end{align}
quantifies the coupling between the two interfaces. When $\epsilon \ll 1$, as
is generally the case in our experiments, the two roots reduce to $\omega_\mix$
and $\omega_\surf$, respectively.

Figure \ref{fig:appendix:B:1}a presents the roots of \eqref{eqn:B1.6} as a
function of \(k\) for \(H \gg h_\init\). Due to the small \(\atwood\), the
upper branch $\omega_0^+ \approx \omega_\surf$ is essentially independent of
$h_\init$, while the lower branch $\omega_0^-$ deviates from $\omega_\mix$ as
\(h_\init\) decreases. By comparison, varying $L$ over the experimentally
relevant range has a much weaker effect on either frequency (figure
\ref{fig:appendix:B:1}b).

\subsubsection{Influence of stratification depth on the linear stability}
\label{section:appendix:stability:3}

Since the forcing is periodic with period \(T=2\pi/\omega\), we may use the
Floquet--Lyapunov theorem to trace the stability chart of \eqref{eqn:B1.2}.
Solutions have the form
\begin{align}
  \label{eqn:B1.10}
  \vec{\alpha}(t) = \sum_{n=-\infty}^\infty e^{\mu_n t} \hat{\vec{\alpha}}_{n}(t)
\end{align}
where \(\mu_n = \gamma + i(\nu + n\omega)\) is the Floquet exponent,
\(\lambda_n = e^{\mu_n T}\) the corresponding multiplier, $\gamma$ the growth
rate, and $\nu$ the base oscillating frequency. Two values are considered,
\(\nu = 0\) (harmonic) and \(\nu = \omega/2\) (sub-harmonic), corresponding to
positive and negative Floquet multipliers \citep{Kumar_1994}. Introducing
\eqref{eqn:B1.10} into \eqref{eqn:B1.2} yields a \(2 \times 2\) problem for
each Fourier mode:
\begin{align}
  \label{eqn:B1.11}
  (\mu_n^2 \mathsfbi{B} + 2\omega \mu_n \mathsfbi{D} + \mathsfbi{M}) \vec{\hat{\alpha}}_n
  + \tfrac{1}{2}F\mathsfbi{M} (\vec{\hat{\alpha}}_{n-1}+\vec{\hat{\alpha}}_{n+1}) = 0.
\end{align}
Retaining the first \(N\) Fourier modes, \eqref{eqn:B1.11} becomes a
\(2N \times 2N\) eigenvalue problem that takes \(k\) and \(\gamma\) as inputs
and returns \(F\) (see, for instance, \cite{Benjamin1954} and
\cite{Huang_2024}).

\begin{figure}
  \begin{center}
    (a) \hfill (c) \hfill~\\
    \includegraphics[width=\linewidth]{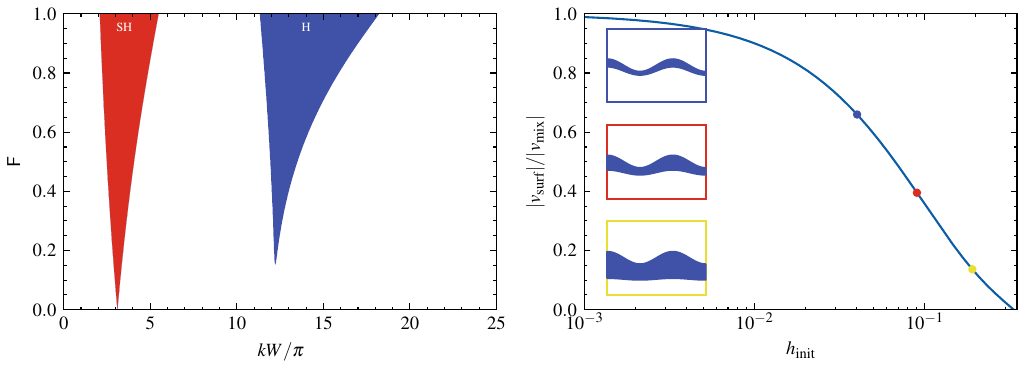}        

    (b) \hfill (d) \hfill~\\
    \includegraphics[width=\linewidth]{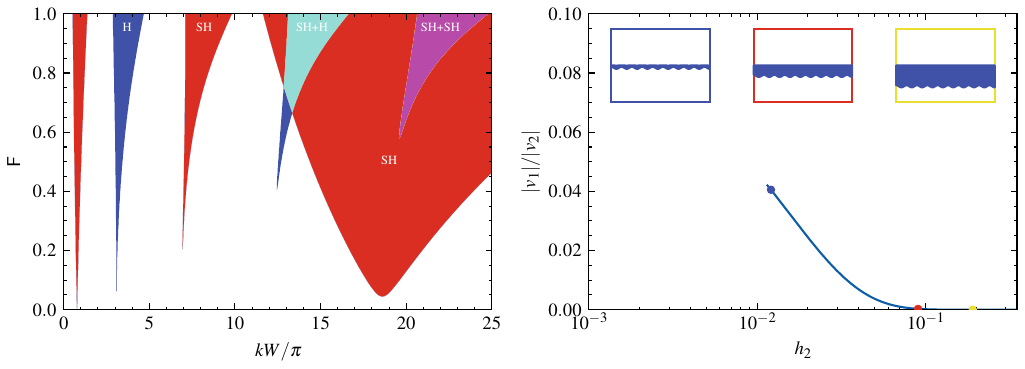}        

  \end{center}
  \caption{
  Linear stability of a three-layer system with \(H = 35~\cm\), matching the
  experimental container height. Left panels show stability charts for
  \(h_\init = 10~\unit{\centi\meter}\); right panels show the relative amplitude
  of the eigenmodes for \(\paramF = 0.5\). (Inset: schematic of the interface
  shapes obtained from the unstable eigenmodes \(v_j\)). Top row: forcing
  frequency \(\omega = 20~\unit{\radian\per\second}\). Bottom row: half the
  forcing frequency, \(\omega/2\).
  }
  \label{fig:appendix:B:2}
\end{figure}

The Floquet multipliers may also be evaluated by numerical integration (see,
for instance, \cite{Kovacic2018}). Introducing \(\vec{\beta}=\vec{\dot{\alpha}}\),
equation \eqref{eqn:B1.2} becomes a system of first-order ODEs
\begin{align}
  \label{eqn:B1.12}
  \dot{\vec{q}} = 
  \begin{bmatrix}
    \vec{\beta} \\ -\mathsfbi{B}^{-1}\mathsfbi{D}\vec{\beta} - (1 + F\cos(\omega t)) \mathsfbi{B}^{-1}\mathsfbi{M}\vec{\alpha}
  \end{bmatrix}
\end{align}
where \(\vec{q}=(\vec{\alpha}, \vec{\beta})^T\) is the state vector.
Integrating \eqref{eqn:B1.12} over a period \(T\) from four different initial
conditions gives a set of trajectories that form a \(4\times4\) monodromy
matrix
\begin{align}
  \label{eqn:B1.13}
  \mathbf{\Phi}(T)=[\vec{q}_{1}(T), \vec{q}_{2}(T),\vec{q}_{3}(T), \vec{q}_{4}(T)].
\end{align}
which maps the state vector from time $t$ to \(t+T\),
\begin{align}
  \label{eqn:B1.14}
  \vec{q}(t+T) = \mathbf{\Phi}(T) \vec{q}(t) = \vec{V}\vec{\Lambda}\vec{V}^{-1} \vec{q}(t),
\end{align}
where the diagonal matrix \(\vec{\Lambda}\) contains the Floquet multipliers
\(\vec{\lambda}=(\lambda_1, \ldots, \lambda_4)\) and \(\vec{V}\) the
eigenvectors of \(\mathbf{\Phi}(T)\).

Figures \ref{fig:appendix:B:2}a,b present typical stability charts for
$\omega=20$ and $10$ \rad, respectively. Blue and red regions have one unstable
mode, indicating primary harmonic and sub-harmonic parametric resonance;
magenta and cyan regions have two, indicating a superposition of two primary
resonances. For primary resonances, the unstable eigenvector is real and gives
the relative displacement of the two interfaces. In figure
\ref{fig:appendix:B:2}c, the free surface is directly excited and, depending on
$h_\init$, may also drive the miscible interface; in figure
\ref{fig:appendix:B:2}d, the miscible interface is directly excited but barely
affects the free surface because of the small \(\atwood\).

\subsection{Weakly nonlinear theory}
\label{section:appendix:amplitude}

Unlike linear theory, where amplitudes increase indefinitely, nonlinear effects
lead to saturation~\citep{Ockendon1973, Miles1984, Nagata1989, Tsai_1990,
Douady_1990, Nesterov_1995}. Here, we use weakly nonlinear theory to derive the
saturation amplitude of a decoupled interface, which, combined with a nonlinear
wave model, gives the shape of the saturated surface wave used in
\S\ref{section:validation} to cross-validate simulations and experiments.

\subsubsection{Mathieu equation with cubic non-linearities}
\label{appendixB:mathieu:cubic1}

Introducing a nonlinear dispersion relation of the
form \(\omega'_0 = \omega_0 (1 + \frac{1}{2} K (k\alpha)^2)\) into the
damped Mathieu equation gives
\begin{align}
  \label{eqn:B2.1}
  \ddot{\alpha} + \gamma \dot{\alpha} + \omega_0^2 (1 + \tfrac{1}{2} K (k\alpha)^2)^2 (1 + F\cos(\omega t))\alpha = 0
\end{align}
where \(\alpha\) is the wave amplitude, dots denote time derivatives,
\(\gamma\) is the linear damping coefficient (distinct from the Floquet
exponent's real part of the same name in
\S\ref{section:appendix:stability:3}), and \(K\) a nonlinear detuning
parameter that depends on the fluid depth, here taken from
\cite{Tadjbakhsh1960}:
\begin{align}
  \label{eqn:B2.2}
  K = \frac{9s^{-4}-12s^{-2}-3-2s}{64}, \quad s\equiv \tanh(kH).
\end{align}
For \(kH \ll 1\), \(K\) tends to \({9}/{[64(kH)^4]}\), while \(K=-1/8\) for
\(kH \gg 1\). The experiments are close to the deep-water limit, with
\(K=-0.124\).

\subsubsection{Solution using the method of multiple scales}
\label{appendixB:mathieu:multiscales}

We now apply the method of multiple scales~\citep{Kevorkian_1996, Bender_1999}
to describe the slow amplitude modulation during the stationary and attenuation
phases. In terms of \(\tau=\omega t\), equation \eqref{eqn:B2.1} reads
\begin{align}  
  \label{eqn:B2.3}
  \frac{d^2\alpha}{d\tau^2} 
  + \frac{2\gamma}{\omega}\frac{d\alpha}{d\tau} 
  + \frac{\omega_0^2}{\omega^2} (1 + K (k\alpha)^2 + \cdots) (1 + F\cos(\tau))\alpha = 0.
\end{align}

In standard fashion, we treat the damping, forcing and nonlinear terms as
small through an ordering parameter \(\epsilon\) (unrelated to the interfacial
coupling ratio of the same name in \eqref{eqn:B1.9}),
\begin{align}
  \label{eqn:B2.4}
  \frac{\gamma}{\omega} = \epsilon\mu, \quad
  \frac{\omega_0^2}{\omega^2} K = \epsilon\gamma, \quad
  \frac{\omega_0^2}{\omega^2} F = 2\epsilon f
\end{align}
where \(\mu\), \(\gamma\), and \(f\) are order-one rescaled coefficients (here
$\gamma$ takes a third, again unrelated, meaning from the linear damping
coefficient of \eqref{eqn:B2.1}). We focus on
the sub-harmonic response near resonance, with a small detuning \(\detune =
\epsilon\delta\) such that ${\omega_0^2}/{\omega^2} = \frac{1}{4} +
\epsilon\delta$. Equation \eqref{eqn:B2.3} then becomes
\begin{align}  
  \label{eqn:B2.5}
  \frac{d^2\alpha}{d\tau^2}   
  + \frac{1}{4} \alpha     
  =
  -\epsilon\left\lbrace 
    2\mu\frac{d\alpha}{d\tau} 
    + \gamma k^2 \alpha^3
    + \delta\alpha   
    + 2f\cos(\tau)\alpha       
  \right\rbrace 
  + O(\epsilon^2)
\end{align}
with initial conditions $\alpha(0)=R_0$ and $\dot{\alpha}(0)=0$, where $R_0$ is
some constant.

We seek solutions of the form
\begin{align}
  \label{eqn:B2.6}
  \alpha = \alpha_0 + \epsilon \alpha_1 + \cdots
\end{align}
which depend on the fast and slow variables
\begin{subequations}
  \begin{align}
    \label{eqn:B2.7a}
    \tf &= \tau
    \\
    \label{eqn:B2.7b}
    \ts &= \epsilon \tau.
  \end{align}
\end{subequations}
Substituting and equating coefficients of powers of $\epsilon$ gives a system
of linear differential equations,
\begin{subequations}
  \begin{align}
    \label{eqn:B2.8a}
    \text{order $\epsilon^0$}:&& \ddfast{\alpha_0} + \frac{1}{4}\alpha_0 &= 0
    \\
    \label{eqn:B2.8b}
    \text{order $\epsilon^1$}:&& \ddfast{\alpha_1} + \frac{1}{4}\alpha_1 &= 
    -\left\lbrace
      2\dmixed{\alpha_0} 
      + 2\mu\dfast{\alpha_0} 
      + \gamma k^2 \alpha_0^3
      + \delta \alpha_0  
      + f\left( e^{i\tf} + e^{-i\tf}\right)\alpha_0
    \right\rbrace
  \end{align}
\end{subequations}
and so on. The initial conditions become $\alpha_0(0)=R_0$, $\alpha_n(0)=0$
for $n\geq1$, and
\begin{subequations}
  \begin{align}
    \label{eqn:B2.9a}
    \text{order $\epsilon^0$}:&& \dfast{\alpha_0}(0) &=  0, &&
    \\
    \label{eqn:B2.9b}
    \text{order $\epsilon^1$}:&& \dfast{\alpha_1}(0) &= -\dslow{\alpha_0}(0). &&
  \end{align}
\end{subequations}

The first of these describes free oscillations, the remainder forced linear
oscillations. Solutions at order $\epsilon^0$ are slowly modulated in amplitude
and phase,
\begin{align}
  \label{eqn:B2.10}
  \alpha_0(\ts,\tf) = A(\ts) e^{i\tf/2} + \conj{A} (\ts) e^{-i\tf/2}
\end{align}
where $\conj{}$ denotes the complex conjugate.

Substituting $\alpha_0$ into the order-$\epsilon^1$ equation gives the
realizability condition
\begin{align}
  \label{eqn:B2.11}
  \Dslow{A} + (\mu - i\delta) A
  - 3 i \gamma k^2\vert A \vert^2 A - i f \conj{A} = 0
\end{align}
Writing $A = R(\ts) e^{i\theta(\ts)}$ in polar form gives the coupled system
\begin{subequations}
\begin{align}
  \label{eqn:B2.12a}
  \Dslow{R} &= f R\sin(2\theta) -\mu R
  \\
  \label{eqn:B2.12b}
  R\Dslow{\theta} &= 
  f R\cos(2\theta) +  \delta R + 3\gamma k^2 R^3
\end{align}
\end{subequations}
Steady-state solutions exist if
\begin{align}
  \label{eqn:B2.13}
  3 \gamma (k R)^2 = - \delta \mp \sqrt{f^2 - \mu^2},
  \quad
  \sin(2\theta) = \frac{\mu}{f}
\end{align}
During the attenuation phase, \(f=0\), with initial conditions $R(0)=R_0/2$
and $\theta(0)=0$, the amplitude decays exponentially:
\begin{subequations}
\begin{align}
  \label{eqn:B2.14a}
  R(t) &= \frac{1}{2}R_0 e^{-\mu\ts},
  \\
  \label{eqn:B2.14b}
  \theta(t) &= \delta\ts + \frac{3\gamma (k R_0)^2}{8\mu} \left( 1 - e^{-2\mu\ts} \right),
\end{align}
\end{subequations}

\begin{figure}
  \quad (a) \hfill (b)\hfill~ \\
  \includegraphics[width=0.5\linewidth]{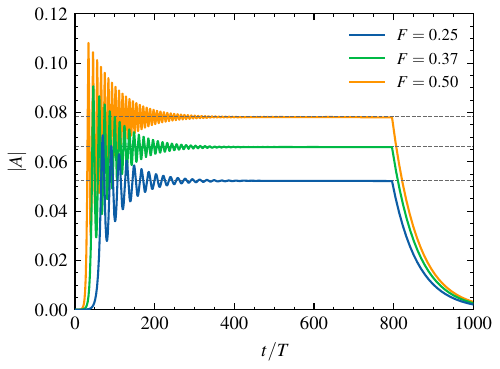}
  \includegraphics[width=0.5\linewidth]{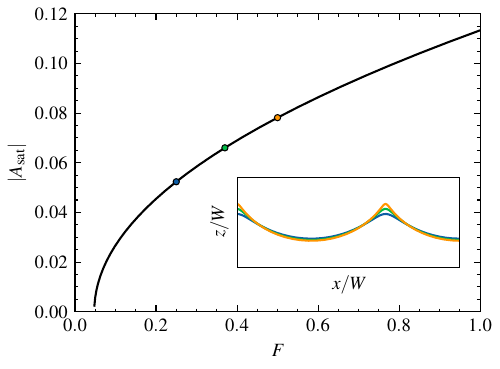}

  \caption{
    Numerical integration of the slow amplitude equations
    \eqref{eqn:B2.12a}--\eqref{eqn:B2.12b} with the measured damping rate
    \(\gamma_\surf\) and \(K\) from \eqref{eqn:B2.2}.
    (a) Time evolution of the slow amplitude \(\vert A \vert\) for different
    forcing amplitudes \(F\); dash-dotted lines indicate the saturation
    amplitude \(\vert A_\steady \vert\) from \eqref{eqn:3.8}.
    (b) \(\vert A_\steady \vert\) as a function of \(F\); symbols mark the
    cases shown in (a). Inset: free surface obtained from the nonlinear wave
    model \eqref{eqn:3.5a}--\eqref{eqn:3.5b} using \(\vert A_\steady \vert\).
  }
  \label{FigureB3:model}
\end{figure}


Figure \ref{FigureB3:model}a shows the numerical integration of
\eqref{eqn:B2.12a}--\eqref{eqn:B2.12b} for different forcing amplitudes \(F\):
the slow amplitude grows, saturates at the value \(\vert A_\steady \vert\)
predicted by \eqref{eqn:3.8}, and decays exponentially once the forcing stops.
Figure \ref{FigureB3:model}b presents the resulting saturation amplitude as a
function of \(F\), together with the free surface reconstructed from the
nonlinear wave model \eqref{eqn:3.5a}--\eqref{eqn:3.5b}. While the model
captures the linear growth, the saturation amplitude, and the exponential
decay, it remains qualitative: the simple cubic nonlinearity neglects
mode-to-mode and surface--interface interactions, so it is not predictive
during the transient phase, where the modulated amplitude shows considerable
variability in the experiments.

\subsubsection{Mathieu-Duffing equation with cubic non-linearities}
\label{appendixB:mathieu:cubic2}

A different type of cubic nonlinearity arises in the Mathieu--Duffing equation
\begin{align}
  \label{eqn:B2.15}
  \frac{d^2\alpha}{d\tau^2}
  + \frac{1}{4} \alpha     
  =
  -\epsilon\left\lbrace 
    2\mu\frac{d\alpha}{d\tau} 
    + \nu k^2 \left(\frac{d\alpha}{d\tau}\right)^3
    + \delta\alpha   
    + 2f\cos(\tau)\alpha       
  \right\rbrace 
  + O(\epsilon^2)
\end{align}
where \(\nu\) is a cubic damping coefficient (unrelated to the Floquet base
frequency of the same name in \S\ref{section:appendix:stability:3}). A similar development leads to
the realizability condition
\begin{align}
  \label{eqn:B2.16}
  \Dslow{A} + (\mu - i\delta) A
  + \tfrac{3}{8} \nu k^2 \vert A \vert^2 A - i f\conj{A} = 0
\end{align}
and the coupled system
\begin{subequations}
\begin{align}
  \label{eqn:B2.17a}
  \Dslow{R} &= f R\sin(2\theta)
  -\mu R - \tfrac{3}{8}\nu k^2 R^3
  \\
  \label{eqn:B2.17b}
  R\Dslow{\theta} &=
  f R\cos(2\theta) +  \delta R.
\end{align}
\end{subequations}
Steady-state solutions now exist if
\begin{align}
  \label{eqn:B2.18}
  \tfrac{3}{8} \nu k^2 R^2 = -\mu \pm \sqrt{f^2 - \delta^2},
  \quad
  \cos(2\theta) = -\frac{\delta}{f}
\end{align}
while during the attenuation phase, in the limit $\mu\to0$, the amplitude
follows the classical Duffing free decay
\begin{subequations}
\begin{align}
  \label{eqn:B2.19a}
  R(t) &= R_0\left(4 + \tfrac{3}{4} \nu k^2 R_0^2 \ts \right)^{-1/2}
  \\
  \label{eqn:B2.19b}
  \theta(t) &= \delta \ts
\end{align}
\end{subequations}
\clearpage
\section{Details on the Modal Decomposition }
\label{section:appendix:POD}
\setcounter{figure}{0}

Proper Orthogonal Decomposition (POD) is used to decompose the flow fields and
interfaces into a set of orthogonal functions. POD is optimal in the sense that
it captures the maximum variance of the data with the fewest
modes~\citep{Holmes2012}. Throughout this work, we used the POD for different
reasons. In \S\ref{section:surface:experiments}, we used it to remove
discontinuities from the reconstructed interfaces and filter out random noise
introduced by the segmentation method. Since POD modes are mutually
uncorrelated, groupings of Fourier harmonics appearing together in a single mode
are statistically coupled, revealing the spatial and temporal harmonic structure
of the interface motion. Additionally, in
\S\ref{section:surface:validation:energy}, we used a POD-based approach to
decompose the velocity and concentration fields into mean, oscillating, and
fluctuating parts.

Some of these analyses were applied to both datasets, while others could only
be applied to one. Whenever a result could be established equally well with
either dataset, we present a single one for concision: the POD of the
interfaces, for instance, was performed on both the experimental and the
numerical data, but since the numerical modes provided no additional insight,
only the experimental ones are shown. The decomposition of the concentration
field, by contrast, is applied only to the DNS. Extracting concentration
fields from the experimental images requires a calibration based on the
intensity of the colorant, which assumes the Beer--Lambert law to hold across
the entire tank width. Close to the surface, however, the three-dimensional
sloshing motion means that the fluid does not always span the full width,
introducing an additional unknown attenuation for which we could not
compensate. For this reason, the interface positions are the only quantities
extracted from the experimental images.

\subsection{POD Methodology}
\label{section:appendix:POD:methodology}

In a domain with periodic boundary conditions, POD modes reduce to Fourier
modes; in our reflecting-wall tank they reduce to a sine/cosine series. 

We express a quantity $q(x,t)$ in terms of a set of orthogonal modes
$\phi^{(n)}$ and corresponding modal coefficients $\alpha^{(n)}$,
\begin{align}
  \label{eq4.1}
  q(x,t) = q_0(t) + \sum_{n=1}^{\infty} \alpha^{(n)} (t) \cdot \phi^{(n)} (x)
\end{align}
where $q_0(t)$ represents the temporal drift of $q(x,t)$. These modes are ranked
by energy and correspond to solutions of the eigenvalue problem:
\begin{align}
  \label{eq4.2}
  \int \mathsfbi{R}_j(x,x') \phi^{(n)}(x')~dx' = \lambda^{(n)} \phi^{(n)}(x)
\end{align}
where 
\begin{align}
  \label{eq4.3}
 \mathsfbi{R}_j(x,x')\equiv (1/M)\sum_{m=1}^{M} q(x,t_m)q(x',t_m) 
\end{align}
is the spatial autocorrelation tensor computed from  \(M\)
snapshots~\citep{Sirovich1987}.

By construction, the modal amplitudes have zero mean, are mutually uncorrelated,
and have a variance equal to $\lambda^{(n)}$. The eigenvalue problem is solved
using the `scikit-learn' and `LAPACK' libraries. We use only snapshots in the
stationary regime and we augment each dataset by applying the reflection
symmetry \(\eg q(x,t) \to q(-x,t)\), which enforces the natural left-right
symmetry of the tank on the resulting modes.

\subsection{Decomposition of the interfaces $\zeta_{\surf}$ and $\zeta_{\mix}$}
\label{section:appendix:POD:surface}

\begin{figure}   
	\begin{minipage}[t]{0.33\linewidth}
		(a) \hfill~\\
		\includegraphics[width=\linewidth]{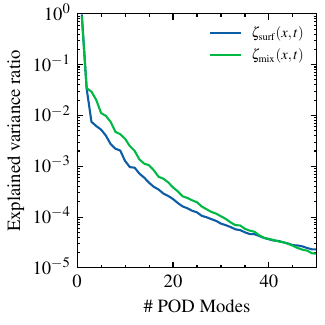}
	\end{minipage}
	\begin{minipage}[t]{0.66\linewidth}
		(b) \hfill (c) \hfill~\\
		\includegraphics[width=0.5\linewidth]{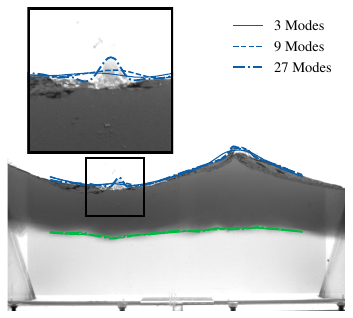}
		\includegraphics[width=0.5\linewidth]{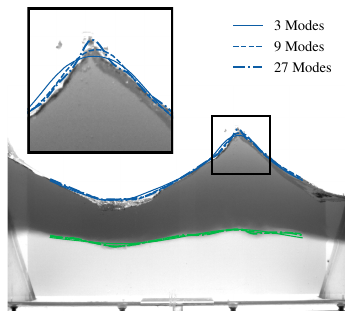}
	\end{minipage}
	\caption{
		Proper Orthogonal Decomposition (POD) for the case EXP-F49-H30
		previously shown in Figure~\ref{Figure3:Hovmoller}. (a) Explained
		variance ratio as a function of the number of modes. Subfigures (b) and
		(c) show a comparison between the recorded images at two representative
		instants and the corresponding low-order reconstruction of the
		interfaces obtained using 3, 9, and 27 modes, respectively.
	}
	\label{FigureC1:POD}
\end{figure}

We apply POD separately to the free surface ($q = \zeta_\surf$, $q_0 = 0$) and
the miscible interface ($q = \zeta_\mix$, $q_0 = h_\mixed$), yielding two
distinct sets of modes identified by the subscripts $\surf$ and $\mix$,
respectively. Treating both interfaces simultaneously is possible but tends to
obscure the resonant modes of the miscible interface, which are less energetic
and not directly coupled to the free surface.

Both interfaces are well-suited for dimensionality reduction: the first 10 modes
capture over 95\% of the total variance (Figure~\ref{FigureC1:POD}a). As shown in
figures~\ref{FigureC1:POD}b-c, a small number of modes suffices to qualitatively
describe the both interfaces, although higher-order modes are required to
correctly describe sharp-crested or splashing waves.

\begin{figure}   
	(a) \nth{1} surface mode \hfill (b) \nth{2} surface mode \hfill (c) \nth{3} surface mode \hfill~\\
	\includegraphics[width=\linewidth]{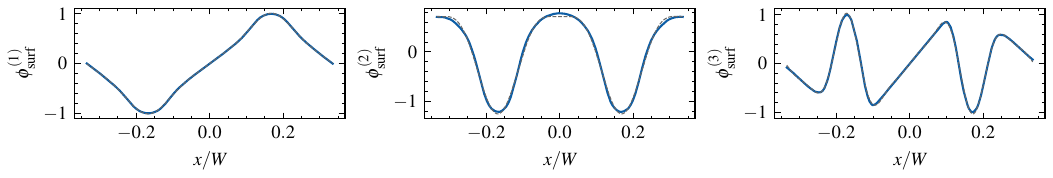}

	(d) \nth{1} interface mode \hfill (e) \nth{2} interface mode \hfill (f) \nth{4} interface mode \hfill~\\
	\includegraphics[width=\linewidth]{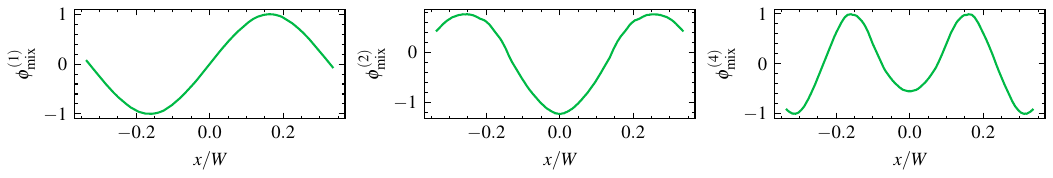}

	\caption{
		Spatial eigenmodes for the case EXP-F49-H30 previously shown in
		Figure~\ref{Figure3:Hovmoller}. Top row: leading POD modes of the
		free surface. Bottom row: leading POD modes of the miscible interface.
	}
	\label{FigureC2:POD:modes}
\end{figure}


The leading POD modes obtained for the case EXP-F49-H30 are shown in
Figure~\ref{FigureC2:POD:modes}. In general, for the same value of $\paramF$,
free surface modes are very similar, and a single common base can be used. This
allows for a more direct comparison between cases, even if the common basis is
not necessarily optimal. The \nth{1} surface mode contains 90\% of the total
variance and conforms well to a Fourier sine series with wavenumbers that are
odd multiples of \(k = 3\pi/W\),
\begin{align}
  \phi^{(1)}_\surf(x) = A_1\sin(kx) + A_3\sin(3kx) + \dots
\end{align} 
while the \nth{2} surface mode conforms well to a Fourier cosine series with
wavenumbers that are even multiples of \(k\) 
\begin{align}
  \phi^{(2)}_\surf = B_0 + B_2\cos(2kx) + B_4\cos(4kx) + \dots. 
\end{align} 
In practice, $\phi^{(2)}_\surf$ corresponds to the spatial harmonic of
$\phi^{(1)}_\surf$, which breaks the top/bottom symmetry and gives the
characteristic cnoidal profile to the primary wave. Higher-order modes
correspond to progressively smaller wavelengths centred around the wave
anti-nodes, capturing features such as wave crests.

Modes of the miscible interface are somewhat different. The \nth{1} (resp.
\nth{2}) interface modes fit well to a sine (resp. cosine) series with the same
wavenumbers as the \nth{1} surface mode. As discussed in
\S\ref{section:surface:experiments}, the miscible interface behaves as a
harmonic oscillator forced by the free surface: the \nth{1} interface mode is
directly forced by the \nth{1} surface mode, while the \nth{1} and \nth{2}
interface modes together generate a travelling wave propagating horizontally.
The \nth{3} and \nth{4} interface modes form another pair with the same
wavenumbers as the \nth{2} surface mode, while additional pairs, represent the
response to higher-order surface modes.

\subsection{Decomposition of fields into large- and small-scales}
\label{section:appendix:decomposition}

\subsubsection{Decomposition of $\vec{u}(\vec{x},t)$ into fluctuating and
oscillating components}

\begin{figure}   
  \begin{center}
    (a) \hfill (b) \hfill (c) \hfill~\\
    \includegraphics[width=0.33\linewidth]{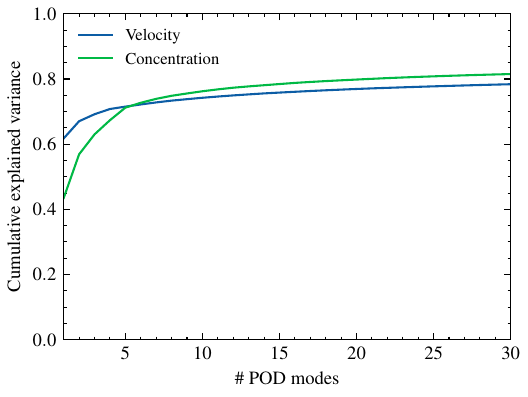}\hfill
    \includegraphics[width=0.33\linewidth]{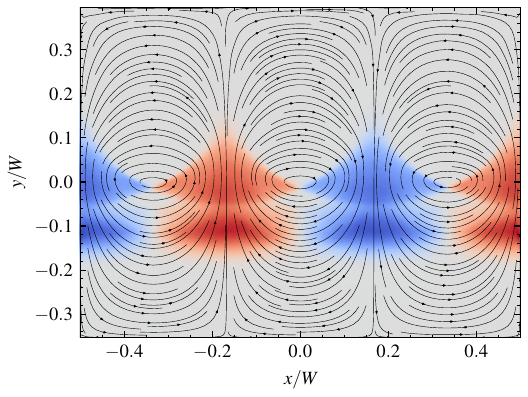}\hfill
    \includegraphics[width=0.33\linewidth]{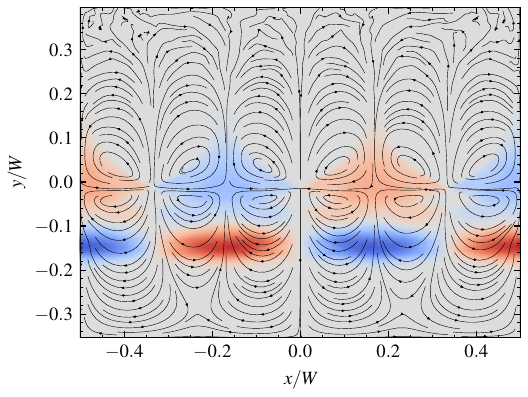}
  \end{center}
  \caption{
    POD of the velocity and concentration fields used for the scale
    decomposition of the DNS-F49-H30 data. Subfigure (a) shows the cumulative
    explained variances as a function of the number of POD modes; Subfigures (b)
    and (c) show the spatial structure of the first and second POD modes for the
    concentration field (in colours) and vertical velocity field (in
    streamlines).
  }
  \label{FigureC3:scales}
\end{figure}

\begin{figure}   
  \begin{center}
    (a) \hfill (b) \hfill (c) \hfill~\\
    \includegraphics[width=0.33\linewidth]{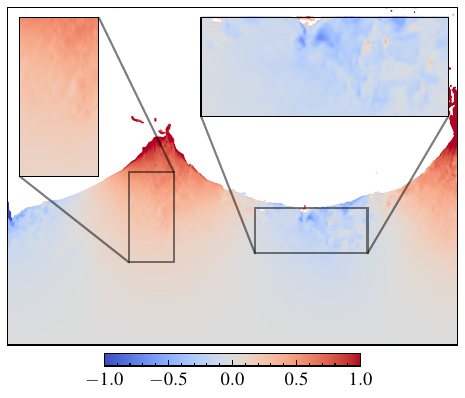}\hfill
    \includegraphics[width=0.33\linewidth]{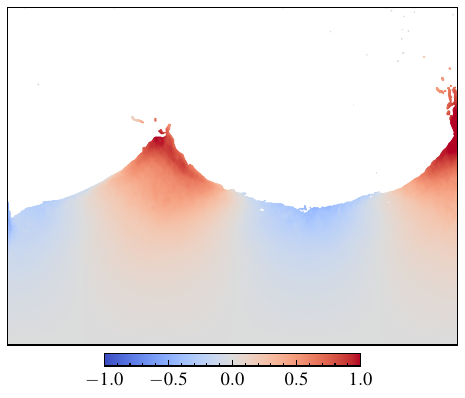}\hfill
    \includegraphics[width=0.33\linewidth]{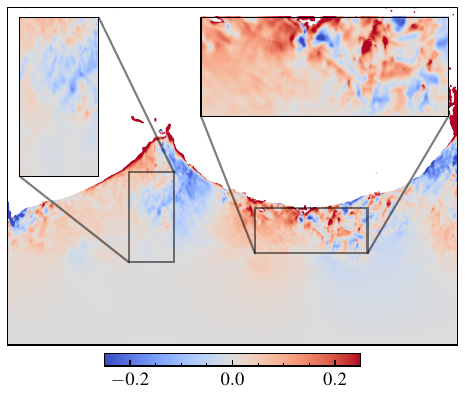}

    (d) \hfill (e) \hfill (f) \hfill~\\
    \includegraphics[width=0.33\linewidth]{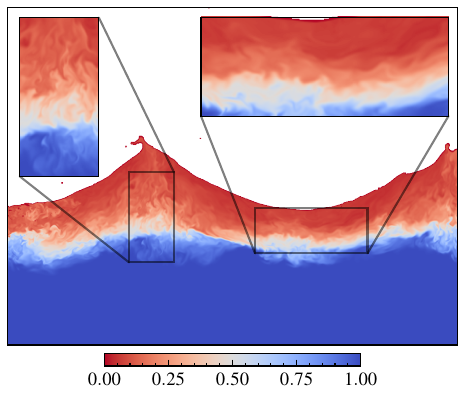}\hfill
    \includegraphics[width=0.33\linewidth]{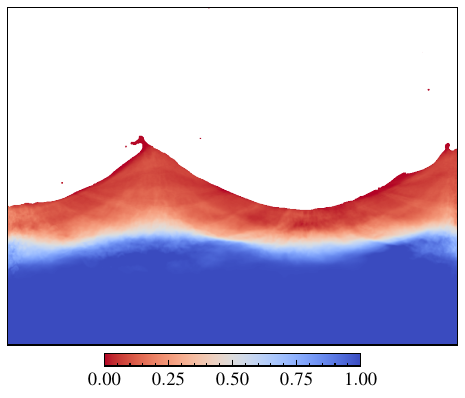}\hfill
    \includegraphics[width=0.33\linewidth]{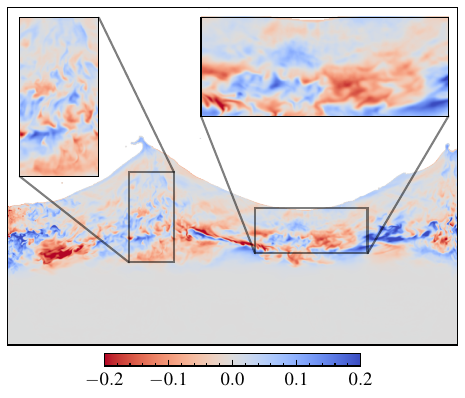}
  \end{center}
  \caption{
    Top row shows an example of the decomposition of the vertical velocity field
    (a) into large- (b) and small-scales (c) based on POD analysis. Bottom row
    shows a similar decomposition of the concentration field (d) into large- (e)
    and small-scales (f).
  }
  \label{FigureC4:scales}
\end{figure}

In \S\ref{section:surface:validation:energy}, we decomposed the velocity and
concentration fields into small- and large-scale components using POD. We
decompose the velocity field as
\begin{align}
  \vec{u}(\vec{x},t) &= \vec{\widetilde{u}}(\vec{x},t) + \vec{u}'(\vec{x},t),
\end{align}
where $\vec{\widetilde{u}}$ captures the large-scale oscillating motions induced
by the Faraday waves, $\vec{u}'$ the turbulent fluctuations, and the
time-averaged velocity is assumed negligible.

To obtain $\vec{\widetilde{u}}(\vec{x},t)$, we perform a POD of the full
velocity field $\vec{u}(\vec{x},t)$. The first few modes capture about 60\% of
the $L^2$-norm and the first 10 modes about 75\%
(figure~\ref{FigureC3:scales}a). The leading mode resembles the irrotational
velocity field of linear Faraday wave theory, and the second mode is its spatial
and temporal harmonic (figures~\ref{FigureC3:scales}b-c). As this is an ad-hoc
decomposition, we choose to reconstruct $\vec{\widetilde{u}}(\vec{x},t)$ using
the first $N=10$ POD modes that capture the wave-driven oscillation. We also
verified that the viscous dissipation associated with
$\vec{\widetilde{u}}(\vec{x},t)$ is negligible compared to the total viscous
dissipation. A typical result of this decomposition is shown in
figure~\ref{FigureC4:scales}a-c.

\subsubsection{Decomposition of $c(\vec{x},t)$ into slow-time, oscillating,
and fluctuating components}

We similarly decompose the concentration field into slow-time, oscillating, and
fluctuating components,
\begin{align}
  c(\vec{x},t) &= \overline{c}(\vec{x},t) + \widetilde{c}(\vec{x},t) + c'(\vec{x},t).
\end{align}
Here, $\overline{c}(\vec{x})$ represents the slow evolution of the mean
stratification and is expected to depend only on the vertical coordinate,
$\widetilde{c}(\vec{x},t)$ represents the oscillating component due to the advection
by $\vec{\widetilde{u}}(\vec{x},t)$, while $c'(\vec{x},t)$ represents the
small-scale fluctuations due to turbulent mixing. 

To obtain this decomposition, we use short-time averaging to separate the
slow-time component, $\overline{c}(\vec{x},t)$, from the oscillating and fluctuating
components, $\widetilde{c}(\vec{x},t) + c'(\vec{x},t)$. Then, we apply the POD to
the remainder to separate the oscillating component, $\widetilde{c}(\vec{x},t)$,
from the fluctuating component, $c'(\vec{x},t)$. As before, we reconstruct
$\widetilde{c}$ using the first $N=10$ POD modes that capture the wave-driven
oscillation. A typical result is shown in figure~\ref{FigureC4:scales}d-f.

A similar decomposition could be achieved by splitting the velocity field into
rotational and irrotational parts, by applying a band-pass filter in the
frequency domain, or by phase-averaging. The POD-based approach, however,
provides a unified and straightforward way to isolate the wave-driven and
turbulent components for both the velocity and concentration fields.
\clearpage
\bibliographystyle{jfm}
\bibliography{faramix}

\end{document}